\documentclass{aastex62}

\def\ie{{\it i.e.}\,}
\def\eg{{\it e.g.}\,}
\def\drm{{\rm d}}

\newcommand{\vbf}       {\mbox{\boldmath$v$}}
\newcommand{\jbf}       {\mbox{\boldmath$j$}}
\newcommand{\Jbf}       {\mbox{\boldmath$J$}}

\newcommand{\Tbf}       {\mbox{\boldmath$T$}}
\newcommand{\Fbf}       {\mbox{\boldmath$F$}}
\newcommand{\Ebf}       {\mbox{\boldmath$E$}}
\newcommand{\Bbf}       {\mbox{\boldmath$B$}}
\newcommand{\bbf}       {\mbox{\boldmath$b$}}

\newcommand{\kbf}       {\mbox{\boldmath$k$}}
\newcommand{\Ubf}       {\mbox{\boldmath$U$}}

\newcommand{\omegabf}       {\mbox{\boldmath$\omega$}}
\newcommand{\xibf}       {\mbox{\boldmath$\xi$}}

\received{March 16, 2018}
\revised{January 7, 2018}
\accepted{\today}
\submitjournal{ApJ}

\shorttitle{Stability of neutron star cores}
\shortauthors{van Eysden and Link}

\begin{document}

\title{HYDRODYNAMIC STABILITY ANALYSIS OF THE NEUTRON STAR CORE}

\correspondingauthor{C. A. van Eysden}
\email{anthonyvaneysden@montana.edu}

\author{C. A. van Eysden}
\affil{Department of Physics \\ Montana State University \\ Bozeman,  MT
59717, USA}

\author{Bennett Link}
\affiliation{Department of Physics \\ Montana State University \\ Bozeman,  MT
59717, USA}



\begin{abstract}

Hydrodynamic instabilities and turbulence in neutron stars have been suggested to be related to observable spin variations in pulsars, such as spin glitches, timing noise, and precession (nutation). 
Accounting for the stabilizing effects of the stellar magnetic field, we revisit the issue of whether the inertial modes of a neutron star can become unstable when the neutron and proton condensates flow with respect to one another. 
The neutron and proton condensates are coupled through the motion of imperfectly pinned vorticity (vortex slippage) and vortex-mediated scattering (mutual friction). 
Two-stream instabilities that occur when the two condensates rotate with respect to one another in the outer core are stabilized by the toroidal component of the magnetic field.
This stabilization occurs when the Alfv\'en speed of the toroidal component of the magnetic field becomes larger than the relative rotational velocity of the condensates, corresponding to toroidal field strengths in excess of $\simeq 10^{10}\,{\rm G}$.  
In contrast with previous studies, we find that spin down of a neutron star under a steady torque is stable.  
The Donnelly--Glaberson instability is not stabilized by the magnetic field, and could play an important role if neutron stars undergo precession.

\end{abstract}



\section{Introduction} \label{sec1}

Pulsars exhibit two varieties of rotational irregularities that are expected to be related to the dynamics of the interior fluid: spin glitches and timing noise.  
Glitches are sudden increases in the rotational frequency $\nu$ of the pulsar, with fractional amplitudes spanning $10^{-11}<\Delta \nu/\nu<10^{-4}$ across the pulsar population (see \eg, \citealt{rad69,esp11}).
The glitch event is unresolved by radio timing data, with a current upper limit of $40\,{\rm s}$ obtained from the 2000 January glitch in the Vela pulsar \citep{dod02}.
Glitches are believed to arise from the global motion of superfluid vorticity in the neutron star crust that is caused by, \eg, a noisy creep process \citep{and75b}, thermal heating induced by star quakes \citep{lin96,lar02}, a self-organized critical process \citep{mel08,war08} or a coherent noise process \citep{mel09}.  
The subsequent glitch recovery occurs over timescales ranging from days to years \citep{mcc87,mcc90,fla90,won01,dod02} and is attributed to dynamical relaxation of the neutron superfluid of the inner crust \citep{alp84b,alp93,alp96,lin14} and of the neutron-proton superfluid mixture of the core \citep{bay69a,eas79a,alp84a,van10,van14,lin14}.

Distinct from glitches is {\sl timing noise}, the stochastic wander of pulse phase, frequency, and frequency derivative. 
This noise process might have many underlying causes and is thought to represent true variations in the star's spin rate \citep{boy72,cor80c,cor85,arz94,dal95,hob06b,hob10}. 
Possible contributing effects include variations in the external spin-down torque (\eg, \citealt{che87a,che87b,ura06,lyn10}), variable torques exerted on the crust by the multiple fluid components
\citep{alp86,jon90}, microglitches \citep{jan06}, and accretion \citep{qia03}.
Variations in the interstellar medium (\eg, \citealt{liu11}), could also play a role in timing noise.
More speculatively, timing noise may be connected with underlying superfluid turbulence, which could produce stochastic variations in the pulsar spin frequency by exerting a variable viscous torque on the rigid crust \citep{mel14}.

\citet{gre70} originally suggested that superfluid turbulence prevails in the core of a spinning-down neutron star. 
Various hydrodynamic instabilities that might lead to turbulence have been proposed as the cause of spin glitches and other timing irregularities.  
The outer core may be unstable to, \eg, a variant of the Kelvin-Helmholtz instability occurring at the interface between the $^1S_0$-- and $^3P_2$--paired neutron superfluids \citep{mas05}.
Two-stream instabilities in the interpenetrating neutron and proton condensates could be present in the rotating outer core, driven by Fermi-liquid interactions \citep{and04} and vortex-mediated processes \citep{gla09,and13}.
\citet{lin12b} argued that slow slippage of vortices induced by relative flow between the neutron superfluid and crust is inherently unstable.  
An analogous instability was identified in the core, driven by the relative motion between the neutron superfluid and the flux tube array \citep{lin12a}.
The Donnelly--Glaberson instability, studied in laboratory superfluid helium, is also expected to have a counterpart in neutron stars if the charged fluid component achieves a critical velocity along the rotation axis \citep{gla74,sid08}.
Such a flow would be produced by precession of the star \citep{gla08}.
\citet{mel12} has argued that if the inner core of the neutron star retains a high rotation rate from its birth, the outer core becomes susceptible to various instabilities in spherical Couette flow \citep{per06a,per08,per09}.

Connecting glitches and timing noise with turbulence in the outer core presents two immediate challenges. 
One challenge is to identify instabilities that can grow to produce a turbulent state. 
A second, and more serious, challenge is to demonstrate how the turbulent state begins and ends.  
Steadily driven classical hydrodynamic systems that become unstable develop a quasi-steady turbulent cascade without global transient behavior.
Some studies \citep{mas05,gla09} find instability growth times short enough to be consistent with the observed glitch rise time of $\lesssim 60$ s (Vela), but a description of how turbulence develops and produces a glitch has not been advanced.
Some studies find evidence that timing irregularities are consistent with a state of underlying turbulence in the outer core \citep{mel07,mel14}, but the origin of this turbulence needs to be rigorously assessed.

An interesting question is whether hydrodynamic instabilities are quenched by magnetic stresses.  
A general feature of magnetic equilibria is a twisted, tangled structure in which the toroidal field is greater than or equal to the poloidal field \citep{bra06,bra09}.  
\citet{van08} demonstrated that poloidal magnetic stresses have a stabilizing effect on a particular class of two-stream instabilities.

In this paper, we evaluate the stability of the relative flow between the interpenetrating neutron and proton fluids.  
Relative flow would arise naturally as the crust and charged components of the star are spun down by the magnetic dipole torque, but vortex pinning prevents the neutron superfluid from corotation with the charged components.
We consider pinning of neutron vortices to flux tubes in the outer core, accounting for slippage of the two lattices with respect to one another (imperfect pinning).
We study the stabilizing effects of the toroidal plus poloidal magnetic field and demonstrate that the magnetic field stabilizes the unstable inertial modes for toroidal magnetic fields greater than $10^{10}\,{\rm G}$.
We find that the instability of \citet{lin12a,lin12b} is not present.  
Instabilities generated by flows along the rotation axis may arise from, \eg, precession, for which we distinguish two instabilities.  
The two-stream instability identified by \citet{gla08} is stabilized by the magnetic field for wobble angles less than $0.1^{ \circ}$, as shown by \citet{van08}.
Under imperfect pinning, the Donnelly--Glaberson instability occurs, which remains present for arbitrary magnetic field strength and which may be excited for wobble angles as small as $10^{-7\; \circ}$.

The paper is structured as follows.
In \S\ref{sec2}, we review the magnetohydrodynamic (MHD) theory of neutron star cores.
We estimate the relevant hydrodynamic parameters in \S\ref{sec3}.
In \S\ref{sec4}, we study two-stream instabilities driven by mutual friction, for rotating fluids (\S\ref{sec4a}) and flows along the rotation axis (\S\ref{sec4b}).
Our conclusions are summarized in \S\ref{sec5}.

\section{Hydrodynamics of a superfluid mixture } \label{sec2}

The core of a neutron star is composed primarily of neutrons, with $\sim 5-10\%$ of the mass in protons; for the electrically neutral medium the number density of electrons is equal to that of the protons.
At the supra-nuclear densities of the outer core, the Fermi energy for protons and neutrons is well above the typical temperature of a mature neutron star, and both the neutrons and protons are expected to condense into BCS superfluids, with $^3PF_2$ and $^1S_0$ Cooper pairing respectively \citep{mig59,bay69a}.
To support rotation, the neutron superfluid forms an array of quantized vortices, filaments of microscopic cross section, each carrying one quantum of circulation.  
The superconductivity of the protons is predicted to be type II, and the magnetic field is supported by an array of quantized flux tubes, each carrying one quantum of magnetic flux.  
Fermi-liquid interactions between the two condensates results in a nondissipative coupling between the mass currents of the two species \citep{and75a,cha06}, so that the neutron vortices are magnetized by entrained proton currents \citep{alp84a}.  
Electron scattering from magnetized vortices and flux tubes produces dissipative and non-dissipative forces on the vortices and flux tubes.  
The magnetic interaction at junctions between magnetized neutron vortices and flux tubes is energetic enough to produce pinning, wherein the neutron vortices pin to the dense array of flux tubes in the outer core \citep{sri90,jon91,cha92,rud98,lin12a}, similar to the predicted pinning of the vortices to the nuclear lattice of the crust \citep{and75b,alp77,eps88,don06,avo07,lin09}.
Thermal fluctuations stochastically excite vortex motion, causing the neutron vortices to slip with respect to the flux tubes \citep{din93,sid09b,lin14}.

In this section, we present the governing MHD equations describing the outer core of a neutron star.
In \S\ref{sec2a}, we describe the equations relevant for this study of unstable inertial modes in the outer core.
The perturbations of the equations about rotational equilibrium are presented in \S\ref{sec2b}.

\subsection{Hydrodynamic treatment} \label{sec2a}

To study the stability of flows much larger than the intervortex spacing $d_n$, it is convenient to perform a smooth-averaging of many vortex lines or flux tubes over scales much larger than $d_n$ \citep{hal56a,hal56b,hal60,kha65,hil77,bay83,cha86,men91a,men91b,gla11}. 
Over length scales that exceed $d_n$, the smooth-averaged vorticity of a rotating neutron condensate is 
\begin{eqnarray}
 \omegabf_n&=&n_{vn} \kappa \, \hat{\omegabf}_n=\nabla \times \vbf_n\,, \label{eq1}
\end{eqnarray}
where $\kappa=\pi \hbar/m$ is the quantum of circulation for neutrons of mass $m$, $n_{vn}$ is the areal density of vortex lines, $\hat{\omegabf}_n$ is the vorticity unit vector directed along the vortex lines, and $\vbf_n$ is the smooth-averaged velocity of the neutron superfluid.
The smooth-averaged magnetic field $\Bbf$ in a type II superconductor is 
\begin{eqnarray}
 \Bbf&=&n_{vp} \phi_0 \, \hat{\bbf} \,, \label{eq2}
\end{eqnarray}
where $\phi_0=\pi \hbar c/e= m c\kappa/e  $ is the quantum of magnetic flux, $n_{vp}$ is the areal density of flux tubes, and $\hat{\bbf}$ is the unit vector directed along the flux tubes.
In the outer core of a neutron star rotating at angular velocity $\Omega_n$ and with magnetic field $B_0$, the flux tubes far outnumber the vortex lines:
\begin{eqnarray}
  \frac{n_{vp}}{n_{vn}}\sim 8\times 10^{13} \left(\frac{B_0}{10^{12}\,{\rm G}}\right)\left(\frac{\Omega_n}{20\pi \, {\rm rad\,s^{-1}}}\right)^{-1}\,. \label{eq3}
\end{eqnarray}
Contributions to the magnetic field arising from the rotation of the proton and neutron condensates are of order $n_{vn}/n_{vp}\sim 10^{-14}$.
We neglect these small corrections.   

In this paper, we focus our attention on the stabilizing effects of the magnetic stresses on the inertial mode instabilities.  
We neglect buoyancy and compressibility restoring forces by assuming constant density flows, which gives
\begin{eqnarray}
  \nabla \cdot \vbf_x&=& 0 \,,  \label{eq4}
\end{eqnarray}
for $x=n,p$.
This assumption neglects g-modes and p-modes, which may be unstable in neutron star cores (see \eg, \citealt{and04,gus13,hab16,pas16}).
We do not study instabilities related to g-modes and p-modes in this paper, but refer the reader to the above works; we return to g-modes and p-modes in the Conclusions.
We also neglect nuclear entrainment in this paper.  Instabilities driven by entrainment coupling do not occur in the parameter range expected in neutron stars \citep{and03}, a result that we have verified using a more comprehensive stability analysis reported in \S\ref{secAe} and discussed further in the Conclusions.  Entrainment has a small effect on the mode frequencies.

The momentum equations for the neutron and proton--electron fluids in the MHD approximation are \citep{men91a,men91b,gla11}
\begin{eqnarray}
  \frac{\partial \vbf_n}{\partial t}  + \left(\nabla \times \vbf_n\right) \times \vbf_n &=&-\nabla p_n  -\Tbf_n+\Fbf_n\,,  \label{eq5} \\
\frac{\partial \vbf_p}{\partial t}+ \left(\nabla \times \vbf_p\right)\times \vbf_p  &=&-\nabla p_p-\Tbf_p -\frac{\rho_n}{\rho_p}\Fbf_n +\nu_{ee} \nabla^2 \vbf_p + \Fbf_{dip} \,,  \label{eq6}
\end{eqnarray}
where $\vbf_p$ is the smooth-averaged velocity of the proton--electron fluid, $\rho_{n,p}$ are the mass densities of the fluids, $p_{n,p}$ are scalar potentials related to thermodynamic variables in Equation (\ref{eqC29}), and $\Fbf_{dip}$ is the external driving force associated with the magnetic dipole torque on the star.
The neutron fluid is inviscid, while the proton--electron fluid has kinematic viscosity $\nu_{ee}$ arising from electron--electron scattering.  
The two fluids are coupled by the mutual friction force $\Fbf_n$, which arises from electron scattering from magnetized neutron vortices and pinning interactions.  
The force acts equally and oppositely on the two fluids and is given by (see \eg, \citealt{hal60,kha65,hil77,bar83,cha86,men91b,per07,gla11}),
\begin{eqnarray}
  \Fbf_n&=& \mathcal{ B}_n \hat{\omegabf}_n \times \left[\omegabf_n \times \left( \vbf_n- \vbf_p \right)+\Tbf_n\right]+\mathcal{B}_n'  \left[ \omegabf_n \times \left(\vbf_n- \vbf_p \right)+\Tbf_n\right]  \,,  \label{eq7}
\end{eqnarray}
where $\mathcal{B}_n$ and $\mathcal{B}'_n$ are the mutual friction coefficients; the first term is dissipative and the second term is nondissipative.  
The mutual friction coefficients are related to scattering and pinning parameters in \S\ref{sec3}.
Electron scattering from flux tubes is connected with the evolution of the magnetic field and describes processes analogous to ohmic and Hall diffusion; see \eg, \citet{gra15}.
These effects are small compared with the inertial modes studied in this paper; see \S\ref{secB} for further discussion.  
The restoring force due to tension of the vortex lines is (see \eg, \citealt{hal60,kha65,hil77,bay83,men91a,per07,gla11}),
\begin{eqnarray}
 \Tbf_n&=& \frac{1}{\rho_n} \omegabf_n \times \left( \nabla  \times \rho_n \nu_n \hat{\omegabf}_n  \right) \,,
\end{eqnarray}
where $\nu_n$ is the vortex line tension parameter, defined in (\ref{eqA24}).
The vortex line tension is negligible compared with other terms in (\ref{eq5}) and (\ref{eq6}); see Equation (\ref{eqC2}).  We set the vortex tension to zero everywhere in this paper except in the analysis of the Donnelly--Glaberson instability in \S\ref{sec4ba}, where it determines the instability condition. 
In a type II superconductor the magnetic stresses arise from the tension of the array of the quantized flux tubes and is given by \citep{eas77}
\begin{eqnarray}
\Tbf_p= \frac{\Bbf}{4\pi \rho_p} \times \nabla \times \left( H_{c1} \hat{\bbf}\right)\,, \label{eq8}
\end{eqnarray}
where $H_{c1} \simeq 10^{15}$ is the lower critical field for type II superconductivity.  
The evolution of the magnetic field is determined by the induction equation
\begin{eqnarray}
  \frac{\partial \Bbf}{\partial t}&=&\nabla \times \left(\vbf_p \times \Bbf \right)\,. \label{eq9} 
\end{eqnarray}

The equations (\ref{eq1})--(\ref{eq9}) suffice to study the stabilizing effects of magnetic fields on the instabilities of interest.  
With $\Tbf_n=0$, the equations do not include vortex line tension forces that produce Kelvin waves, which are small compared with the Coriolis force.
The evolution of the magnetic field is slow with respect to the timescales for oscillation modes.  
Magnetic stresses generated by rotation of charged fluid components, \ie, the London field, are also negligible.  
For a detailed discussion of the magnetohydrodynamic theory of \citet{gla11}, and a scaling analysis determining the relevant terms, the reader is referred to \S\ref{secA}--\ref{secB}.

\subsection{Perturbation equations} \label{sec2b}

Consider a neutron star comprising a neutron and a proton--electron fluid rotating as rigid bodies with angular velocities $\Omega_{n,p}$.
The star is spinning down under a constant external torque that acts only on the proton--electron fluid over the spin-down time of the star.  
Meanwhile, the proton--electron fluid spins down the neutrons via the vortex-mediated mutual friction force, $\Fbf_n$. 
As a consequence, the neutron fluid is rotating faster than the proton--electron fluid by an amount $\Delta \Omega =\left(\Omega_n-\Omega_p\right)$.
Taking $\hat{z}$ to be the rotation axis and denoting the unperturbed state with subscript $0$, we write the unperturbed velocities in the inertial frame as
\begin{eqnarray}
  \vbf_{n0}&=& \Omega_n \hat{z}  \times {\bf r}+ \Delta v_z \hat{z} \,,\\
  \vbf_{p0}&=&\left(\Omega_n-\Delta \Omega\right) \hat{z}  \times {\bf r} \,,
\end{eqnarray}
where the parameter $\Delta v_{z}$ is introduced to study the two-stream instabilities arising from relative velocity between the two fluids along the rotation axis.  
The lag $\Delta \Omega$ in the unperturbed state is determined by the momentum equations (\ref{eq5}) and (\ref{eq6}). 
Assuming that the spin-down rate ($\dot{\Omega}_p/\Omega_p$) is much slower than the rotation frequency, in cylindrical coordinates $(r,\phi,z)$ the azimuthal components of (\ref{eq5}) and (\ref{eq6}) give
\begin{eqnarray}
  \dot{\Omega}_n &=& -2\Omega_n \mathcal{B}_n \Delta \Omega \,,  \label{eq12aa} \\
  \dot{\Omega}_p &=&  \frac{2\Omega_n \rho_n \mathcal{B}_n \Delta \Omega}{\rho_p} +\frac{F_{dip,\phi}}{r}\,. \label{eq12ab}
\end{eqnarray}
Defining the pulsar spin-down time $\tau_{sd}=\Omega_p/(2 | \dot{\Omega}_p | )$, where $| \dot{\Omega}_n |/2 \pi = | \dot{\nu}|$ is the magnitude of the frequency derivative of the the pulsar's observed spin rate, and assuming that the spin-down rates of the neutron fluid and proton--electron fluid is equal ($\dot\Omega_n=\dot\Omega_p$) and $\Delta \Omega/\Omega_n \ll 1$, we find that Equation (\ref{eq12aa}) gives the lag
\begin{eqnarray}
  \Delta \Omega  &=& \left( 4 \tau_{sd} \,  \mathcal{B}_n \right)^{-1} \,. \label{eq14z}
\end{eqnarray}
The lag depends on the dissipative mutual friction coupling between the two fluids. 
The coefficient $\mathcal{B}_n$ depends on the scattering of electrons with vortices and the pinning between vortices and flux tubes in the outer core and is discussed further in \S\ref{sec3}.
Combining (\ref{eq12aa}) and (\ref{eq12ab}), multiplying by $r$ and integrating over the volume of the star gives the spin-down equation
\begin{eqnarray}
  I \dot\Omega_p  &=&-N_{dip} \,,
\end{eqnarray}
where $I=8\pi (\rho_n+\rho_p)  R^5/15$ is the moment of inertia, $N_{dip}=-\int r \rho_p F_{dip,\phi} dV =2 B^2 R^6 \Omega_p^3/3c^3$ is the external dipole torque, and $R$ is the stellar radius.

We now study the stability of the state described by Equations (\ref{eq12aa}) and (\ref{eq12ab}).
We use a local plane wave analysis, taking $x,y$ to be the local radial and azimuthal coordinates, respectively.  
The local plane wave analysis is adequate for wavenumbers $k R \gg 1$.  Recall that the hydrodynamic approximation is valid for $k d_n \ll 1$ where $d_n$ is the neutron vortex spacing.
These conditions restrict the treatment to wavenumbers in the range $d_n \ll k^{-1}\ll R$.
In this coordinate system, the velocities in the inertial frame are
\begin{eqnarray}
  \vbf_{n0}&=&R \Omega_n \hat{y}+ \Delta v_z \hat{z} \,, \label{eq12} \\
  \vbf_{p0}&=&R  \left(\Omega_n-\Delta \Omega \right) \hat{y} \,. \label{eq13}
\end{eqnarray}
The unperturbed magnetic field has poloidal $\hat{z}$ and toroidal $\hat{y}$ components and is given by 
\begin{eqnarray}
  \Bbf_0&=&B_0 \hat{\bbf}_0=B_{0y} \hat{y}+ B_{0z} \hat{z} \,.
\end{eqnarray}

Denoting the perturbed quantities by $\delta$, the perturbed momentum equations for the neutron and proton--electron fluids are 
\begin{eqnarray}
  \frac{\partial \delta \vbf_n}{\partial t}  + 2\Omega_n \hat{z} \times \delta \vbf_n + \left(\nabla \times \delta \vbf_n\right) \times \vbf_{n0}&=&-\nabla \delta p_n - \delta \Tbf_n + \delta \Fbf_n\,,  \label{eq14} \\
\frac{\partial \delta \vbf_p}{\partial t}+ 2\left(\Omega_n-\Delta \Omega \right) \hat{z} \times \delta \vbf_p + \left(\nabla \times \delta \vbf_p\right) \times \vbf_{p0}  &=&-\nabla \delta p_p- \delta \Tbf_p -\frac{\rho_n}{\rho_p} \delta \Fbf_n +\nu_{ee} \nabla^2 \delta \vbf_p \,.  \label{eq15}
\end{eqnarray}
The perturbations of the vortex line tension are 
\begin{eqnarray}
  \delta \Tbf_n &=& - \nu_n \hat{z}\cdot\nabla \left[\delta \omegabf_n - \hat{z} \left(\hat{z}\cdot  \delta \omegabf_n\right)\right]\,.
\end{eqnarray} 
The perturbed flux tube tension is
\begin{eqnarray}
\delta \Tbf_p= -\frac{H_{c1} }{4\pi \rho_p} \hat{\bbf}_0 \cdot \nabla  \left[ \delta \Bbf- \hat{\bbf}_0 \left( \hat{\bbf}_0 \cdot \delta \Bbf \right) \right]\,,
\end{eqnarray}
and the mutual friction force is
\begin{eqnarray}
  \delta \Fbf_n&=& \mathcal{ B}_n R\Delta \Omega \hat{x} \times \left(\nabla \times \delta \vbf_n \right) -  \mathcal{ B}_n \Delta v_z \hat{z}\times \left[ \hat{z}\times \left(\nabla \times \delta \vbf_n \right) \right] +\mathcal{ B}_n  2\Omega_n \hat{z}\times\left[ \hat{z}\times \left(\delta \vbf_n -\delta \vbf_p\right) \right]  + \mathcal{ B}_n \hat{z} \times \delta \Tbf_n \nonumber \\
&+& \mathcal{B}'_n \left(\nabla \times \delta \vbf_n \right)\times \left( R\Delta \Omega \hat{y} + \Delta v_z \hat{z} \right)  +  \mathcal{B}'_n 2\Omega_n \hat{z} \times \left(\delta \vbf_n -\delta \vbf_p\right) + \mathcal{ B}'_n \delta \Tbf_n\,. \label{eq23z}
\end{eqnarray}
Here we ignore dependence of $\mathcal{B}_n$ and $\mathcal{B}'_n$ on fluid velocity; see \S\ref{sec3} and \S\ref{secD} for further discussion of this point.
The induction equation for the perturbations is
\begin{eqnarray}
  \frac{\partial \delta \Bbf}{\partial t}&=&\nabla \times \left( \vbf_{p0} \times \delta \Bbf +\delta \vbf_p \times \Bbf_0 \right)\,. \label{eq18} 
\end{eqnarray}
The spin-down rate ($\dot{\Omega}_p/\Omega_p$) is much slower than the frequency of any hydrodynamic mode in the system, and perturbations of the external torque $\Fbf_{dip}$ are negligible.

To satisfy the continuity equations for the perturbations, we introduce the potential
\begin{eqnarray}
  \delta\vbf_n=\nabla \times \left( \psi_{n x}\hat{x}+ \psi_{ny}\hat{y} + \psi_{nz}\hat{z} \right)\,, \label{eq19}
\end{eqnarray}
and similarly for proton--electron fluid. For the magnetic field, we write
\begin{eqnarray}
  \delta \Bbf=\nabla \times \left( A_{x}\hat{x}+ A_{y}\hat{y}+ A_{z}\hat{z}\right)\,. \label{eq20}
\end{eqnarray}
To solve the system, we assume solutions of the form $e^{i {\bf k}\cdot{\bf x}-i \omega t}$ for all parameters.
One component of the potentials in Equations (\ref{eq19}) and (\ref{eq20}) is redundant, and we take $ \psi_{nz}=A_{z}=0$.
Eliminating $\delta p_{n,p}$ using the $\hat{z}$ components of (\ref{eq14}) and (\ref{eq15}), the $x$ and $y$ components of the (\ref{eq14}) and (\ref{eq15}) and the induction equation (\ref{eq18}) give a  matrix system of six equations in the unknowns
$\psi_{nx}$, $\psi_{ny}$, $\psi_{px}$, $\psi_{py}$, $A_x$ and
$A_y$. 
The complete dispersion relation is extremely lengthy, and we do not present it here. 
In \S\ref{sec4} we consider limits of the full dispersion relation that elucidate each of the instabilities present in the system.

\section{Neutron star parameters and relevant terms} \label{sec3}

Before solving the perturbation equations, we obtain numerical
estimates of the quantities that appear.

An approximate expression for the electron--electron scattering contribution to the viscosity is provided by \citet{cut87}.
More recent calculations by \citet{sht08} account for transverse Landau damping in charged particle collisions and find a viscosity approximately a factor of three smaller than that of \citet{cut87}.  
The kinematic viscosity $v_{ee}$ is defined in terms of the shear viscosity $\eta$ by
\begin{equation} \label{eq50}
 \nu_{ee}=\frac{\eta}{\rho_p}=6\times 10^{5}\, \left(\frac{\rho}{3\times 10^{14} \, \rm{g\,cm^{-3}}}\right) \left(\frac{x_p}{0.1}\right)^{-1} \left(\frac{\it{T}}{10^8 \, \rm{K}}\right)^{-2}\, \rm{cm^{2} s^{-1} }\,.
\end{equation}
The relative size of the viscous forces and Coriolis force is parameterized by the Ekman number, $E=\nu_{ee}/(\Omega_n R^2)$.
Equation (\ref{eq50}) gives
\begin{eqnarray} 
E=  10^{-9}  \left(\frac{\rho}{3\times 10^{14} \, \rm{g\,cm^{-3}}}\right)^{-1}\left(\frac{x_p}{0.1}\right) \left(\frac{\it{T}}{10^8 \, \rm{K}}\right)^{2} \left(\frac{R}{10^6\,{\rm cm}}\right)^{-2} \left(\frac{\Omega_n}{20\pi\, {\rm rad\,s^{-1}}}\right)^{-1}\,. \label{eq76}
\end{eqnarray}
Viscosity plays an important role in damping of high-wavenumber perturbations.

To estimate the importance of magnetic stresses, we note that magnetic stresses dominate the inertial forces when the vortex-cyclotron crossing time becomes shorter than the rotational period, \ie, for wavenumbers satisfying $\left| \vbf_{vc} \cdot \kbf  \right|  \gg 2\Omega_n$ where $|\vbf_{vc}|=\sqrt{H_{c1} B_0/(4 \pi \rho_p)} $ is the vortex-cyclotron wave speed.
The magnetic stress dominates the inertial force when
\begin{eqnarray}
  k R & \gg& 10^2 \left(\frac{H_{c1}}{4\times 10^{14} \,{\rm G}}\right)^{-1/2} \left(\frac{B_0}{10^{12} \,{\rm G}}\right)^{-1/2} \left(\frac{x_p}{0.1}\right)^{1/2} \left(\frac{\rho}{3\times 10^{14} \,{\rm g\,cm^{-3}}}\right)^{1/2} \left(\frac{\Omega }{20 \pi \,{\rm rad\,s^{-1}}}\right)\left(\frac{R}{10^6\,{\rm cm}}\right) \,. \label{eq114a}
\end{eqnarray} 
In this limit, the flux tube array appears infinitely rigid to the neutron fluid, and the neutron fluid decouples from the proton--electron fluid.
For low wavenumbers with $k R\sim 1$, magnetic stresses are negligible.

To estimate the mutual friction coefficients when the vortex lines and flux tubes are pinned together, we consider the rotational equilibrium described in \S\ref{sec2b}, for which $\dot{\Omega}_n=\dot{\Omega}_p$.  
Pinning forces can sustain a relative angular velocity $\Delta\Omega$ between the neutron and proton--electron fluids of up to the critical angular velocity for unpinning $\Delta \Omega_{crit}$.  
Numerical estimates for conditions in the outer core give $\Delta \Omega_{crit} \approx 0.1 \,{\rm rad\,s^{-1}}$ \citep{lin14}.
From Equation (\ref{eq14z}), ${\cal B}_n$ is related to $\Delta\Omega$ by
\begin{eqnarray}
   \mathcal{B}_n &=& \left( 4 \tau_{sd} \,\Delta \Omega \right)^{-1} \label{eq33} \,.
\end{eqnarray}
In the microscopic treatment of thermally activated vortex motion, the
mutual friction coefficients take the form \citep{lin14} [see Equation (51) therein]
\begin{eqnarray}
  \mathcal{B}_n&=&  \frac{\gamma \mathcal{R}_{n}}{1+\mathcal{R}_{n}^2}  \label{eq24a} \,,  \\
  1-\mathcal{B}'_n&=& \frac{\gamma}{1+\mathcal{R}_{n}^2} \,, \label{eq24}
\end{eqnarray}
where $\mathcal{R}_n$ is a scattering coefficient related to electron
scattering from magnetized vortex lines, $\gamma={\rm e}^{-A/k_B T}<<1$ is
the fraction of unpinned vorticity, $A$ is the activation energy for
unpinning, $k_B$ is Boltzmann's constant, and $T$ is the temperature. 
The activation energy depends on the lag $\Delta\Omega$. 
For a given ${\cal R}_n$ and $T$, the value of the activation energy adjusts so that (\ref{eq33}) holds. 
For typical parameters of a neutron star, the equilibrium lag is very close to
the critical value; see \citet{lin14} and \S\ref{secD} for a detailed calculation.  
We take $\Delta\Omega=\Delta\Omega_{crit}$ in (\ref{eq33}) and below when making numerical estimates.

Recall that the mutual friction force takes the form (\ref{eq7}),
\begin{eqnarray}
  \Fbf_n&=& \mathcal{ B}_n \hat{\omegabf}_n \times \left[\omegabf_n \times \left( \vbf_n- \vbf_p \right)+\Tbf_n\right]+\mathcal{B}_n'  \left[ \omegabf_n \times \left(\vbf_n- \vbf_p \right)+\Tbf_n\right]  \,,  \label{eq7z}
\end{eqnarray}
In perturbing this force, we took the mutual friction coefficients to
be constant; see Equation (\ref{eq23z}).
Thermally activated vortex motion causes the mutual friction coefficients to depend on $\vert \omegabf_n\times (\vbf_n-\vbf_p)\vert$ through the activation energy, which must be included when perturbing (\ref{eq7z}).  
This is explored in detail in  \S\ref{secD}.

The scattering coefficient $\mathcal{R}_n$ is calculated from the relaxation time for the electron distribution function due to relativistic electron scattering from a magnetized neutron vortex.  
The coefficient is related to the scattering time $\tau_{sn}$ by $\mathcal{R}_{n} = ( |\omegabf_n| \tau_{sn})^{-1}$ and is given by \citep{alp84a,har86,jon87}
\begin{eqnarray}
  \mathcal{R}_{n}&=& \frac{\rho_{p}}{ \rho_n}\left(\frac{\rho_{np} }{\rho_{pp}}\right)^2 \frac{3 \pi  e^2 \phi_0^2}{64 m_p c E_F \Lambda \kappa}   \label{eq26}\,, 
\end{eqnarray}
where $E_{Fe}=\hbar c (3 \pi^2 \rho_p/m)^{1/3}$ is the Fermi energy of the electrons. 
Based on the results of \citet{alp84a}, \citet{men91b} obtained the approximate expression
\begin{eqnarray}
   \mathcal{R}_{n}&=&0.011 \left(\frac{m_p^*-m_p}{m_p}\right)^2\left(\frac{m_p}{m_p^*}\right)^{1/2} \left(\frac{x_p^{7/6}}{1-x_p}\right)\left(\frac{\rho}{10^{14}\,{\rm g\,cm^{-3}}}\right)^{1/6}\,. \label{eq35z}
\end{eqnarray}
The scattering coefficient $\mathcal{R}_n$ is related to the drag coefficient $\eta_n$ and dissipation angle $\theta_{d}$ used by other authors \citep{alp84a,har86,jon87,lin14} by
\begin{eqnarray}
  \mathcal{R}_{n}=\frac{\eta_n}{\rho_n \kappa} = \tan \theta_{d}\,.
\end{eqnarray}

From (\ref{eq33}), (\ref{eq24a}) and (\ref{eq24}), the nondissipative mutual friction coefficient is 
\begin{eqnarray}
  1-  \mathcal{B}'_n &=& \left( 4 \tau_{sd} \,\Delta \Omega \mathcal{R}_{n} \right)^{-1} \label{eq34} \,. 
\end{eqnarray}
Using estimates for the critical velocity for unpinning in the outer core obtained by \citet{lin14}, we find Equations (\ref{eq33}) and (\ref{eq34}) give
\begin{eqnarray}
  \mathcal{B}_n &=& 8\times 10^{-12} \left( \frac{\Delta \Omega_{crit}}{0.1 \,{\rm rad\,s^{-1}}} \right)^{-1} \left(\frac{\tau_{sd}}{10\,{\rm kyr}}\right)^{-1} \,, \label{eq32a} \\ 
 1-\mathcal{B}_n'&=&2\times 10^{-8} \left( \frac{ \mathcal{R}_{n} }{4\times10^{-4}} \right)^{-1}\left( \frac{\Delta \Omega_{crit} }{0.1 \,{\rm rad\,s^{-1}}} \right)^{-1}\left(\frac{\tau_{sd}}{10\,{\rm kyr}}\right)^{-1}\,. \label{eq32b}
\end{eqnarray} 
We stress that these are crude estimates; for thermally activated vortex motion, these coefficients depend on the fluid velocities.

\section{Two-stream instabilities driven by mutual friction} \label{sec4}

\subsection{Rotational Lag during Spin-down} \label{sec4a}

As a neutron spins down under the magnetic dipole torque, pinning forces produce a rotational lag  between the neutron and proton--electron fluids; see \S\ref{sec2b}.
Two instabilities appear in this system: a fast two-stream instability with a growth time of seconds, and a slow two-stream instability with a growth time of days..  
Instabilities of this nature have been studied by  \citet{gla09} and \citet{and13} respectively, by looking at selective modes in spherical geometry and neglecting the magnetic field.  
We consider these instabilities in \S\ref{sec4aa} and \S\ref{sec4ab} and demonstrate that both are stabilized by the toroidal component of the magnetic field.  
In \S\ref{sec4ac} the instabilities of \citet{lin12a,lin12b} are revisited in the full two-fluid hydrodynamic theory.  
An algebraic error in those papers is corrected and the system is shown to be stable.

\subsubsection{A Fast Two-stream instability} \label{sec4aa}

The dispersion relation derived in \S\ref{sec2b} has significant algebraic complexity and we begin by exploring the parameter space numerically.  We identify an instability with a growth time of seconds.  This instability is stabilized by the toroidal component of the magnetic field $B_{0y}$.

To understand this instability, we explore the numerical solutions to the dispersion relation further. We find that approximating the mutual friction coefficients (\ref{eq32a}) and (\ref{eq32b}) by $\mathcal{B}_n=1-\mathcal{B}'_n=0$ has no significant effect on the instability.  For simplicity, we set $\mathcal{B}_n=1-\mathcal{B}'_n=0$ in the calculation presented here; this is discussed further later.  This approximation implies that the vortices and flux tubes move together, an approximation referred to as `perfect pinning' elsewhere.
The poloidal field has no significant affect on the instability and we assume $B_{0z}=0$. 
Only the toroidal field field $B_{0y}$ plays an essential role in this instability.
We ignore the vortex line tension and take $\nu_n=0$.
The dispersion relation under these assumptions reduces to  
\begin{eqnarray}
  \omega_n^2\left(A \omega_n^4+ B \omega_n^3+C \omega_n^2+D\omega_n +E\right)=0\,, \label{eq42z}
\end{eqnarray}
where
\begin{eqnarray}
  A &=& |k|^4 x_p^2 \,, \nonumber  \\
  B &=& 2 i x_p^2 |k|^6 \nu_{ee} \,, \nonumber  \\
  C &=& -|k|^2\left\{ 4 k_z^2 \left[\Omega_n^2+2 x_p \Omega_n  \Delta\Omega +\left(\Omega_n-\Delta\Omega\right)^2 x_p^2 \right] + x_p^2 k_y^2 \left(|k|^2+k_y^2\right)v^2_{vcy} + x_p^2 |k|^6 \nu_{ee}^2 \right\} \,, \nonumber  \\
  D &=& 8 k_y k_z^2|k|^2  \Omega_n  \Delta\Omega R \left( \Omega_n+\Delta\Omega x_p-\Omega_n x_p\right)    - \nu_{ee} |k|^4 \left[8 \Omega_n^2 k_z^2+x_p k_y^2 \left(|k|^2+k_y^2\right) v_{vcy}^2 \right] \,, \nonumber \\
  E &=&4 k_z^2 \Omega_n^2 \left(4 k_z^2 \Omega_n^2+x_p k_y^4 v^2_{vcy}-\Delta \Omega_n^2 k_y^2 |k|^2 R^2\right) + x_p k_y^2 |k|^2 v_{vcy}^2\left(4k_z^2 \Omega_n^2+k_y^4 x_p v^2_{vcy} \right) \,, 
\end{eqnarray}
and $v_{vcy}=\sqrt{H_{c1} B_{0y}/(4 \pi \rho_p)}$ is the speed of vortex-cyclotron waves.
In the unperturbed state, the neutron vortex lines move with the proton--electron fluid.  The frequency in this frame $\omega_n$ is related to the frequency in the inertial frame $\omega$ by 
\begin{eqnarray}
  \omega = \left( \Omega_n-\Delta\Omega\right) k_y R +\omega_n \,.
\end{eqnarray}
Therefore the dispersion relation (\ref{eq42z}) has two solutions that are zero in the rotating frame, which become $\left(\Omega_n-\Delta \Omega\right) k_y R $ after transforming back into the inertial frame.

\begin{figure*}
\centering
\includegraphics[width=.4\linewidth]{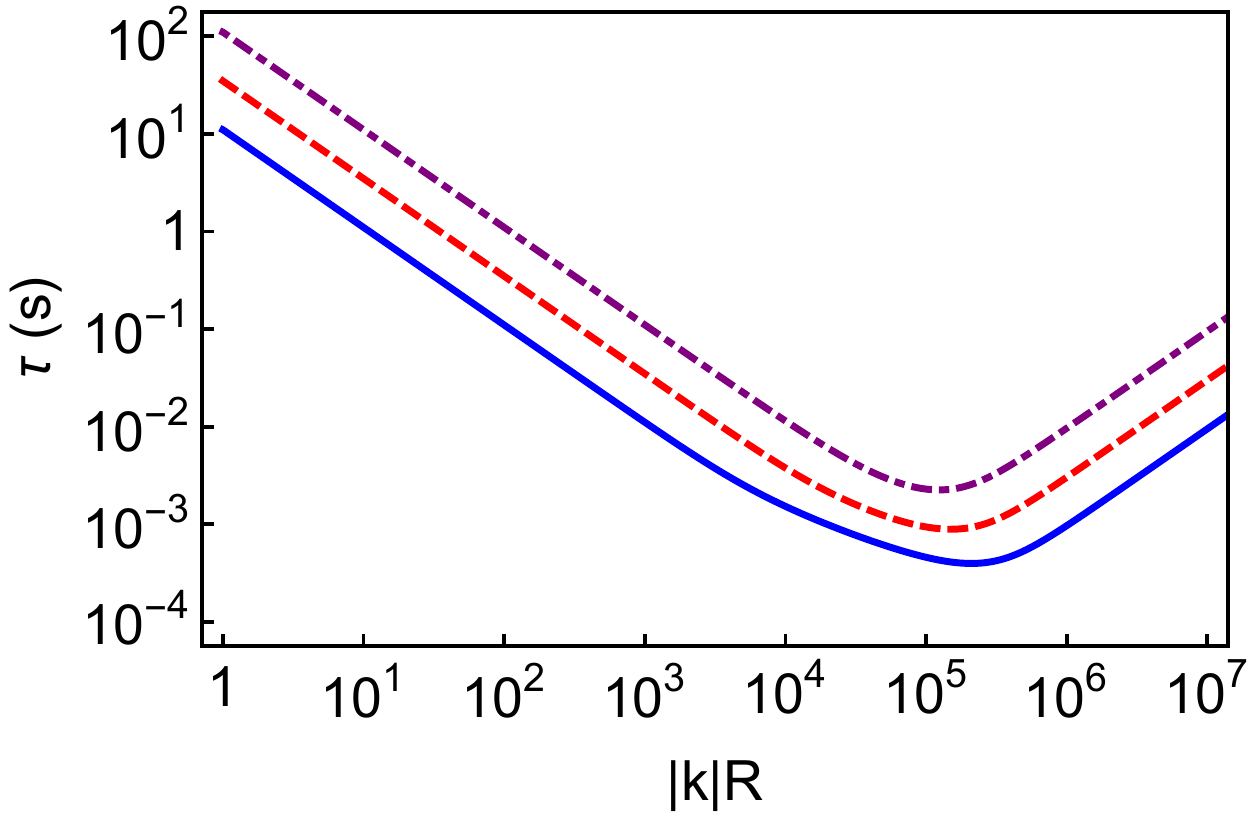} 
\caption{\footnotesize Growth time $\tau$ of the unstable solution of (\ref{eq42z}) as a function of dimensionless wavenumber $|k| R$ for no magnetic field. 
Growth time is plotted for $\theta_c$ given by (\ref{eq101}).
Three values of $\Delta \Omega$ are plotted: $10^{-2}\,{\rm rad\,s^{-1}}$ (dot-dashed), $10^{-3/2}\,{\rm rad\,s^{-1}}$ (dashed), and $10^{-1}\,{\rm rad\,s^{-1}}$ (solid).  Viscous forces suppress the instability at high wavenumber.  } 
\label{fig3a}
\end{figure*}
First, we examine the instability in the absence of the magnetic field.
Writing the wavenumber in spherical coordinates as $k_x=|k| \sin\theta \cos \phi $, $k_y=|k| \sin\theta \sin\phi$ and $k_z=|k| \cos\theta$, the dispersion relation is
\begin{eqnarray}
\omega_n^2 \left[\omega_n^2+\left(B_+ + i B_i\right) \omega_n + C_+ \right]\left[\omega_n^2+\left(B_-+i B_i \right) \omega_n + C_-\right]=0\,, \label{eq45z}
\end{eqnarray}
where
\begin{eqnarray}
  B_\pm &=& \pm \frac{2 \cos\theta }{x_p} \left(\Omega_n - \Omega_n x_p+\Delta \Omega x_p \right)  \,, \nonumber  \\ 
B_i&=& |k|^2 \nu_{ee} \,, \nonumber  \\ 
  C_\pm &=& -\frac{2 \Omega_n \cos\theta }{x_p} \left(2 \Omega_n \cos\theta \pm \Delta \Omega |k| R \sin\theta \sin\phi \right)   \,.
\end{eqnarray}
Separating out the real and imaginary parts, the unstable solutions to (\ref{eq45z}) can be written
\begin{eqnarray}
  \omega_n&=&-\frac{B_\pm}{2} \pm \frac{1}{2\sqrt{2}}\sqrt{\sqrt{\left(B_\pm^2-B_i^2-4C_\pm\right)^2+\left(2 B_\pm B_i\right)^2}+\left(B_\pm^2-B_i^2-4C_\pm\right)} \nonumber \\
&-& i \left[ \frac{B_i}{2} - \frac{1}{2\sqrt{2}}\sqrt{\sqrt{\left(B_\pm^2-B_i^2-4C_\pm\right)^2+\left(2 B_\pm B_i\right)^2}-\left(B_\pm^2-B_i^2-4C_\pm\right)}  \right] \,. \label{eq47z} 
\end{eqnarray}
The solution (\ref{eq47z}) is unstable when the term in the square braces is negative.  
This occurs for $C_\pm > 0$, yielding the instability condition
\begin{eqnarray}
 \pm |k| R \tan\theta  \sin\phi > \frac{2 \Omega_n }{  \Delta\Omega } \,. \label{eq116f}
\end{eqnarray}
Generally, $\Delta \Omega \ll \Omega_n$ and $x_p\ll 1$.  Viscous stresses are negligible compared to the inertial forces when $B_i \ll B_\pm$, which occurs for wavenumbers satisfying $|k| \ll \sqrt{2\Omega_n/\nu_{ee} x_p}$.  Using the neutron star parameters in \S\ref{sec3}, this gives 
\begin{eqnarray}
  |k| R \ll 10^5 \left(\frac{\Omega_n }{20 \pi \, {\rm rad\,s^{-1}} }\right)^{1/2} \left(\frac{R }{10^6 \, {\rm cm} }\right) \left(\frac{\rho}{3\times 10^{14} \, \rm{g\,cm^{-3}}}\right)^{-1/2} \left(\frac{\it{T}}{10^8 \, \rm{K}}\right) \,. \label{eq43z}
\end{eqnarray}
Under these assumptions, the `$-$' solution of (\ref{eq47z}) reduces to 
\begin{eqnarray}
 \omega_n &=& \frac{\cos\theta}{x_p}\left[\Omega_n + i \sqrt{\Omega_n \left(2\Delta \Omega |k|R x_p \tan\theta\sin\phi -\Omega_n\right) } \right] \nonumber \\
&-&\frac{|k|^2 \nu_{ee} }{2}\left[i + \frac{ \Omega_n }{\sqrt{\Omega_n \left(2\Delta \Omega |k|R x_p \tan\theta\sin\phi -\Omega_n\right) } }\right]\,. \label{eq50z}
\end{eqnarray}
The solution (\ref{eq50z}) has two distinct growth times depending on the sign of the term under each square root.  
If $0<2\Delta \Omega |k|R x_p \tan\theta\sin\phi<\Omega_n$, the term under each square root is negative, and there is an instability with growth time determined by the second term, namely $\tau\sim (|k|^2 \nu_{ee})^{-1}$. 
The growth time of this instability is 
\begin{equation} \label{eq102z}
 \tau =5\,  \left(\frac{|k| R}{2 \pi }\right)^{-2} \left(\frac{\rho}{3\times 10^{14} \, \rm{g\,cm^{-3}}}\right)^{-1}\left(\frac{x_p}{0.1}\right) \left(\frac{\it{T}}{10^8 \, \rm{K}}\right)^{2}   \, {\rm days }  \,. \label{eq50g}
\end{equation} 
If $2\Delta \Omega |k|R x_p \tan\theta\sin\phi>\Omega_n$, the term under the square root in (\ref{eq50z}) is positive and the first term dominates the growth time.  
The quickest growth time occurs for $\cos\phi=0$ and an angle $\theta_c$ given by
\begin{eqnarray}
   \tan 2 \theta_c  \approx  \pm \frac{2 x_p  \Delta\Omega |k|R }{\Omega_n } \,. \label{eq101}
\end{eqnarray}
Substituting this result into (\ref{eq50z}), and noting that $\sin\theta_c \approx 1$ and $\cos\theta_c=\Delta \Omega |k|R x_p/\Omega_n \ll1$, we find the growth time is approximately $\tau \approx \left( \Delta \Omega |k| R \right)^{-1}$.
Typical neutron star numbers in \S\ref{sec3} give
\begin{equation} \label{eq102}
 \tau =2\,  \left(\frac{|k| R}{2 \pi }\right)^{-1} \left(\frac{\Delta \Omega}{0.1\, {\rm rad\,s^{-1}}}\right)^{-1} \, {\rm s }  \,. 
\end{equation}
This growth time for this instability is much faster than (\ref{eq102z}).
Similar arguments apply to the `$+$' solution of (\ref{eq47z}), which can be obtained by making the replacement $\theta\rightarrow -\theta$.  
Therefore the instability condition for the fast instability is
\begin{eqnarray}
 \pm |k| R \tan\theta  \sin\phi > \frac{\Omega_n }{ 2 x_p \Delta\Omega } \,. \label{eq51z}
\end{eqnarray}
For wavenumbers satisfying (\ref{eq116f}) and not (\ref{eq51z}), the slow instability with growth time (\ref{eq50g}) occurs. 

In Figure \ref{fig3a}, we plot the growth time (in seconds) of the unstable solution of (\ref{eq47z}) as a function of the dimensionless wavenumber $|k|R$.
The orientation of the wave vector is chosen to give the quickest growth time, given by (\ref{eq101}).
At low wavenumbers, defined by (\ref{eq43z}), the growth time is well approximated by (\ref{eq102}).
In this regime, the growth time becomes faster as the wavenumber increases.
At high wavenumbers, viscous forces slow the growth time of the instability.
The growth time becomes infinitely long as the wavenumber approaches infinity.

\begin{figure*}
\centering
\includegraphics[width=.4\linewidth]{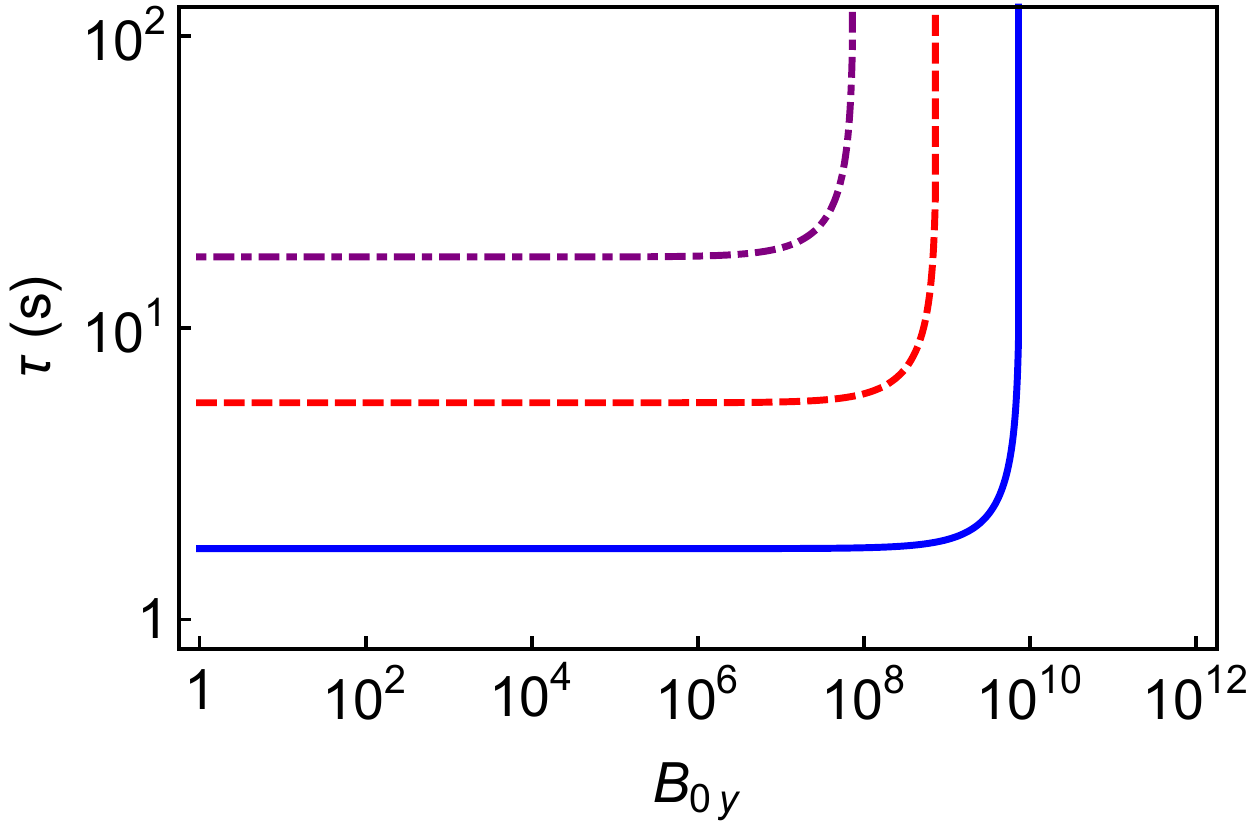} 
\includegraphics[width=.4\linewidth]{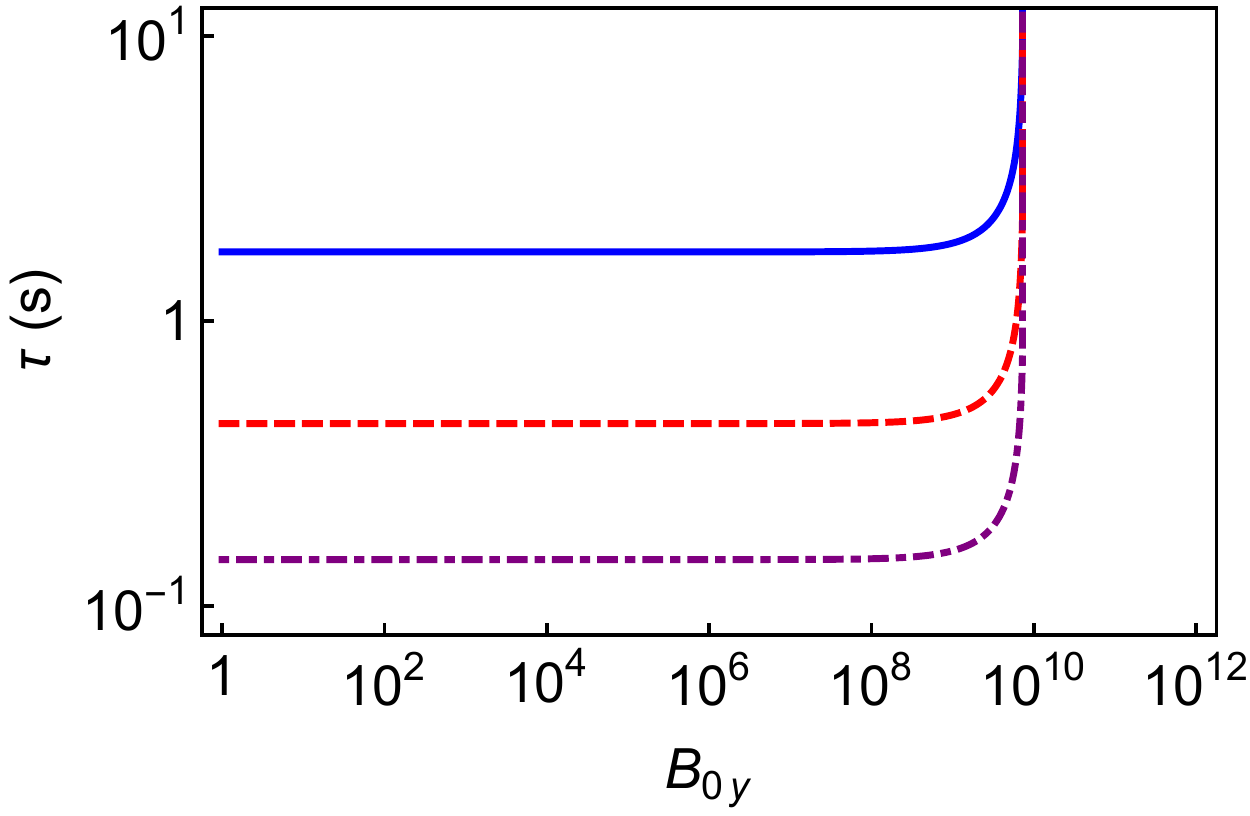} \\
\caption{\footnotesize Growth time $\tau$ of the unstable solution of (\ref{eq42z}) as a function of the azimuthal field $B_{0y}$ for $\theta_c$ given by (\ref{eq101}).
Left-hand panel shows three values of $\Delta \Omega$: $10^{-2}\,{\rm rad\,s^{-1}}$ (dot-dashed curve), $10^{-3/2}\,{\rm rad\,s^{-1}}$ (dashed), $10^{-1}\,{\rm rad\,s^{-1}}$ (solid), and $|k|R=2 \pi$. Right-hand shows three values of $|k|R$: $2 \pi$ (solid curve), $8 \pi$ (dashed curve) and $24\pi$ (dot-dashed curve) for $\Delta \Omega=10^{-1}\,{\rm rad\,s^{-1}}$. The instability is stabilized for magnetic fields above the critical value (\ref{eq53z}).  } 
\label{fig3}
\end{figure*}
We now turn on the azimuthal magnetic field $B_{0y}$ and examine the growth time as a function of magnetic field strength.
As before, we examine the instability when the growth time is quickest, given by (\ref{eq101}).
In Figure \ref{fig3} we plot the growth time of the unstable solution of (\ref{eq42z}) as a function of $B_{0y}$. 
In the left-hand panel, we plot three values of $\Delta \Omega$: $10^{-2}\,{\rm rad\,s^{-1}}$ (dot-dashed curve), $10^{-3/2}\,{\rm rad\,s^{-1}}$ (dashed), and $10^{-1}\,{\rm rad\,s^{-1}}$ (solid) for $|k|R=2 \pi$.  
All remaining parameters correspond to those given in \S\ref{sec3}.
The instability is present below a critical value of $B_{0y}$, at which it abruptly disappears.  This panel shows that the critical value of $B_{0y}$ scales as $\Delta \Omega^2$.  
In the right-hand panel, we plot three values of $|k|R$: $2 \pi$ (solid curve), $8 \pi$ (dashed), and $24\pi$ (dot-dashed) for $\Delta \Omega=10^{-1}\,{\rm rad\,s^{-1}}$.  
This panel shows that the critical value of $B_{0y}$ is independent of the dimensionless wavenumber $|k|R$.  
Further exploration of the parameter space demonstrates that the critical value of $B_{0y}$ depends weakly on all other parameters except $R$.  
These findings suggest that the critical $B_{0y}$ scales as $B_{0y} \sim R^2 \Delta \Omega^2$.
To obtain the proportionality factor, we assume that the turnover occurs when vortex-cyclotron velocity satisfies $v_{vcy}^2=B_{0y}H_{c1}/4 \pi \rho_p\sim R^2 \Delta \Omega^2$.  
The critical azimuthal field is then
\begin{eqnarray} 
 B_{0ycrit}=9\times 10^{9}\, {\rm G }\left(\frac{\rho}{3\times 10^{14} \, \rm{g\,cm^{-3}}}\right) \left(\frac{x_p}{0.1}\right) \left(\frac{R}{10^6\, {\rm cm}}\right)^2 \left(\frac{\Delta \Omega}{0.1\, {\rm rad\,s^{-1}}}\right)^2 \left(\frac{H_{c1}}{4 \times 10^{14}\, {\rm G}}\right)^{-1} \,.  \label{eq53z}
\end{eqnarray}
This result has agrees well with the critical values obtained numerically in Figure \ref{fig3}. 

Stable magnetic field configurations in a neutron star require that the toroidal field component exceed the poloidal component \citep{bra06,bra09}.
For a typical neutron star magnetic field of $10^{12}\,{\rm G}$, the lag in the outer core must exceed $1\,{\rm rad\,s^{-1}}$ for instability to occur; see Equation (\ref{eq53z}).  
However, \citet{lin14} estimates $\Delta\Omega \lesssim 0.1 \,{\rm rad\,s^{-1}}$, so the toroidal field component will quench this instability.

The instability identified here occurs when the wave vector for the perturbations has a component  oriented parallel to the relative background flow, in this case the azimuthal direction.
Therefore, the perturbations must be nonaxisymmetric for the inertial mode instability to operate.
The instability is stabilized by a sufficiently large component of the magnetic field that is also oriented parallel to the relative flow.
These generic properties for stabilizing two-stream inertial mode instabilities by magnetic stresses are also found in the later sections \S\ref{sec4ab} and \S\ref{sec4ba}. 

We now compare our findings with those of \citet{gla09}.
In their paper, \citet{gla09} solved the governing equations (\ref{eq5}) and (\ref{eq6}) neglecting viscous and magnetic stresses.
In contrast to the present plane wave analysis, \citet{gla09} solved for the unstable inertial modes in spherical geometry.
By assuming a power-law radial dependence and an r-mode angular dependence for the modes, \citet{gla09} showed that the $l=m$ mode is unstable when 
\begin{eqnarray}
m>\sqrt{\frac{\Omega_n}{2 x_p \Delta \Omega}}\,. \label{eq110f}
\end{eqnarray}
This instability condition has qualitative agreement with the result
of this paper; see Equation (\ref{eq51z}).  Because the instabilities are both
solutions to the same governing equations perturbed about the same
background, we expect a similar result for the instability condition.
Because of the similar nature of the problem, and
because we expect the stabilizing of unstable inertial modes in
two-fluid systems by the magnetic field to be a generic result, we
expect that the instability studied in \citet{gla09} is also
stabilized by the toroidal component of the magnetic field for
realistic neutron star configurations, in which the toroidal component of the magnetic field is comparable to or larger than the poloidal field component; see \eg, \citet{bra06} and \citet{bra09}. 

The instability in this section has been derived by approximating $\mathcal{B}_n=1-\mathcal{B}'_n=0$. Numerical solutions to the complete dispersion relation derived in \S\ref{sec2b} show that the stability criteria and growth times of the instability considered in this section are not significantly changed for the realistic neutron star parameters (\ref{eq32a}) and (\ref{eq32b}).  Exploring the numerical solutions to the dispersion relation accounting for thermal activation, presented in \S\ref{secD}, we find that thermal activation of pinned vorticity does not significantly alter instability.

\subsubsection{A slow two-stream instability} \label{sec4ab}

Exploring the solutions to the dispersion relation in \S\ref{sec2b} further, we find a second instability with a growth time of days. This instability is also stabilized by the toroidal component of the magnetic field $B_{0y}$. 

\begin{figure*}
\centering
\includegraphics[width=.4\linewidth]{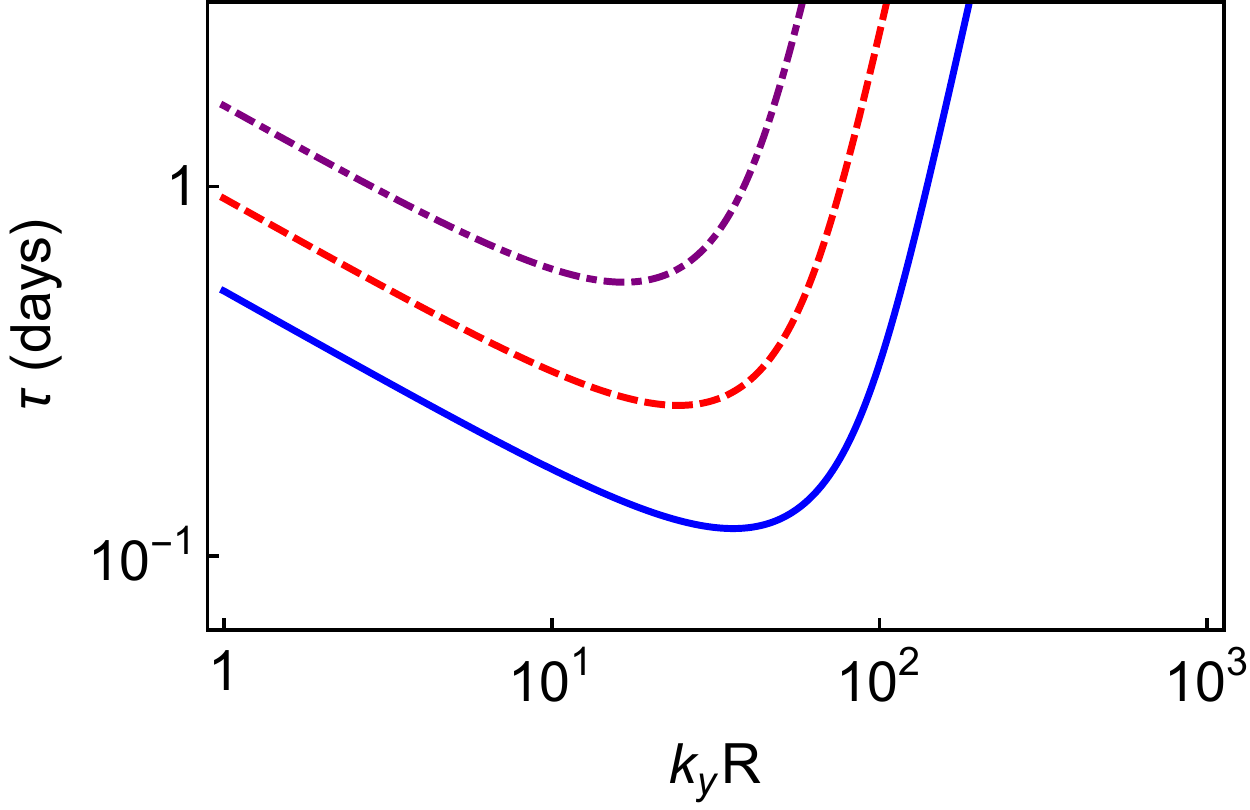} 
\caption{\footnotesize Growth time of the unstable solution to (\ref{eq121g}) as a function of dimensionless wavenumber $k_y R$ for no magnetic field. Curves correspond to $\Delta\Omega=10^{-1}$ (solid curve), $10^{-3/2}$ (dashed curve), and $ 10^{-2}$ (dot-dashed curve).
At high wavenumbers, the instability is suppressed by viscous forces. }
\label{fig4b}
\end{figure*}

To understand this instability, we explore the parameter space numerically and find that this instability occurs when the wavenumber is oriented in the azimuthal direction, and we take $k_x=k_z=0$.
The poloidal field has no significant effect  on the instability,  and we take $B_{0z}=0$. 
Only the toroidal field $B_{0y}$ plays an essential role in this instability.
We neglect the vortex tension and take $\nu_n=0$.
The dispersion relation in Section \S\ref{sec2b} reduces to
\begin{eqnarray}
 \left[ \omega_p - \Delta\Omega k_y R\left(1-\mathcal{B}_n'\right) \right] \left(\omega_p^2 + i \nu_{ee} k_y^2 \omega_p -  k_y^2 v_{vcy}^2\right) \left( \omega_p^3 + B \omega_p^2 + C \omega_p   + D \right)=0\,. \label{eq121g}
\end{eqnarray}
where
\begin{eqnarray}
 B&=&-\Delta\Omega  \left(1-\mathcal{B}'_n\right) k_y R + \frac{2 i \Omega_n }{x_p}\left(1 + x_p\right)\mathcal{B}_n  + i \nu_{ee} k_y^2 \nonumber \,, \\
 C&=&- 
      2  \Omega_n \mathcal{B}_n \nu_{ee} k_y^2  - v_{vcy}^2 k_y^2  -\frac{ 2 i \Omega_n \Delta\Omega }{x_p}\mathcal{B}_n   k_y R- i \Delta\Omega  \left(1-\mathcal{B}'_n \right) \nu_{ee} k_y^3 R    \nonumber  \,, \\
D&=& k_y^2 v_{vcy}^2 \left[ -2 i \Omega_n \mathcal{B}_n + \Delta\Omega  \left(1-\mathcal{B}'_n \right) k_y R\right]\,,
\end{eqnarray}
and $v_{vcy}=\sqrt{H_{c1} B_{0y}/(4 \pi \rho_p)}$.
The frequency in the frame rotating with the proton--electron fluid is related to the frequency in the inertial frame by
\begin{eqnarray}
  \omega = \left( \Omega_n-\Delta\Omega\right) k_y R +\omega_p \,.
\end{eqnarray}
The cubic factor in (\ref{eq121g}) gives unstable modes.  The instability is identified in the limit $v_{vcy}=\nu_{ee}=0$, reducing this factor to a quadratic in $\omega_p$.  
After separating out the real and imaginary parts, the unstable solution is
\begin{eqnarray}
  \omega_p&=&-\frac{B_r}{2} - \frac{1}{2\sqrt{2}}\sqrt{\sqrt{\left(B_r^2-B_i^2\right)^2+\left(2 B_r B_i-4C_i\right)^2}+\left(B_r^2-B_i^2\right)} \nonumber \\
&-& i \left[ \frac{B_i}{2} - \frac{1}{2\sqrt{2}}\sqrt{\sqrt{\left(B_r^2-B_i^2\right)^2+\left(2 B_r B_i-4C_i\right)^2}-\left(B_r^2-B_i^2\right)}  \right] \,, \label{eq126g}
\end{eqnarray}
where the subscripts $r$ and $i$ denote the real and imaginary components of $B$ and $C$ for $v_{vcy}=\nu_{ee}=0$.
The solution (\ref{eq126g}) is unstable for $C_i\left(C_i-B_r B_i\right)>0$, which gives
\begin{eqnarray}
  \mathcal{B}_n \Omega_n \Delta \Omega k_y R \left[1 - \left(1-\mathcal{B}'_n\right)\left(1+x_p\right)\right]>0 \,.
\end{eqnarray}
For the neutron star parameters in \S\ref{sec3}, the solution (\ref{eq126g}) is unstable for $k_y>0$.
The imaginary component in (\ref{eq126g}) is dominated by $C_i$, giving the growth time $\tau\approx \sqrt{2/|C_i|}=\sqrt{x_p/\mathcal{B}_n \Omega_n \Delta \Omega k_y R}$.  
Using the scaling (\ref{eq33}) for the dissipative mutual friction coefficient yields
\begin{eqnarray}
  \tau &=& 0.2  \,  \left(\frac{x_p}{0.1} \right)^{1/2}\left(\frac{k_y R}{2\pi }\right)^{-1/2} \left( \frac{\Omega_n}{20\pi  \,{\rm rad\,s^{-1}}} \right)^{-1/2}\left(\frac{\tau_{sd}}{10\,{\rm kyr}}\right)^{1/2} \, {\rm days} \,. \label{eq128g}
\end{eqnarray}
The growth time shortens with increasing with wavenumber according to (\ref{eq128g}) until viscous forces become important.  Viscous stresses are negligible when the square of the imaginary component of $B$  in (\ref{eq126g}) is much less than $C_i$, or $\nu_{ee}^2 k_y^3 \ll  4 \Omega_n \Delta \Omega \mathcal{B}_n R/x_p$.  Using the scalings (\ref{eq50}) and (\ref{eq33}) gives
\begin{eqnarray}
  k_y R &\ll & 10^2  \,  \left(\frac{x_p}{0.1} \right)^{1/3} \left( \frac{\Omega_n}{20\pi  \,{\rm rad\,s^{-1}}} \right)^{1/3} \left(\frac{\tau_{sd}}{10\,{\rm kyr}}\right)^{-1/3} \left( \frac{R}{10^6  \,{\rm cm}} \right)^{4/3} \left(\frac{\rho}{3\times 10^{14} \, \rm{g\,cm^{-3}}}\right)^{-1/3} \left(\frac{\it{T}}{10^8 \, \rm{K}}\right)^{4/3} \,. \label{eq62z}
\end{eqnarray}

In Figure \ref{fig4b}, we plot the growth time of the unstable solution to (\ref{eq121g}) as a function of dimensionless wavenumber $k_y R$ for zero magnetic field, $v_{vcy}=0$.
Three values of $\Delta \Omega$ are plotted: $10^{-1}$ (solid curve), $10^{-3/2}$ (dashed curve), and $10^{-2}$ (dot-dashed curve).
For low wavenumbers, defined by (\ref{eq62z}), the growth time is given by (\ref{eq128g}).
At high wavenumbers, the instability is suppressed by viscous forces.

\begin{figure*}
\centering
\includegraphics[width=.4\linewidth]{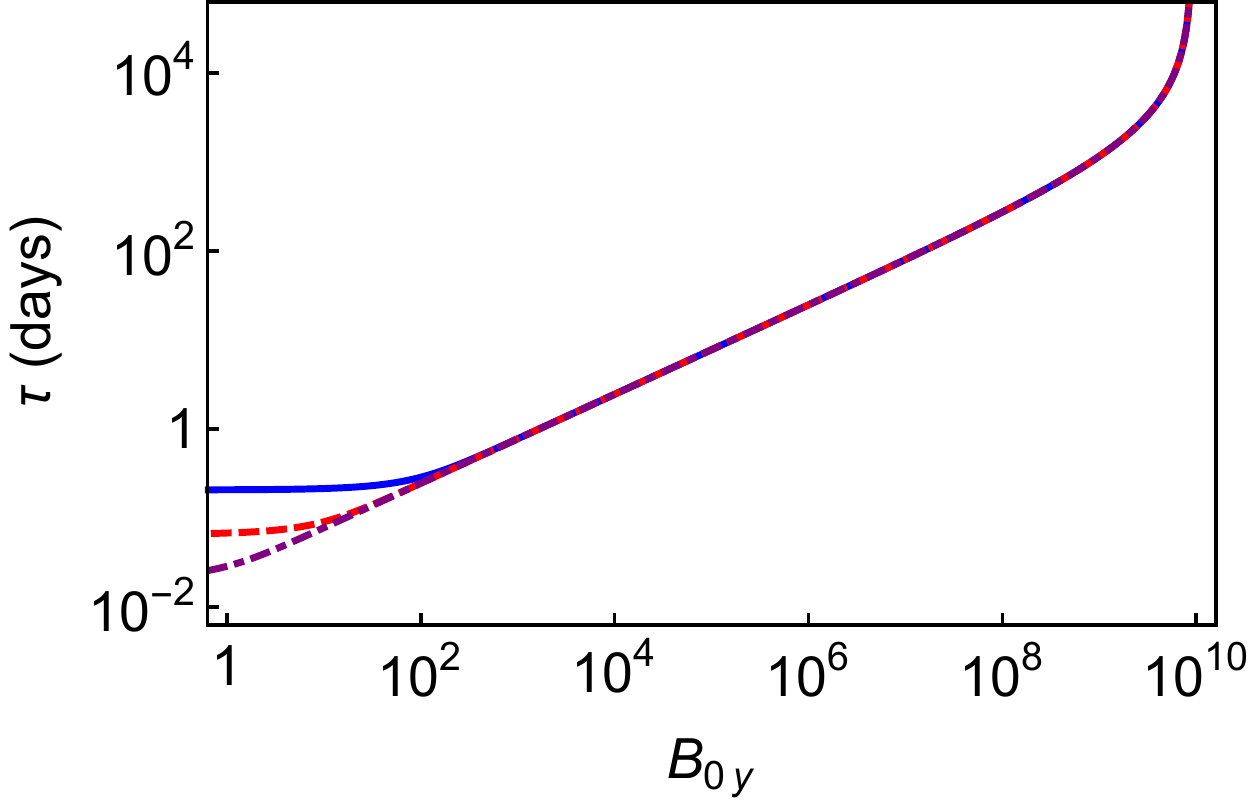}  
\caption{\footnotesize Growth time of the unstable solution to (\ref{eq121g}) as a function of azimuthal magnetic field. Curves correspond to $k_y=2\pi/R$ (solid curve), $20\pi/R$ (dashed curve), and $ 200\pi/R$ (dot-dashed curve).  
The instability is stabilized for magnetic fields above the critical value (\ref{eq102}), as for Figure \ref{fig3}.  }
\label{fig4a}
\end{figure*}
We now turn on the azimuthal magnetic field $B_{0y}$ and examine the growth time.  
Figure \ref{fig4a} shows the growth time of the unstable solution to (\ref{eq121g}) as a function of $B_{0y}$.  
Three values of $k_y$ are plotted: $2\pi/R$ (solid), $20\pi/R$ (dashed), and $200\pi/R$ (dot-dashed).
For small $B_{0y}$, the growth time is nearly independent of $B_{0y}$ and given approximately by (\ref{eq128g}).  
At larger $B_{0y}$ the magnetic field begins to influence the growth time, which becomes independent of $k_y$. 
In this regime, the growth time is approximately $\tau \approx v_{vcy} x_p/\mathcal{B}_n \Omega_n \Delta \Omega R$.
Using the mutual friction scaling (\ref{eq33}) yields 
\begin{eqnarray}
  \tau &=& 30  \, \left(\frac{H_{c1}}{3.8\times 10^{14} \,{\rm G}}\right)^{1/2} \left(\frac{B_{0y}}{10^{6} \,{\rm G}}\right)^{1/2} \left(\frac{x_p}{0.1}\right)^{1/2} \left(\frac{\rho}{3\times 10^{14} \,{\rm g\,cm^{-3}}}\right)^{-1/2} \nonumber \\
 &\times& \left(\frac{\tau_{sd}}{10\,{\rm kyr}}\right) \left( \frac{\Omega_n}{20\pi  \,{\rm rad\,s^{-1}}} \right)^{-1}\left( \frac{R}{10^6  \,{\rm cm}} \right)^{-1} \,  {\rm days} \,. \label{eq129g}
\end{eqnarray}
Comparing the growth times (\ref{eq128g}) and (\ref{eq129g}), we see the turnover between the two solutions for the growth times occurs at $v^2_{vcy}\sim \mathcal{B}_n \Omega_n \Delta \Omega R/x_p k_y$.  Using the scalings for the mutual friction coefficients derived in the pinning regime in \S\ref{sec3} gives 
\begin{eqnarray}
  B_{0y} &=& 70  \, \left(\frac{H_{c1}}{3.8\times 10^{14} \,{\rm G}}\right)^{-1}  \left(\frac{\rho}{3\times 10^{14} \,{\rm g\,cm^{-3}}}\right)^{-1/2} \left( \frac{k_y R}{2\pi } \right)^{-1} \left(\frac{\tau_{sd}}{10\,{\rm kyr}}\right)^{-1} \left( \frac{\Omega_n}{20\pi  \,{\rm rad\,s^{-1}}} \right)\left( \frac{R}{10^6  \,{\rm cm}} \right)\, {\rm G}  \,. \label{eq10g}
\end{eqnarray}
At a field of $\sim 10^{12}\,{\rm G}$, the growth time becomes infinite and the instability is quenched.
This occurs at $v_{vcy}^2\sim R^2 \Delta \Omega^2$, identical to the result obtained in \S\ref{sec4aa}.

The findings in \S\ref{sec4aa} and this section suggest that when there is no magnetic field, inertial modes coupled by mutual friction become unstable when the background fluids rotate relative to each other.  
However, these instabilities are stabilized by the azimuthal (toroidal) magnetic field $B_{0y}$.  
We conclude that there are no instabilities in neutron stars when the neutron and proton--electron fluids rotate with respect to one another in realistic magnetic field configurations.
These findings are verified by a thorough numerical search of the parameter space of the  complete dispersion obtained using the equations derived in \S\ref{sec2b} and \S\ref{secAe}.  

For the instabilities considered in \S\ref{sec4aa} and \S\ref{sec4ab}, the mode must have a nonvanishing projection of the wavenumber in the azimuthal direction for the instability to operate.  
The unstable mode is stabilized for a sufficiently large component of the magnetic field oriented in the same direction.  
For realistic neutron star configurations, in which the toroidal field component is greater than or equal to the poloidal field component, the instabilities in \S\ref{sec4aa} and \S\ref{sec4ab} are stabilized by the toroidal field.
The poloidal magnetic field has no effect on the instability.  

\citet{and13} studied the unstable inertial modes in two fluids rotating with respect to each other and coupled by mutual friction, the same problem considered here but neglecting the magnetic field.  
\citet{and13} generalized the study of \citet{gla09} to consider arbitrary mutual friction coefficients, assuming a power-law radial dependence and an r-mode angular dependence for the modes, and focusing on the $l=m$ mode as before.
In \S\ref{sec4aa}, we showed that the growth times are qualitatively similar to those found by \citet{gla09}.
Similarly, the secular growth times for the instability studied in this section arise from the dissipative mutual friction in a manner similar to that of \citet{and13}.  
Because \citet{gla09} and \citet{and13} solve the same equations as those in this study but in a different coordinate system, we expect that the instabilities found in \citet{gla09} and \citet{and13} will also be stabilized by the toroidal magnetic field.
In general, we find no unstable modes for neutron stars in which the condensates rotate relative to one another.
In summary, we expect that all such instabilities are stabilized by the toroidal component of the magnetic field.
In \S\ref{secD}, we show that thermal activation does  not alter the results of this section.

\subsubsection{\citet{lin12b,lin12a} Instabilities} \label{sec4ac}

In \S\ref{sec4aa} and \S\ref{sec4ab}, we showed that all unstable inertial modes in condensates rotating relative to one another are stabilized by the toroidal magnetic field.  
This finding contradicts that of \citet{lin12a}, who reported an instability in the neutron superfluid when the pinned neutron vortices undergo slow slippage with respect to the rigid flux tube lattice due to thermal activation in the outer core.  
An analogous instability was reported in the neutron star crust, where the slow slippage of vortices with respect to the nuclear lattice was shown to be unstable \citep{lin12b}.
We revisit the calculations of \citet{lin12a,lin12b} and show that these results are in error and that there is no instability.

In \citet{lin12b,lin12a}, it was assumed that the pinned vortices in the neutron superfluid undergo slippage with respect to a rigid lattice due to thermal activation. 
In \citet{lin12b} the lattice is the crust; in \citet{lin12a} the lattice is the dense array of flux tubes in the outer core.
To reproduce the latter calculation, we take the limit of infinite flux tube tension in the outer core, $v_{vc}\rightarrow\infty$.  
In this limit, the neutron superfluid decouples from the proton--electron fluid.  
The dispersion relation is found by solving (\ref{eq14}) and (\ref{eq23z}) and neglecting perturbations in the proton--electron fluid ($\delta \vbf_p=0$).  
The resulting dispersion relation is equivalent to that obtained for the neutron superfluid modes in the limit $v_{vc}\rightarrow\infty$.
We take the vortex line tension to be negligible ($\Tbf_n=0$).

After defining the wave-vector components $k_x=|k| \cos\phi\sin\theta$, $k_y=|k| \sin\theta\sin\phi$ and $k_z=|k| \cos\theta$, the dispersion relation is
\begin{eqnarray}
  \omega_n^2&+&2 i  \Omega_n  \mathcal{B}_n \left(1+\cos ^2\theta \right) \omega_n-\left(2 \Omega_n \cos\theta\right)^2  \left[\left(1-\mathcal{B}_n'\right)^2+\mathcal{B}_n^2\right]=0\,, \label{eq109}
\end{eqnarray}
where $\omega_n$ is the frequency in the frame rotating with the neutron vortices, related to the frequency in the inertial frame by
\begin{equation}
  \omega= \vbf_{Ln0}\cdot \kbf +\omega_n = \mathcal{B}_n \Delta \Omega k_x R+\left(\Omega_n - \Delta\Omega \mathcal {B}'_n \right)k_y R +\omega_n \,. \label{eq110}
\end{equation}
The solutions to (\ref{eq109}) are
\begin{eqnarray}
  \omega_n &=& -i \Omega_n  \mathcal{B}_n \left(1+\cos ^2\theta \right) \pm i \Omega_n \left\{  \mathcal{B}_n^2 \left(1+\cos ^2\theta \right) ^2 - \left(2 \Omega_n \cos\theta\right)^2   \left[\left(1-\mathcal{B}_n'\right)^2+\mathcal{B}_n^2\right] \right\}^{1/2} \,. \label{eq111}
\end{eqnarray}
The imaginary component of (\ref{eq111}) is always negative, so there are no unstable inertial modes. 
The error in \citet{lin12a,lin12b} can be traced to an incorrect perturbation of the neutron superfluid vorticity unit vector.  

We also revisit the assumption that the flux tube array provides an infinitely rigid pinning lattice for neutron vortices using the two-fluid magnetohydrodynamic theory in \S\ref{sec2}.
Scaling arguments in \S\ref{sec3} demonstrate that the magnetic stresses only dominate the inertial forces for large wavenumber; see Equation (\ref{eq114a}). 
Therefore, the flux tube array only appears infinitely rigid to the neutron superfluid for large wavenumbers satisfying (\ref{eq114a}), and not for small wavenumbers with $k R \sim 1$.

In \S\ref{secD}, we account for the effects of thermal
activation.  We find that no new instabilities are present.

\subsection{Relative Flow along the Rotation Axis} \label{sec4b}

In \S\ref{sec4a}, we studied instabilities that arise when condensates in the outer core rotate relative to one another.  
The condensates may also develop relative flow along the rotation axis, which may drive additional instabilities.  We examine two possibilities under which this may occur:  (1) the Ekman flow induced by the spin-down of the pulsar and (2) precession.

First, we examine the possibility of the development of a flow along the rotation axis arising from the spin down of a pulsar.  
If the magnetic field penetrates the entire star, the crust and the proton--electron fluid in the outer core are coupled by the magnetic field during spin-down.  
However, if the magnetic field does not penetrate the outer core, the fluid there will respond via viscous forces.
Rapidly rotating fluids respond to changes in the angular velocity of their container via {\it Ekman pumping}, wherein a secondary meridional flow transports angular momentum from viscous boundary layers into the interior on a timescale $E^{-1/2}\Omega_n^{-1}$, where $E$ is the Ekman number defined in (\ref{eq76}) (see, \eg, \citealt{gre63,gre68,van10,van14}).
The component of secondary flow along the rotation axis scales as $Ro E^{1/2} \Omega_n R$, where the Rossby number $Ro$ is a dimensionless angular velocity change of the container, typically the fractional increase in angular velocity for impulsive spin-up problems.  
For steady spin-down, the relevant timescale for the Rossby number is set by the external torque, and the velocity of the secondary flow scales as $ (\tau_{sd} \Omega_n)^{-1} E^{1/2} \Omega_n R$, where $\tau_{sd}$ is the spin-down time defined in (\ref{eq14z}).  
Compared with the rotational velocity of the star, the secondary flow along the rotation axis induced by Ekman pumping scales as
\begin{eqnarray}
  Ro E^{1/2} & \sim &  10^{-18}  \left(\frac{\tau_{sd}}{10\,{\rm kyr}}\right)^{-1} \left(\frac{\Omega_n}{20\pi\, {\rm rad\,s^{-1}}}\right)^{-3/2}   \left(\frac{\rho}{3\times 10^{14} \, \rm{g\,cm^{-3}}}\right)^{-1/2}\left(\frac{\it{T}}{10^8 \, \rm{K}}\right) \left(\frac{R}{10^6\,{\rm cm}}\right) \,. \label{eq69z}
\end{eqnarray}
We show below that such a tiny Ekman flow cannot induce instability.

The second possibility for developing relative flow along the rotation axis is precession of a neutron star.
During precession, the neutron and proton--electron angular velocity vectors are misaligned, inducing a relative flow along the proton--electron fluid along the rotation axis of the neutron fluid that can be directly related to the wobble angle of the precession.
 \citet{gla08} found an unstable mode in this context with a growth time of fractions of a second at small wavelengths, however the \citet{gla08} do not account for the magnetic field, which significantly modifies the instability.  
\citet{van08} included the magnetic field in their analysis.
Assuming perfect pinning, they show that the magnetic field stabilizes the instability.

In the following sections, we revisit the instabilities driven by relative flow along the rotation axis.  
We distinguish two distinct instabilities in this system: a two-stream instability and the Donnelly--Glaberson instability.
The two-stream instability develops in both fluids and is stabilized by sufficiently large magnetic fields.
This instability is studied by \citet{van08}.
The second instability is the Donnelly--Glaberson instability, which is also driven by relative flow along the rotation axis, but only develops in the neutron superfluid and is unaffected by magnetic stresses.  

To investigate instabilities arising from relative flow along the rotating axis, we consider flows with nonzero $\Delta v_z$ in \S\ref{sec2b}.  
The only relevant component of the wave vector is along the vortex lines, and we take $k_x=k_y=0$.
The neutron vortex line tension $\nu_n$ is retained because it plays an important role in the Donnelly--Glaberson instability. 
Under these assumptions, the equations in \S\ref{sec2b} give the dispersion relation
\begin{eqnarray}
\left( \omega^3+A_+ \omega^2 + B_+ \omega+C_+\right)\left( \omega^3+A_- \omega^2 + B_- \omega+C_-\right)=0\,, \label{eq122g}
\end{eqnarray}
where
\begin{eqnarray}
  A_\pm &=& \pm \frac{2\Omega_n}{x_p} \left(1 - x_p\right) \pm 2 \Delta \Omega+ 
 i k_z^2 \nu_{ee} + \left[i \mathcal{B}_n \mp \left(1-\mathcal{B}'_n \right) \right] \left[\frac{2 \Omega_n}{x_p}+ \left(2 \Omega_n + k_z^2 \nu_n \pm k_z \Delta v_z \right) \right] \,, \nonumber  \\ 
  B_\pm &=& \left\{-\frac{2 \Omega_n}{x_p} + \left[i \mathcal{B}_n \mp \left(1-\mathcal{B}'_n\right) \right] \left[\mp \frac{2 \Omega_n}{x_p} \left(1 + x_p\right) \pm  2 \Delta \Omega + i k_z^2 \nu_{ee}\right] \right\}  \left(2 \Omega_n + k_z^2 \nu_n \pm k_z \Delta v_z\right) - k_z^2 v_{cvz}^2 \,, \nonumber \\ 
  C_\pm &=& -v_{cvz}^2 k_z^2 \left[i  \mathcal{B}_n \mp\left(1-\mathcal{B}'_n\right)\right] \left(2 \Omega_n  + k_z^2 \nu_n \pm k_z \Delta v_z\right) \,,
\end{eqnarray}
and $v_{cvz}=\sqrt{H_{c1} B_{0z}/(4 \pi \rho_p)}$ is the vortex-cyclotron wave speed.
The `$+$' factor in (\ref{eq122g}) is identical to the dispersion relation obtained by \citet{van08} (see Appendix A therein), with the addition of the lag $\Delta \Omega$ and vortex tension $\nu_n$ terms.  
Analytic solutions to the cubics in (\ref{eq122g}) can be obtained but are cumbersome and uninformative, so we do not present them here.
We now study the two distinct instabilities in this system in turn.

\subsubsection{Two-stream instability} \label{sec4ba}  

To study the two-stream instability in this system, we approximate the mutual friction coefficients (\ref{eq32a}) and (\ref{eq32b}) with $\mathcal{B}_n=1-\mathcal{B}'_n=0$.
This instability was studied by \citet{gla08} neglecting magnetic fields, and by \citet{van08} including magnetic fields.
To put this instability in context with additional results in this paper, we summarize the results of \citet{van08} here, expanding on the discussion of the role of viscosity and growth times.  

Exploring the instability numerically, we find that the vortex tension and lag are negligible, and we set $\Delta \Omega=\nu_n=0$.
Assuming $\mathcal{B}_n=1-\mathcal{B}'_n=0$, the dispersion relation (\ref{eq122g}) reduces to   
\begin{eqnarray}
\omega^2 \left[\omega^2+\left(B_+ + i B_i\right) \omega + C_+ \right]\left[\omega^2+\left(B_-+i B_i \right) \omega + C_-\right]=0\,, \label{eq122}
\end{eqnarray}
where
\begin{eqnarray}
  B_\pm &=& \pm \frac{2\Omega_n}{x_p} \left(1 - x_p\right)   \,, \nonumber  \\ 
B_i&=& k_z^2 \nu_{ee} \,, \nonumber  \\ 
  C_\pm &=& -\frac{2 \Omega_n}{x_p} \left(2 \Omega_n  \pm k_z \Delta v_z\right) - k_z^2 v_{vcz}^2  \,.
\end{eqnarray}
Separating out the real and imaginary parts, the unstable solutions to (\ref{eq122}) can be written
\begin{eqnarray}
  \omega&=&-\frac{B_\pm}{2} \pm \frac{1}{2\sqrt{2}}\sqrt{\sqrt{\left(B_\pm^2-B_i^2-4C_\pm\right)^2+\left(2 B_\pm B_i\right)^2}+\left(B_\pm^2-B_i^2-4C_\pm\right)} \nonumber \\
&-& i \left[ \frac{B_i}{2} - \frac{1}{2\sqrt{2}}\sqrt{\sqrt{\left(B_\pm^2-B_i^2-4C_\pm\right)^2+\left(2 B_\pm B_i\right)^2}-\left(B_\pm^2-B_i^2-4C_\pm\right)}  \right] \,. \label{eq105f}
\end{eqnarray}
The solution is unstable for $C_\pm > 0$.  Focusing on the `$-$' solution, we find (\ref{eq105f}) is unstable for wavenumbers in the range $k_-<k_z<k_+$ where 
\begin{eqnarray}
  k_\pm =\frac{\Omega_n}{v_{vcz}^2 x_p} \left[ \Delta v_z \pm \sqrt{ \Delta v_z^2 -4 x_p v_{vcz}^2 } \right] \,, \label{eq120a}
\end{eqnarray}
which has real and distinct bounds when
\begin{eqnarray}
   \Delta v_z \geq 2 \sqrt{x_p} v_{vcz} \,. \label{eq2a}
\end{eqnarray}
This is the condition for instability, as found by \citet{van08}.

Viscous stresses are negligible when the viscous damping time is much longer than the vortex-cyclotron crossing time, \ie, $\nu_{ee} k_z^2 \ll v_{cvz} k_z$.  Using the results in \S\ref{sec3}, this occurs for wavenumbers satisfying
\begin{eqnarray}
 k_z R &\ll& 2\times 10^7  \left(\frac{H_{c1}}{4\times 10^{14} \,{\rm G}}\right)^{1/2} \left(\frac{B_0}{10^{12} \,{\rm G}}\right)^{1/2} \left(\frac{x_p}{0.1}\right)^{1/2}   \left(\frac{\it{T}}{10^8 \, \rm{K}}\right)^{2} \left(\frac{\rho}{3\times 10^{14} \,{\rm g\,cm^{-3}}}\right)^{-3/2} \left(\frac{R}{10^6\,{\rm cm}}\right) \label{eq77z} \,.
\end{eqnarray}
In this regime, we can approximate (\ref{eq105f}) by taking the limit $B_i^2 \ll B_\pm^2-4 C_\pm $.  
The unstable `$-$' solution is
\begin{eqnarray}
  \omega&=&\Omega_n \left( \frac{1-x_p}{x_p} \right) + \frac{i}{x_p}\sqrt{2 \Omega_n x_p \Delta v_z k_z  -\Omega_n^2\left(1+x_p\right)^2-x_p^2 v_{vcz}^2 k_z^2 } \nonumber \\
&-&\frac{i \nu_{ee} k_z^2 }{2} \left[1- \frac{ i \Omega_n \left(1-x_p\right)}{\sqrt{2 \Omega_n x_p \Delta v_z k_z -\Omega_n^2\left(1+x_p\right)^2-x_p^2 v_{vcz}^2 k_z^2 }}\right]\,. \label{eq119a}
\end{eqnarray}
The instability can be separated into two distinct regions depending on the sign of the quantity under the square root. 
For $k'_-<k_z<k'_+$ where
\begin{eqnarray}
  k'_\pm =\frac{\Omega_n}{v_{vcz}^2 x_p} \left[ \Delta v_z \pm \sqrt{ \Delta v_z^2 -v_{vcz}^2 \left(1+x_p\right)^2} \right] \,, \label{eq120b}
\end{eqnarray}
the expression under the square root is positive and the second term in (\ref{eq119a}) is imaginary.  
The third term is negligible compared with the first and second, and the growth time is determined by the second term.  
For wavenumbers within the bounds given by (\ref{eq120a}) but not those given by (\ref{eq120b}), the second term is real, and the third term is imaginary and determines the growth time.
These results agree with those obtained in Appendix B of \citet{van08}.
The instability criterion obtained by \citet{gla08} is recovered by taking $v_{vcz}\rightarrow 0$ in (\ref{eq119a}) and (\ref{eq120b}).

We now estimate the instability condition in a neutron star using the typical neutron star parameters in \S\ref{sec3}.
The instability condition (\ref{eq2a}) requires
\begin{eqnarray}
  \frac{\Delta v_z}{\Omega_n R} & > &  10^{-2} \left(\frac{H_{c1}}{4\times 10^{14} \,{\rm G}}\right)^{1/2} \left(\frac{B_0}{10^{12} \,{\rm G}}\right)^{1/2}\left(\frac{\rho}{3\times 10^{14} \,{\rm g\,cm^{-3}}}\right)^{-1/2}  \left(\frac{\Omega_n }{20 \pi \,{\rm rad\,s^{-1}}}\right)^{-1}\left(\frac{R}{10^6\,{\rm cm}}\right)^{-1} \,. \label{eq122b}
\end{eqnarray}
Therefore a relative velocity along the rotation axis greater than the vortex-cyclotron speed, or approximately one-hundredth of the equatorial velocity of the star, is required for instability.
This critical velocity is too large to be achieved by Ekman pumping during spin-down, which only induces a relative flow of $10^{-18}$; see Equation (\ref{eq69z}).  
In a freely precessing neutron star in which the neutron condensate is strongly pinned to the flux tubes, the wobble angle is related to the relative flow along the rotation axis by \citep{gla08}
\begin{eqnarray}
  \Delta v_z= \frac{ \theta_w \Omega_n R }{x_p}\,. \label{eq81z}
\end{eqnarray}
From (\ref{eq122b}) we find the critical wobble angle (in degrees) for instability is
\begin{eqnarray}
  \theta_w & > & 0.06^{\circ} \left(\frac{H_{c1}}{4\times 10^{14} \,{\rm G}}\right)^{1/2} \left(\frac{B_0}{10^{12} \,{\rm G}}\right)^{1/2}\left(\frac{\rho}{3\times 10^{14} \,{\rm g\,cm^{-3}}}\right)^{-1/2}  \left(\frac{x_p}{0.1}\right)  \left(\frac{\Omega_n }{20 \pi \,{\rm rad\,s^{-1}}}\right)^{-1}\left(\frac{R}{10^6\,{\rm cm}}\right)^{-1} \,. \label{eq122z}
\end{eqnarray}
The strongest precession candidate, PSR B1828-11, has an estimated wobble angle of $3^\circ$  \citep{sta00,cut03,akg06,lin07}.
Therefore this instability is likely to play a role in that object if the putative precession is real.

We now estimate the growth time of the instability in a neutron star.  
For wavenumbers between the bounds (\ref{eq120b}), the second term in (\ref{eq119a}) yields the approximate growth time $\tau\approx\sqrt{x_p/2 \Delta v_z k_z \Omega_n}$. 
Using the neutron star numbers in \S\ref{sec3} this gives
\begin{eqnarray}
  \tau & = & 1 \times 10^{-4} \,   \left(\frac{\Delta v_z}{\Omega_n R} \right)^{-1/2}\left(\frac{k_z R}{10^4}\right)^{-1/2} \left(\frac{x_p}{0.1}\right)^{1/2}  \left(\frac{\Omega_n }{20 \pi \,{\rm rad\,s^{-1}}}\right)^{1/2} \, {\rm s}  \,. \label{eq15a}
\end{eqnarray}
For wavenumbers outside the bounds (\ref{eq120b}), but within the bounds (\ref{eq120a}), the third term in (\ref{eq119a}) yields the approximate growth time $\tau\approx 2/ \nu_{ee} k_z^2$.
Using the scaling (\ref{eq50}), this gives
\begin{eqnarray}
  \tau & = & 2 \times 10^2  \, \left(\frac{\rho}{3\times 10^{14} \, \rm{g\,cm^{-3}}}\right)^{-1} \left(\frac{x_p}{0.1}\right) \left(\frac{\it{T}}{10^8 \, \rm{K}}\right)^{2} \left(\frac{k_z R}{10^{1/2}}\right)^{-2} \, {\rm days} \,. \label{eq15b}
\end{eqnarray}

\begin{figure*}
\centering
\includegraphics[width=.4\linewidth]{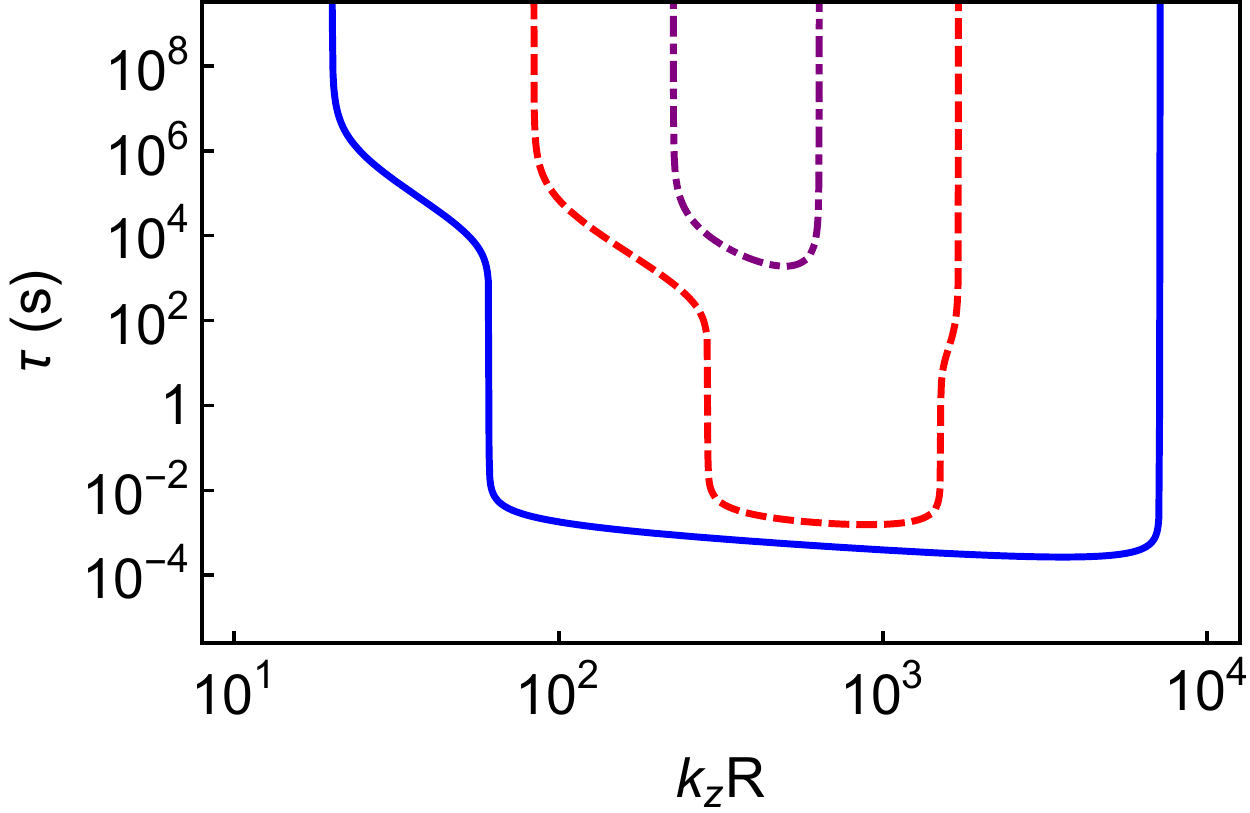} 
\caption{Growth time for the unstable solution of (\ref{eq122}) as a function of the dimensionless wavenumber for $\Delta v_z/\Omega_n R=0.1$ (solid curve), $0.025$ (dashed curve), and $0.012$ (dot-dashed curve).   } 
\label{fig5}
\end{figure*} 
In Figure \ref{fig5}, we plot the growth time of the two-stream instability, as determined from the `$-$' solution (\ref {eq105f}).
Curves are plotted for a poloidal field $B_{0z}=10^{12}\,{\rm G}$ and three values of relative flow along the rotation axis: $\Delta v_z/\Omega_n R=0.1$ (solid curve), $0.025$ (dashed curve), and $0.012$ (dot-dashed curve).
Using the relation (\ref{eq81z}), these correspond to wobble angles of $0.57^\circ$, $0.14^\circ$ and $0.07^\circ$ respectively.
For wavenumbers between the bounds (\ref{eq120b}), the growth time is quick and  given approximately by (\ref{eq15a}).
For wavenumbers outside the bounds (\ref{eq120b}) but within the bounds (\ref{eq120a}), the growth time determined by the viscosity and given approximately by (\ref{eq15b}).  
The dot-dashed curve has $\Delta v_z< (1+x_p) v_{vcz}$, and the bounds (\ref{eq120b}) are imaginary.  
In this case, only the slow instability with growth time (\ref{eq15b}) operates.
The instability window broadens as $\Delta v_z$ increases.  
Even for the relatively large $\Delta v_z$, the instability window occurs for $k_z$ much less than the condition (\ref{eq77z}), so viscosity has a negligible effect on the growth time.

We now compare the characteristics of the instability studied in this section with those of the two-stream instability in \S\ref{sec4aa} and note some similar features.  
Both instabilities operate when a component of the wave vector for the perturbations is oriented parallel to the relative background flow.  
In this case, the wave vector is along the rotation axis, whereas in \S\ref{sec4aa} the wave vector requires an azimuthal component.  In both cases, the instability is suppressed by a sufficiently large component of the magnetic field oriented in the same direction as the relative flow.
We find that these are general characteristics of the two-stream
instabilities considered in this paper.  
We emphasize again that the instability considered in this section is two-stream in nature and develops in both the neutron and proton--electron fluids.  
This distinguishes the instability from the Donnelly--Glaberson instability, as the latter only develops in the neutron superfluid.

\subsubsection{Donnelly--Glaberson instability} \label{sec4bb}

The second instability present in the dispersion relation (\ref{eq122g}) is the Donnelly--Glaberson instability.  We find that, in contrast with other instabilities considered in this paper, it is not suppressed by the magnetic field. This instability only occurs for $\mathcal{B}_n\neq \left(1-\mathcal{B}'_n\right)\neq 0$ and was not studied by \citet{van08} who derived the general dispersion relation (\ref{eq122g}) but only studied instabilities for $\mathcal{B}_n = \left(1-\mathcal{B}'_n\right) = 0$.
\citet{gla08} studied this instability, but did not consider the effects of the magnetic field.

The Donnelly--Glaberson instability is present in rotating superfluids such as terrestrial helium II.
The instability is excited when a normal fluid component, comprising thermal excitations, flows parallel to the vortex lines in the superfluid.
For a single vortex in an external flow, the critical velocity is given by the product of the vortex line tension and the wavenumber of the perturbed Kelvin waves, \ie, $\Delta v_z>\nu_n k_z$.
In the hydrodynamic limit for many vortices, the instability criterion becomes $\Delta v_z > 2\sqrt{2\Omega_n\nu_n}$ \citep{gla74,don05}.  
This instability has an analog in neutron stars, where the charged fluid component plays the role of the normal fluid component driving the instability \citep{sid08}.

To study the Donnelly--Glaberson instability in this system, we consider the high wavenumber limit (\ref{eq114a}).  
In this limit, the magnetic stresses in the proton--electron fluid dominate the inertial forces, and the neutron fluid decouples from the proton--electron fluid.  
This is equivalent to considering the problem of a neutron fluid coupled to a rigid lattice; see also \S\ref{sec4ac}.
In the limit (\ref{eq114a}), $v_{vcz}\rightarrow \infty $, and the dispersion relation for the neutron modes is a quadratic in $\omega$:
\begin{eqnarray}
  \left(\omega+C_+\right)\left(\omega+C_-\right) = 0 \,, \label{eq85z}
\end{eqnarray}
where
\begin{eqnarray}
  C_\pm = \left[\mp \left(1-\mathcal{B}'_n\right)+ i \mathcal{B}_n\right] \left[ \left( 2\Omega_n + \nu_n k_z^2\right) \pm k_z \Delta v_z\right] \,. \label{eq86z}
\end{eqnarray}
Let us consider the stability of the `$-$' solution of (\ref{eq85z}), given by
\begin{eqnarray}
  \omega= \left[\left(1-\mathcal{B}'_n\right)+ i \mathcal{B}_n\right] \left[k_z \Delta v_z - \left( 2\Omega_n + \nu_n k_z^2\right) \right]\,. \label{eq105}
\end{eqnarray}
Because the neutron fluid is decoupled from the proton--electron fluid in this limit, the viscosity does not affect the mode (\ref{eq105}).
For instability, we require that the imaginary component of (\ref{eq105}) is positive, which occurs for $k_z$ between in the range $k_-<k_z<k_+$ , where
\begin{eqnarray}
  k_\pm = \frac{1}{2\nu_n} \left( \Delta v_z \pm \sqrt{\Delta v_z^2 -8 \Omega_n \nu_n}\right) \,. \label{eq106b}
\end{eqnarray}
For two real, distinct bounds, we must have 
\begin{eqnarray}
  \Delta v_z > 2\sqrt{2 \Omega_n \nu_n} \,, \label{eq106c}
\end{eqnarray}
which recovers the condition for the Donnelly--Glaberson instability \citep{gla74}.

We now estimate the instability condition in neutron stars using the numbers in \S\ref{sec3}.
The condition (\ref{eq106c}) requires 
\begin{eqnarray}
  \frac{\Delta v_z}{\Omega_n R} & > & 2\times 10^{-8} \left(\frac{\nu_n}{4\times 10^{-3} \,{\rm cm^2\,s^{-1}}}\right)^{1/2}  \left(\frac{\Omega_n }{20 \pi \,{\rm rad\,s^{-1}}}\right)^{-1/2}\left(\frac{R}{10^6\,{\rm cm}}\right)^{-1} \,. \label{eq122a}
\end{eqnarray}
The relative flow along the rotation axis induced by Ekman pumping during spin-down is  $10^{-18}$; see Equation (\ref{eq69z}).  
Therefore this instability is not excited during spin-down.
Next, we consider whether this relative velocity is likely to occur in a neutron star precessing with wobble angle $\theta_w$. 
Using the previous result to relate the wobble angle to the relative flow along the rotation axis (\ref{eq81z}), the critical wobble angle (in degrees) for instability is
\begin{eqnarray}
  \theta_w & > & 10^{-7 \; \circ} \left(\frac{\nu_n}{4\times 10^{-3} \,{\rm cm^2\,s^{-1}}}\right)^{1/2}  \left(\frac{\Omega_n }{20 \pi \,{\rm rad\,s^{-1}}}\right)^{-1/2}\left(\frac{R}{10^6\,{\rm cm}}\right)^{-1} \left(\frac{x_p}{0.1}\right) \,. \label{eq122f}
\end{eqnarray}
Therefore the Donnelly--Glaberson instability is likely to be relevant in precessing neutron stars with relatively small wobble angles.

We now estimate the critical wavenumber and growth time for instability.
Assuming the relative flow along the rotation axis greatly exceeds the critical velocity $\Delta v_z \gg 2\sqrt{2 \Omega_n \nu_n}$, we find the lower bound in (\ref{eq106b}) is approximately $2\Omega_n/\Delta v_z$, which gives
\begin{eqnarray}
  k_- R >  2 \left(\frac{\Delta v_z}{\Omega_n R} \right)^{-1}\,. \label{eq123a}
\end{eqnarray}
This lower bound  becomes larger as $\Delta v_z$ becomes smaller.
For $\Delta v_z< 0.02 \, \Omega_n R $, the lower critical wavenumber for instability satisfies the assumption that the flux tube lattice appears infinitely rigid to the neutron superfluid, given by (\ref{eq114a}). 
The upper bound in (\ref{eq106b}) is approximately $\Delta v_z/\nu_n$, which is the critical wavenumber for instability on an individual vortex filament.
Using the neutron star parameters in \S\ref{sec3}, we find
\begin{eqnarray}
  k_+ R <  2\times 10^{16} \left(\frac{\Delta v_z}{\Omega_n R} \right) \left(\frac{\nu_n}{4\times 10^{-3} \,{\rm cm^2\,s^{-1}}}\right)^{-1}  \left(\frac{\Omega_n }{20 \pi \,{\rm rad\,s^{-1}}}\right)\left(\frac{R}{10^6\,{\rm cm}}\right)^{2} \,. \label{eq123b}
\end{eqnarray}
The hydrodynamic approximation breaks down for wavenumbers greater than $2\pi /d_n$, which occurs for
\begin{eqnarray}
  k_z R> 2\times 10^9 \left(\frac{\Omega}{20 \pi  \,{\rm rad\,s^{-1}}}\right)^{1/2} \left(\frac{R}{10^6\,{\rm cm}}\right) \,. \label{eq94z}
\end{eqnarray}
Therefore the upper limit (\ref{eq123b}) is outside the range of validity of the hydrodynamic approximation.  
For wavenumbers greater than (\ref{eq94z}) and less than (\ref{eq123b}),  individual vortex filaments are unstable to the Donnelly--Glaberson instability.
The growth time for the Donnelly--Glaberson instability is $\tau\approx \left(\mathcal{B}_n \Delta v_z k_z\right)^{-1} $, yielding
\begin{eqnarray}
  \tau &=&  2 \, \left(\frac{\Delta v_z}{\Omega_n R} \right)^{-1}\left(\frac{k_z R}{10^4}\right)^{-1} \left( \frac{\Delta \Omega_{crit}}{0.1 \,{\rm rad\,s^{-1}}} \right)\left(\frac{\tau_{sd}}{10\,{\rm kyr}}\right) \,  {\rm days} \,. \label{eq10a}
\end{eqnarray}

\subsubsection{General Instability for Relative Flow along the Rotation Axis}

\begin{figure*}
\centering
\includegraphics[width=.4\linewidth]{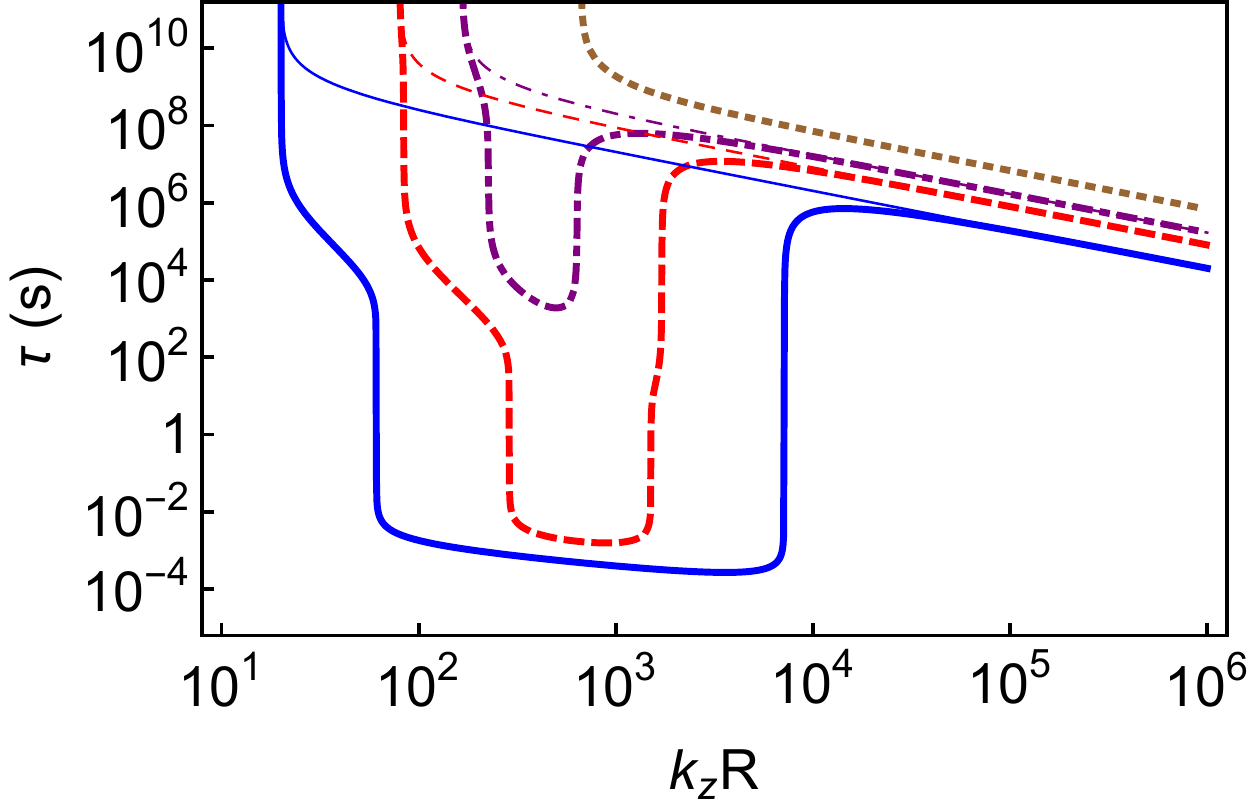} 
\caption{Growth time for the unstable solution of (\ref{eq122g}) as a function of the dimensionless wavenumber for $\Delta v_z/\Omega_n R=0.1$ (solid curve), $0.025$ (dashed curve), $0.012$ (dot-dashed curve), and $0.003$ (dotted curve).  
The corresponding growth time for the Donnelly--Glaberson instability (\ref{eq86z}) is plotted as thin lines for comparison.  At low wavenumbers the two-stream instability dominates.  At larger wavenumbers the two-stream instability is stabilized by the magnetic field, and the Donnelly--Glaberson instability operates.  The dotted curve has the smallest $\Delta v_z$, for which the two-stream instability is always stabilized by the magnetic field.}
\label{fig6}
\end{figure*}

We now combine the results of \S\ref{sec4ba} and \S\ref{sec4bb} to study the unstable solution of the complete dispersion relation (\ref{eq122g}). 
In Figure \ref{fig6}, the growth time of the unstable mode of (\ref{eq122g}) is plotted for the typical pulsar parameters in \S\ref{sec3}, taking the mutual friction coefficients (\ref{eq32a}) and (\ref{eq32b}) and poloidal magnetic field $B_{0z}=10^{12}\,{\rm G}$.
The growth time is plotted as a function of the dimensionless wavenumber for four values of relative flow along the rotation axis: $\Delta v_z/\Omega_n R=0.1$ (heavy solid curve), $0.025$ (heavy dashed curve), $0.012$ (heavy dot-dashed curve), and $0.003$ (heavy dotted curve).
Using the relation (\ref{eq81z}), these correspond to wobble angles of $0.57^\circ$, $0.14^\circ$, $0.07^\circ$ and $0.02^\circ$ respectively. 
Plotted for comparison in corresponding thin lines is the growth time of the unstable Donnelly--Glaberson solution (\ref{eq86z}).
Comparing Figures \ref{fig5} and \ref{fig6}, we see that both instabilities in \S\ref{sec4ba} and \S\ref{sec4bb} manifest in the same unstable mode.
At low wavenumbers, the two-stream instability studied in \S\ref{sec4ba} dominates the Donnelly--Glaberson instability.  
There is a negligible different between the results for $\mathcal{B}_n$ and $1-\mathcal{B}'_n$ given by (\ref{eq32a}) and (\ref{eq32b}), and $\mathcal{B}_n=1-\mathcal{B}'_n=0$.
At wavenumbers exceeding the upper bound (\ref{eq120a}), the two-stream instability is quenched, and the growth time of the unstable mode is determined by the Donnelly--Glaberson instability.  
The Donnelly--Glaberson instability is never suppressed by the magnetic field, and its growth time continues to shorten until the hydrodynamic approximation breaks down at the wavenumber given by (\ref{eq94z}), not shown in Figure \ref{fig6}.
At higher wavenumbers, the growth time continues to shorten until the critical wavenumber for instability of an individual vortex line (\ref{eq123b}) is reached.
The instability window for the two-stream instability decreases as $\Delta v_z$ decreases.
For $\Delta v_z = 0.003\, \Omega_n R$ (dotted curve) and below, the two-stream instability does not operate, and the growth time is determined by the Donnelly--Glaberson instability.
When the growth time of the unstable mode is determined by the Donnelly--Glaberson instability, the solution (\ref{eq105}) is a good approximation to the unstable solution of (\ref{eq122g}).

A distinguishing characteristic of the Donnelly--Glaberson instability is that it only develops in the superfluid; there is no velocity perturbation in the normal fluid.
Therefore the development of the Donnelly--Glaberson instability is not inhibited even when magnetic stresses in the proton--electron fluid become large.

Figure \ref{fig6} demonstrates that, for relative flows along the rotation axis exceeding the critical value (\ref{eq122a}) or wobble angle (\ref{eq122f}), the outer core of a neutron star is always unstable to the Donnelly--Glaberson instability at high wavenumbers.
This suggests that the dynamics in the outer core of precessing neutron stars may be very different from nonprecessing stars.

In \S\ref{secD} we include the effects of thermal activation
in the stability analysis. 
We find that only the Donnelly--Glaberson instability is significantly modified by this effect.  
The instability growth time is lengthened from days to decades in the regime where the hydrodynamic approximation is valid.  The upper bound for instability (\ref{eq123b}) is also increased. The lower bound (\ref{eq123a}), critical velocity (\ref{eq122a}), and hence the critical wobble angle (\ref{eq122f}) are unchanged.  A comparison of the growth time as a function of wavenumber with and without thermal activation is shown in Figure \ref{fig7}.

\section{Discussion and Conclusions} \label{sec5}

Hydrodynamic instabilities in neutron stars are of interest for their possible role in spin glitches, timing noise, and precession. 
In this connection, transitions in and out of states of superfluid turbulence, driven by relative flow between the neutron and proton--electron fluids, have been hypothesized to be responsible for spin glitches \citep{and04,gla09,and13}. 
The purpose of this study was to determine whether magnetic stresses stabilize the candidate instabilities. A summary of our conclusions for instabilities driven by relative rotation and relative flow along the rotation axis is presented in Table \ref{tab1}.

As a neutron star spins down due to the external torque, an angular velocity difference develops between the neutron and proton--electron fluids.  
Our chief conclusion is that this state possesses no unstable inertial modes.  
The two-stream instabilities in this system are stabilized by the toroidal magnetic field when the vortex-cyclotron speed becomes larger than the relative velocity of two condensates; this stabilization occurs for toroidal field strengths of order $10^{10}\,{\rm G}$ or higher. 
Calculations of magnetostatic neutron star equilibria give toroidal fields that are at least as strong as the poloidal component \citep{bra06,bra09}.
Therefore we expect that a neutron star should be stable against the instabilities found by \citet{gla09} and \citet{and13}.  \citet{lin12a,lin12b} erroneously found that relative flow between the neutron and proton--electron fluids is unstable when the neutron vortices slip with respect to the flux tubes through thermal activation. 
We have ascertained that this process is actually stable.

If relative flow along the rotation axis is produced by precession, for example, there are two instabilities of possible relevance. 
At low wavenumber, a two-stream instability operates with growth time shorter than a second.  
This instability is suppressed by the magnetic field at high wavenumbers, as shown by \citet{van08}.
However, at high wavenumbers the Donnelly--Glaberson instability occurs, which is not suppressed by the magnetic field.
In contrast with the two-stream instabilities considered in this paper, the Donnelly--Glaberson instability only develops in the neutron superfluid and can therefore operate even when the magnetic stresses in the proton--electron fluid become large.
In precessing neutron stars, the two-stream instability is excited for wobble angles of a fraction of a degree, while the Donnelly--Glaberson instability can be excited by wobble angles as small as $10^{-7}$ degrees.  
The wobble angle for PSR B1828-11 within a precession interpretation is much larger than these critical values \citep{sta00,cut03,akg06,lin07}, so hydrodynamic instability and turbulence could be important in this object.

The local conditions for which instability occurs depend on the
local density, and hence upon the dense-matter equation of state to some extent.  
For the two-stream instabilities studied in \S\ref{sec4aa} and
\S\ref{sec4ab}, the critical value of the toroidal field for is relatively
low ($\sim 10^{10}$ G) for any reasonable equation of state. 
For the two-stream instability of \S\ref{sec4ba}, stability depends upon the proton fraction and the vortex-cyclotron velocity. 
Our estimate for the critical flow along the $z$ axis thus depends on density, but more importantly on the strength of the dipole field, which varies significantly from star to star. 
The critical velocity for the Donnelly--Glaberson instability (\S\ref{sec4bb}) depends on only the spin rate of the star and on the vortex line tension; the latter has a dependence on local density that is fairly weak. 
More accurate numbers for the onset of these instabilities could
be obtained with a stellar-structure model, but we expect our
estimates to be reliable.

\begin{table}
\begin{tabular}{|llllll}
\hline
Relative flow & Instability type & Growth time & Toroidal field & Poloidal field & Ref. \\
\hline
$\Delta \Omega$ & Two-stream & Seconds & $R\Delta \Omega< v_{vcy}$ & No effect & \S\ref{sec4aa}\\
$\Delta \Omega$ & Two-stream & Days & $R\Delta \Omega< v_{vcy}$ & No effect &  \S\ref{sec4ab}\\
$\Delta v_z$ & Two-stream & Seconds & No effect & $\Delta v_z < 2\sqrt{x_p} v_{vcz}$ & \S\ref{sec4ba}\\
$\Delta v_z$ & Donnelly--Glaberson & Days & No effect & No effect &  \S\ref{sec4bb} 
\end{tabular}
\caption{\footnotesize Summary of the results obtained in this paper.  The first and second columns give the relative flow and type of instability.  The third column gives the characteristic growth time.  The fourth and fifth columns list the stabilization condition for the toroidal and poloidal fields.  The final column gives the section of this paper in which the instability is studied.  }
\label{tab1}
\end{table}
 
Two-stream instabilities have also been reported for the Fermi-liquid entrainment coupling between the two fluids in the absence of mutual friction ($\rho_{np}\neq 0$ and $\Fbf_n=0$). See \eg, \citet{and04}, but they find no instability for values of the entrainment parameter in the expected range for a realistic neutron star.  
We verify this result using a more general study reported in \S\ref{secAe}, and we do not present results for two-stream instabilities driven by entrainment coupling.  
Entrainment has a small effect on the stability analysis in this paper and modifies the inertial mode frequencies by a factor of less than two.

It appears unlikely that the effects of compressibility and buoyancy would alter our conclusions.
\citet{gus13} have noted that low-frequency thermal g-modes in a star composed of superfluid neutrons, superconducting protons, and normal electrons is unstable at low densities.
However, \citet{pas16} have shown that this instability is weak and likely to only operate just below the crust in young neutron stars, where only very short wavelengths are unstable.
Deeper in the core, g-modes are restored by muon composition gradients and are expected to be stable \citep{kan14}.
These modes have kilohertz frequencies and are unlikely to be modified by the magnetic stresses, entrainment, or mutual friction forces, which are much smaller than the buoyancy restoring forces.
\citet{and04} showed that relative flow between two chemically coupled superfluids produces unstable sound modes. 
The instability is shown to operate in the outer core just below the crust, where the required relative flow is a significant fraction of the speed of sound of the neutron gas.
Such a large relative flow, of order $10^8\,{\rm cm \, s^{-1}}$, is unlikely in a realistic neutron star, making this instability difficult to excite.

For the conditions that prevail in a spinning-down neutron star, we conclude that the hydrodynamic flow is stable; in particular, hydrodynamic turbulence does not develop and therefore is not the cause of spin glitches as postulated by, \eg, \citet{gla09} and \citet{and13}. 
Should an instability with sufficiently fast rise time exist, however, two challenges still remain.  The first challenge is to demonstrate how the instability develops to produce a glitch.
The second challenge is to demonstrate that this turbulent state ends and resets the system for the next glitch.
Steadily driven classical systems susceptible to instability develop a quasi-steady turbulent cascade without global transient behavior; therefore, if the spin-down under an external torque is unstable, we expect the turbulent state to persist.
On the other hand, hydrodynamic instabilities could indeed play a role in precessing neutron stars.
The development of such an instability and its effects is an interesting problem for future study.

For completeness, the complete MHD theory including
entrainment, Kelvin waves, and magnetic field evolution, is presented in \S\ref{secAe}.  
Stability is studied assuming constant density flow, and the dispersion relation is explored numerically over the relevant range of parameters in neutron stars.
We find no additional instabilities. 

The effects of thermal activation are studied in \S\ref{secD}.  
The growth time of the Donnelly--Glaberson instability is lengthened to decades for wavenumbers in the hydrodynamic regime. 
The other instabilities considered in this paper are unaffected by thermal activation.

\acknowledgments

We thank Y. Levin for helpful comments on this work.  This
work was supported by NSF award AST-1211391 and NASA award NNX12AF88G.

\appendix

The governing equations in \S\ref{sec2} are derived from the MHD theory developed by previous authors; see \eg, \citet{men91a,men91b} and \citet{gla11}.
In the Appendices, we reduce the equations of previous authors to those presented in \S\ref{sec2}.
In \S\ref{secA}, we show that a self-consistent hydrodynamic theory should not contain  the London depth.  In a type II superconductor, charged currents are screened over the London depth, an effect that occurs on length scales much smaller than that at which the hydrodynamic approximation applies. 
In \S\ref{secB}, we reduce the full MHD equations presented in \S\ref{secA} to those presented in \S\ref{sec2}. 
Using scaling arguments, we show that Kelvin waves and magnetic field evolution are negligible for studying the stability of the inertial modes considered in this paper.

\section{Self-consistent hydrodynamic theory} \label{secA}

In this appendix, we review the hydrodynamic approximation applied to the outer core of neutron stars by previous authors. In \S\ref{secAa}, we review the relevant length scales for the vortex and flux tube arrays in the outer core.
In \S\ref{secAb}, we define the smooth-averaged quantities in the hydrodynamic theory.
In \S\ref{secAc}, we show that the charge current is screened over the London depth in a type II superconductor.
The full equations of the self-consistent hydrodynamic theory are presented in \S\ref{secAd}.
The perturbation equations used to study the complete dispersion relation for this system are presented in \S\ref{secAe}.

\subsection{Length scales} \label{secAa}

The smooth-averaging over the vortex and flux tube arrays must be performed over a length scale much larger than size or separation of the vortices or flux tubes.  We calculate each of these scales below.

The cross-sectional areas of the vortices and flux tubes are determined by the coherence lengths of the condensates, given by $\xi_x=\hbar p_{Fx}/(\pi m_x^* \Delta)$, where $p_{Fx}=\hbar (3 \pi^2 \rho_x/m_x)^{1/3}$ is the Fermi momentum, $m_x^*$ is the effective mass for neutrons ($x=n$) and protons ($x=p$) due to entrainment, and $\Delta=1.76 k_B T_c$ and $\Delta=2.4 k_B T_c$ are the energy gaps for singlet and triplet pairing, respectively, for a critical temperature $T_c$.  The coherence lengths are
\begin{eqnarray}
  \xi_n &=& 30 \left(\frac{\rho}{3\times 10^{14} \,{\rm g\,cm^{-3}}}\right)^{1/3}\left(\frac{1-x_p}{0.9}\right)^{1/3}\left(\frac{m_n/m^*_n}{0.9}\right)\left(\frac{T_c}{5\times 10^9\,{\rm K}}\right)^{-1} \,{\rm fm}\,, \label{eqB1} \\
 \xi_p &=& 30 \left(\frac{\rho}{3\times 10^{14} \,{\rm g\,cm^{-3}}}\right)^{-1/2}\left(\frac{x_p}{0.1}\right)^{-1/2}\left(\frac{m_p/m^*_p}{2}\right)\left(\frac{T_c}{5\times 10^9\,{\rm K}}\right)^{-1} \,{\rm fm}\,. \label{eqB2}
\end{eqnarray}
where $\rho$ is the mass density and $x_p$ is the proton fraction.
The magnetic field around a flux tube decays over a characteristic length scale of the London depth and is given by [see (\ref{eqB9}) below] 
\begin{eqnarray}
  \Lambda = 40 \left(\frac{\rho}{3\times 10^{14} \,{\rm g\,cm^{-3}}}\right)^{-1/2}\left(\frac{x_p}{0.1}\right)^{-1/2}\left(\frac{m_p/m^*_p}{2}\right)^{-1/2} \,{\rm fm}\,. \label{eqB3}
\end{eqnarray}
Type II superconductivity occurs when the proton coherence length and the London depth obey $\xi_p/\sqrt{2} \Lambda <1$, which is satisfied in the outer core.

For a fluid rotating at angular velocity $\Omega$, the neutron condensate forms an array with vortex areal density $n_{nv}=2\Omega/\kappa$ where $\kappa =\pi \hbar/m$ is the quantized circulation per vortex. 
For a triangular lattice, the intervortex spacing is 
$d_n=(\kappa /\sqrt{3} \Omega)^{1/2}$, giving 
\begin{eqnarray}
  d_n &=& 4\times 10^{-3} \left(\frac{\Omega}{20 \pi  \,{\rm rad\,s^{-1}}}\right)^{-1/2}\,{\rm cm} \,. \label{eqB4}
\end{eqnarray}
Similarly, in an external magnetic field $B_0$, the areal number density of flux tubes is $n_{vp}=B_0/\phi_0$, where $\phi_0=\pi \hbar c/e=mc\kappa/e$ is the quantized magnetic flux per flux tube.  
The spacing for a triangular flux tube lattice is $d_p=(2\phi_0 /\sqrt{3} B_0 )^{1/2}$, giving
\begin{eqnarray}
  d_p &=& 4\times 10^{3} \left(\frac{B_0}{10^{12}\,{\rm G}}\right)^{-1/2}\,{\rm fm} \,. \label{eqB5}
\end{eqnarray}
Comparing (\ref{eqB1})--(\ref{eqB5}), we find the largest length scale in a mixture of a type II superconducting protons and a superfluid neutrons is the intervortex spacing $d_n$. 
The length scales are ordered as
\begin{eqnarray}
  \Lambda\simeq \xi_n \simeq \xi_p<<d_p<<d_n\,. \label{eqB6}
\end{eqnarray}
The hydrodynamic approximation for the electron gas applies for length scales much larger than the electron mean free path, which is given by 
\begin{eqnarray}
 \lambda_e = 3\times 10^{-2} \,  \left(\frac{\rho}{3\times 10^{14} \,{\rm g\,cm^{-3}}}\right) \left(\frac{x_p}{0.1}\right)^{-1}\left(\frac{T}{10^8\,{\rm K}}\right)^{-2} \,{\rm cm}\,, \label{eqB7}
\end{eqnarray}
where $T$ is the temperature.

\subsection{Smooth-averaged Vorticity and Magnetic Field in Fermi Mixures} \label{secAb}

In this section, we present the theory of the smooth-averaged vorticity and magnetic field in a Fermi-liquid mixture.  The results presented in this section are the same as in previous works (see \eg, \citealt{men91a,men91b,gla11}), but some notation differs.

In the outer core of a neutron star, the mixed proton and neutron condensates experience a nondissipative interaction, or entrainment.
The mass densities  of the neutron ($\rho_n$) and proton ($\rho_p$) condensates satisfy
\begin{eqnarray}
 \rho_n&=&\rho_{nn} +\rho_{np} \,, \label{eqA3}\\
 \rho_p&=&\rho_{pp} +\rho_{np} \,, \label{eqA4}
\end{eqnarray}
where $\rho_{np}$ parameterizes the entrainment interaction.
The entrainment densities (\ref{eqA3}) and (\ref{eqA4}) are related by the effective mass of the proton, see \eg, \citet{alp84a} and \citet{men91a}.  
For proton density fraction $x_p=\rho_p/\rho\approx \rho_p/\rho_n \ll 1$ where $\rho$ is the total mass density, the densities are related by
\begin{eqnarray}
  \rho_{pp} &=& \rho_p \left(\frac{m_p}{m^*_p}\right)=\rho x_p \left(\frac{m_p}{m^*_p}\right) \,, \nonumber \\
  \rho_{np} &=&  \rho_p\left(1-\frac{m_p}{m^*_p}\right)= \rho_n\left(1-\frac{m_n}{m^*_n}\right)= \rho x_p\left(1-\frac{m_p}{m^*_p}\right) \,, \nonumber \\
  \rho_{nn}&=&\rho_n \left(\frac{m_n}{m^*_n}\right)= \rho \left(1-x_p+ x_p \frac{m_p}{m^*_p}\right)  \,. \label{eqA5}
\end{eqnarray}
The mass currents obey the continuity equations
\begin{eqnarray}
 \frac{\partial \rho_x}{\partial t}+\nabla\cdot\jbf_x&=&0\,, \label{eqA6} 
\end{eqnarray}
where the mass currents are defined as
\begin{eqnarray}
 \jbf_n&=&\rho_{nn}\vbf_n +\rho_{np} \vbf_p\,, \label{eqA7}\\
 \jbf_p&=&\rho_{pp} \vbf_p+\rho_{np} \vbf_n\,.\label{eqA8}
\end{eqnarray}
There are two conventions for the definition of the velocity for entrained systems.  In this paper, and in that of \citet{alp84a} and \citet{men91a,men91b}, the velocity is directly related to the wave function of the condensate, i.e., $\vbf_x=\hbar/m_x\nabla \phi_x$.  
However, recent works define the conjugate momenta in terms of the wave-function phase, and the velocity in terms of the mass currents, taking $\jbf_x=\rho_x \vbf_x$ \citep{gla11}.  
The latter formulation can be obtained by making the replacement
\begin{eqnarray}
  \vbf_n\rightarrow \vbf_n +\varepsilon_n\left(\vbf_p-\vbf_n\right) \,, \nonumber \\
  \vbf_p\rightarrow \vbf_p +\varepsilon_p\left(\vbf_n-\vbf_p\right) \,, 
\end{eqnarray}
where
\begin{eqnarray}
 \rho_n \varepsilon_n=\rho_p \varepsilon_p = \frac{\rho_n \rho_p \rho_{np} }{\rho_{np}^2-\rho_{nn} \rho_{pp} } \,.
\end{eqnarray}

Smooth-averaging over length scales much larger than the intervortex spacing $d_n$, we define the quantities
\begin{eqnarray}
 \omegabf_n&=&\kappa n_{vn} \hat{\omegabf}_n =\nabla \times \vbf_n\,, \label{eqA1}\\
 \omegabf_p&=&\kappa n_{vp} \hat{\omegabf}_p =\nabla \times \vbf_p+\frac{e}{mc} \Bbf \label{eqA2} \,,
\end{eqnarray}
where $\vbf_n$ and $\vbf_p$ are the smooth-averaged neutron and proton fluid velocities respectively, $\Bbf$ is the smooth-averaged magnetic field, and $\hat{\omegabf}_x=\omegabf_x/|\omegabf_x|$ are the vorticity unit vectors.
Note that (\ref{eqA2}) is defined as in \citet{gla11}, whereas \citet{men91a,men91b} defines $\omegabf_p =\nabla \times \vbf_p$.
As the number of vortex lines is conserved, the smooth-averaged quantities (\ref{eqA1}) and (\ref{eqA2}) obey the conservation law
\begin{eqnarray}
  \frac{\partial \omegabf_x}{\partial t}+\nabla \times \left( \omegabf_x\times \vbf_{Lx}  \right) =0\,, \label{eqA34}
\end{eqnarray}
where $\vbf_{Ln,p}$ are the neutron vortex line and flux tube velocities.

The entrainment of the proton current by the neutron current magnetizes the neutron vortices with magnetic flux $\rho_{np} \phi_0/\rho_{pp}$.
The smooth-averaged magnetic field is given by
\begin{eqnarray}
 \Bbf=\frac{m c}{e}\left(\omegabf_p+\frac{\rho_{np}}{\rho_{pp}}\omegabf_n\right) +\Bbf_L= \phi_0 n_{vp} \hat{\omegabf}_p+\frac{\rho_{np}}{\rho_{pp}} \phi_0 n_{vn}\hat{\omegabf}_n +\Bbf_L  \,, \label{eqA9}
\end{eqnarray}
where the first term on the right side of (\ref{eqA9}) is the contribution to the magnetic field from the flux tubes and the second term is the contribution from the neutron vortices.
The third term is the contribution to the magnetic field from macroscopic rotation and is called the London field.
Assuming that the ratio $\rho_{np}/\rho_{pp}$ is constant and using definitions of vorticity, (\ref{eqA1}) and (\ref{eqA2}), and the proton mass current (\ref{eqA8}) yield
\begin{equation}
 \Bbf_L=-\frac{m c}{e}\nabla \times \left(\frac{\jbf_p}{\rho_{pp}}\right)\,. \label{eqA10}
\end{equation}

\subsection{Charge current screening} \label{secAc}

The relativistically degenerate electrons have a very high electrical conductivity.
We therefore employ the MHD approximation and treat the protons and electrons as a single fluid. In this limit, the displacement current is negligible, and \citet{gla11} obtain Ampere's law [see (85) therein]
\begin{eqnarray}
 \nabla\times\Bbf_L&=&\frac{4\pi}{c} \Jbf\,,  \label{eqA13}
\end{eqnarray}
where the charge current density is given by
\begin{eqnarray}
  \Jbf&=&\frac{e \rho_p}{m}\left( \frac{\jbf_p}{\rho_p}- \vbf_e \right)\,, \label{eqA14}
\end{eqnarray}
and $\vbf_e$ is the electron velocity. 
This result differs from that of \citet{men91a,men91b}, who had $\Bbf$ instead of $\Bbf_L$ in (\ref{eqA13}).

We now consider the consequences of the law (\ref{eqA13}) in the smooth-averaged hydrodynamic theory.
Taking the curl of (\ref{eqA14}), and using (\ref{eqA10}) and (\ref{eqA13}) yield
\begin{eqnarray}
  \Lambda^2 \nabla^2 \Bbf_L-\Bbf_L=\frac{mc}{e\rho_{pp}} \nabla \times \vbf_e\,, \label{eqB8}
\end{eqnarray}
where 
\begin{equation}
 \Lambda=\frac{m c}{e} \frac{1}{\sqrt{ 4\pi \rho_{pp} }}\,, \label{eqB9}
\end{equation}
is the London depth.  Equation (\ref{eqB8}) is the London equation in a type II superconductor.  The solutions have 
\begin{eqnarray}
  \Bbf_L=-\frac{mc}{e\rho_{pp}} \nabla \times \vbf_e\,, \label{eqB8a}
\end{eqnarray}
everywhere except in boundary layer regions of length scale $\Lambda$.  
Combining (\ref{eqA10}) and (\ref{eqB8a}), implies  
\begin{eqnarray}
 \Jbf =\frac{e \rho_p}{m}\left( \frac{\jbf_p}{\rho_p}- \vbf_e \right)=0\,.
\end{eqnarray}
This shows that charged currents are screened over length scale $\Lambda$ and vanish in the bulk of a type II superconductor.

In \S\ref{secAa}, we showed that the smooth-averaged hydrodynamic approximation applies for length scales much larger than the intervortex spacing $d_n$.  Scales smaller than $d_n$ are smooth-averaged and do not appear in the hydrodynamic approximation.  This includes the London depth, which determines the size of the flux tubes. Therefore, the term on the left in (\ref{eqB8}) should be neglected in the hydrodynamic approximation.

These arguments make the MHD approximation in a type II superconductor different from the classical case.
In classical MHD, the velocity difference $\vbf_p-\vbf_e=m \Jbf/(\rho_p e) = m c \nabla \times \Bbf /(4\pi \rho e)$ determines the magnetic field.
However, in a type II superconductor, the charge current $\Jbf $ vanishes in the bulk and the electrons move with the proton mass current; Equation (\ref{eqB8}) gives $\vbf_e=\jbf_p/\rho_p$.  The magnetic field is determined by the smooth-averaged magnetic field of the flux tubes and vortex lines, given in (\ref{eqA9}).

\subsection{Final MHD Equations} \label{secAd}

We now list the governing equations for the MHD in neutron star cores.  
The equations are obtained from \citet{gla11}, neglecting the charge current, \ie, taking $\vbf_e=\jbf_p/\rho_p$, for reasons discussed in the previous section.  
The full set of equations includes those in \S\ref{secAb}.
We show that the resulting system satisfies conservation of momentum, energy, vortex lines, and flux tubes as required.

The momentum equations for the neutron and proton--electron fluids are 
\begin{eqnarray}
  \frac{\partial \vbf_n}{\partial t}  + \left(\nabla \times \vbf_n\right) \times \frac{\jbf_n}{\rho_n} &=&-\nabla \left(\tilde{\mu}_n +\frac{1}{2} v_n^2\right)  -\Tbf_n+\Fbf_n\,,  \label{eqB15} \\
\frac{\partial \vbf_p}{\partial t}+ \left(\nabla \times \vbf_p\right)\times \frac{\jbf_p}{\rho_p}  &=&-\nabla \left(\tilde{\mu}_p +\frac{1}{2} v_p^2\right) -\frac{s}{\rho_p}\nabla T-\Tbf_p -\frac{\rho_n}{\rho_p}\Fbf_n +\frac{1}{\rho_p} \nabla_j T^e_{ij} \,,  \label{eqB16}
\end{eqnarray}
where the $\tilde{\mu}_n=\mu_n/m$ is the chemical potential per unit mass for a neutron, $\tilde{\mu}_p\approx(\mu_p+\mu_e)/m$ is the chemical potential per unit mass of a proton--electron pair, $s$ is the specific entropy, and $T$ is the temperature.  
The tension forces are
\begin{eqnarray}
 \Tbf_x&=& \frac{1}{\rho_x} \omegabf_x \times \left( \nabla  \times \rho_x \nu_x \hat{\omegabf}_x  \right) \,, \label{eqB12} 
\end{eqnarray}
where the vortex line tension parameters are defined
\begin{eqnarray}
  \nu_n&=& \frac{\kappa}{4 \pi \rho_n}  \left(\rho_{nn}-\frac{ \rho_{np}^2}{\rho_{pp}} \right) \log\left(\frac{d_n}{\xi_n}\right)\,, \label{eqA24} \\
 \nu_p&=&\frac{\kappa \rho_{pp} }{4 \pi \rho_{p} } \log\left(\frac{\Lambda}{\xi_p}\right)\,. \label{eqA25}
\end{eqnarray}
The tension force $\Tbf_p$ is analogous to magnetic tension in classical MHD, where the tension parameter $\nu_p$ is related to the lower critical field $H_{c1}$ by  (see, \eg, \citealt{eas77})
\begin{equation}
H_{\rm cl}=  \frac{4\pi \rho_{p} e \nu_p}{m c}  \,. \label{eqA30}
\end{equation}
The mutual friction force arises from the scattering of electrons with magnetized vortex lines and pinning interactions and acts equally and oppositely on the two fluids. 
It is given by
\begin{eqnarray}
  \Fbf_n&=& \mathcal{ B}_n \hat{\omegabf}_n \times \left[\omegabf_n \times \left( \frac{\jbf_n}{\rho_n}-\frac{\jbf_p}{\rho_p}  \right)+\Tbf_n\right]+\mathcal{B}_n'  \left[ \omegabf_n \times \left( \frac{\jbf_n}{\rho_n}-\frac{\jbf_p}{\rho_p}  \right)+\Tbf_n\right]  \,,  \label{eqB13}
\end{eqnarray}
where $\mathcal{B}_n$ and $\mathcal{B}'_n$ are the mutual friction coefficients, discussed further in \S\ref{sec3}.
The viscous stress tensor is
\begin{equation}
  T_{ij}^e=\eta \left( \nabla_j \frac{j_{pi}}{\rho_p} +\nabla_i \frac{j_{pj}}{\rho_p} -\frac{2}{3} \delta_{ij} \nabla\cdot \frac{\jbf_p}{\rho_p} \right) +\zeta \delta_{ij} \nabla\cdot \frac{\jbf_p}{\rho_p} \,, \label{eqB17}
\end{equation}
where $\eta$ and $\zeta$ are the shear and bulk viscosities arising from electron--electron scattering.
The energy functional takes the form
\begin{equation}
  \mathcal{E}\left(\rho_n,\rho_p,s,\omegabf_n,\omegabf_p\right)=\mathcal{E}_0 (\rho_n,\rho_p,s)-\frac{\rho_{np}}{2}\left(\vbf_n-\vbf_p\right)^2+\rho_n \nu_n |\omegabf_n| +\rho_p \nu_p |\omegabf_p | \,, \label{eqB18}
\end{equation}
where $\mathcal{E}_0$ is the energy functional in the absence of entrainment, vortices, and flux tubes.
The first law of thermodynamics is
\begin{eqnarray}
 \drm \mathcal{E}&=& T \drm s+ \tilde{\mu}_n \drm \rho_n +\tilde{\mu}_p \drm\rho_p-\rho_{np} \left(\vbf_n-\vbf_p\right) \cdot \drm \left(\vbf_n-\vbf_p\right) \nonumber \\
&+&  \rho_n \nu_n \hat{\omegabf}_n \cdot \drm\omegabf_n+ \rho_p \nu_p \hat{\omegabf}_p \cdot \drm\omegabf_p \,, \label{eqAd31}
\end{eqnarray}
which defines $T$, $\tilde{\mu}_n$, and $\tilde{\mu}_p$.
Note that $\rho_{np}$, $\nu_n$, and $\nu_p$ are functions of $\rho_n$ and $\rho_p$ when calculating $\tilde{\mu}_x$.
The magnetic field evolution is governed by the equations
\begin{eqnarray}
 \frac{\partial \Bbf}{\partial t} &=&-c \nabla \times \Ebf \,, \label{eqA11} \\
 \nabla \cdot \Bbf&=&0 \,, \label{eqA12}
\end{eqnarray}
where the electric field is
\begin{eqnarray}
  \Ebf&=&-\frac{m s}{e \rho_p} \nabla T-\frac{\jbf_p}{c \rho_p} \times \Bbf  - \frac{m}{e} \left(\Fbf_p + \frac{\rho_n}{\rho_p} \Fbf_n\right)+\frac{m}{e \rho_p} \nabla_j  T^e_{ij}\,. \label{eqB20}
\end{eqnarray}
In (\ref{eqB20}) the scattering of electrons from the flux tubes is described by the force
\begin{eqnarray}
  \Fbf_p&=& \mathcal{ B}_p \hat{\omegabf}_p \times \Tbf_p +\mathcal{B}_p'  \Tbf_p \,,  \label{eqB14}
\end{eqnarray}
where $\mathcal{ B}_p$ and $\mathcal{ B}'_p$ are scattering coefficients that relate to the evolution timescales for the magnetic field; see \S\ref{secB}.  

Laws for energy, momentum, vortex line, and flux tube conservation can be derived by combining the above results with the equations in \S\ref{secAb}.
Taking the curl of (\ref{eqB15}) and (\ref{eqB16}), and using (\ref{eqA1}), (\ref{eqA2}), (\ref{eqA11}), and (\ref{eqB20}), we obtain 
\begin{eqnarray}
  \frac{\partial \omegabf_x}{\partial t}+\nabla \times \left( \omegabf_x\times \frac{\jbf_x}{\rho_x}+\Tbf_x-\Fbf_x\right) =0\,, \label{eqA33}
\end{eqnarray}
for $x=n,p$.
Vortex line and flux tube conservation (\ref{eqA34}) is satisfied if the vortex lines and flux tubes obey the equations of motion
\begin{eqnarray}
  \omegabf_x \times \left( \vbf_{Lx}- \frac{\jbf_x}{\rho_x}  \right) =\Tbf_x - \Fbf_x\,. \label{eqB21} 
\end{eqnarray}
Combining (\ref{eqB15}), (\ref{eqB16}), (\ref{eqA6}), (\ref{eqB18}) and (\ref{eqAd31}) gives the momentum conservation law
\begin{equation}
 \frac{\partial }{\partial t}\left(\rho_n \vbf_n+\rho_p \vbf_p\right)+\nabla_j\cdot\left(j_{nj} v_{ni}+ j_{pj}v_{pi}-T^{vp}_{ij}-T^{vn}_{ij}-T^e_{ij}+p\delta_{ij}\right)=0\,, 
\end{equation}
where the pressure is defined as
\begin{eqnarray} 
  p= -\mathcal{E} +\rho_n \tilde{\mu}_n+\rho_p \tilde{\mu}_p + s T \,,
\end{eqnarray} 
and the vorticity stress tensors are
\begin{eqnarray} 
  T_{ij}^{vx}&=& \rho_x \nu_x |\omegabf_x| \left(\hat{\omega}_{xi}\hat{\omega}_{xj} -\delta_{ij} \right) \,.
\end{eqnarray}
Similarly, the conservation of energy equation is
\begin{eqnarray}
 &&\frac{\partial}{\partial t}\left[\frac{1}{2}\rho_n v_n^2+\frac{1}{2}\rho_p v_p^2+\mathcal{E}  \right] \nonumber \\
 &&\,\,+\nabla\cdot\left[\jbf_n\left(\frac{1}{2}v_n^2+\tilde{\mu}_n\right) +\jbf_p\left(\frac{1}{2}v_p^2+\tilde{\mu}_p +\frac{sT}{\rho_p}  \right) -v_{Lnj}T^{vn}_{ij} -  v_{Lpj}T^{vp}_{ij} - \frac{j_{pj}}{\rho_p} T^e_{ij}\right]=0\,, 
\end{eqnarray} 
where entropy equation is
\begin{eqnarray}
  T \left[\frac{\partial s }{\partial t}+ \nabla \cdot \left(  s \frac{\jbf_p}{\rho_p}  \right)\right]&=& \left(\nabla_i \frac{j_{pj}}{\rho_p} \right) T^{ij}_e -\rho_n \Fbf_n\cdot\left(\vbf_{Ln}-\frac{\jbf_p}{\rho_p}\right) -\rho_p \Fbf_p\cdot\left(\vbf_{Lp}-\frac{\jbf_p}{\rho_p}\right) \nonumber \\
&=& \frac{\eta}{2}\left( \nabla_j \frac{j_{pi}}{\rho_p} +\nabla_i \frac{j_{pj}}{\rho_p} -\frac{2}{3} \delta_{ij} \nabla\cdot \frac{\jbf_p}{\rho_p} \right)^2 +\zeta \left( \nabla\cdot \frac{\jbf_p}{\rho_p} \right)^2  \nonumber\\
&+&\rho_n \mathcal{B}_n|\omegabf_n| \left[ \Ubf_n^2 - \left(\hat{\omegabf}_n\cdot \Ubf_n \right)^2 \right] +\rho_p \mathcal{B}_p|\omegabf_p| \left[ \Ubf_p^2  - \left(\hat{\omegabf}_p\cdot\Ubf_p \right)^2 \right]   \,, \label{eqB24}
\end{eqnarray}
and
\begin{eqnarray}
  \Ubf_n&=& \frac{\jbf_n}{\rho_n}- \frac{\jbf_p}{\rho_p}+\frac{1}{\rho_p }\nabla \times \left( \rho_n \nu_n \hat{\omegabf}_n  \right) \,, \\
  \Ubf_p&=& \frac{1}{\rho_p }\nabla \times\left( \rho_p \nu_p \hat{\omegabf}_p  \right)\,.
\end{eqnarray}

In the case of constant density, $\drm \rho_n =\drm \rho_p=0$ and 
\begin{eqnarray}
 \drm \mathcal{E}_0= T \drm s \,, \label{eqC27}
\end{eqnarray}
and therefore
\begin{eqnarray}
 \drm \left(\mathcal{E}_0 -sT\right)= -s \drm T \,. \label{eqC28}
\end{eqnarray}
It is convenient to define 
\begin{eqnarray}
  p_n &=&\tilde{\mu}_n +\frac{1}{2} v_n^2 \,, \nonumber \\
  p_p &=& \tilde{\mu}_p +\frac{1}{2} v_p^2 +\frac{s T}{\rho_p}-\frac{\mathcal{E}_0}{\rho_p} \,, \label{eqC29}
\end{eqnarray}
which are used to obtain (\ref{eq5}) and (\ref{eq6}).

\subsection{Perturbation equations for constant density} \label{secAe}

In this section, we present the perturbation equations for the hydrodynamic theory presented in the previous section, assuming constant density flow.  
These equations have been used in a more general search for instabilities than those considered in the main body of the paper.
We find no new instabilities of interest for this more general set of equations, but we present the analysis here for completeness.

As described in \S\ref{sec2}, we restrict our study to constant density flows satisfying (\ref{eq4}): 
\begin{eqnarray}
  \nabla \cdot \vbf_x&=& 0 \,,  \label{eqB29}
\end{eqnarray}
for $x=n,p$. 
It is convenient to solve the the system using the equations of vortex line conservation (\ref{eqA34}):
\begin{equation}
  \frac{\partial \omegabf_x}{\partial t}+\nabla \times \left( \omegabf_x\times \vbf_{Lx}  \right) =0\,,\label{eqB30} 
\end{equation}
where the solutions to the vortex line equations of motion (\ref{eqB21}), given the mutual friction forces (\ref{eqB13}) and (\ref{eqB14}), are
\begin{eqnarray} 
   \vbf_{Ln}  &=&\frac{\jbf_p}{\rho_p}+\left(1-\mathcal{B}'_n\right) \left[ \left( \frac{\jbf_n}{\rho_n}-\frac{\jbf_p}{\rho_p} \right) - \hat{\omegabf}_n \times \frac{\Tbf_n}{|\omegabf_n|}  \right]-\mathcal{B}_n \left[ \hat{\omegabf}_n \times \left( \frac{\jbf_n}{\rho_n}-\frac{\jbf_p}{\rho_p}  \right)+\frac{\Tbf_n}{|\omegabf_n|} \right]  \, \nonumber \\
&+& \mbox{additional terms along $\hat{\omegabf}_n$}  \,,  \label{eqB31} \\
  \vbf_{Lp}&=&\frac{\jbf_p}{\rho_p}-\left(1-\mathcal{B}'_p\right)\hat{\omegabf}_p\times \frac{\Tbf_p}{|\omegabf_p|} -\mathcal{B}_p\frac{\Tbf_p}{|\omegabf_p|}  \nonumber \\
&+& \mbox{additional terms along $\hat{\omegabf}_p$} \,.  \label{eqB32}
\end{eqnarray}
The additional terms along $\hat{\omegabf}_x$ are inconsequential to the dynamics and are henceforth omitted for simplicity and clarity.  Equations (\ref{eqB31}) and (\ref{eqB32}) satisfy the equilibrium vortex line equations of motion (\ref{eqB21}) as required. 
The magnetic field evolves according to the induction equation:
\begin{eqnarray}
 \frac{\partial \Bbf}{\partial t} &=&-c \nabla \times \Ebf \,, \label{eqB33} 
\end{eqnarray}
Using the vortex line velocity equations (\ref{eqB21}), Ohm's law can be expressed in terms of the vortex line velocities, which gives \begin{eqnarray}
 \Ebf&=&-\frac{\vbf_{Lp}}{c} \times \Bbf  + \frac{m}{e} \left\{\left( \nabla\times \vbf_p \right) \times \left( \vbf_{Lp}-\frac{\jbf_p}{\rho_p} \right) + \frac{\rho_n}{\rho_p} \omegabf_n \times \left( \vbf_{Ln}-\frac{\jbf_n}{\rho_n} \right) -\Tbf_p-\frac{\rho_n}{\rho_p}\Tbf_n + \frac{ \eta }{ \rho_p} \nabla^2 \frac{\jbf_p}{\rho_p} \right\} \,. \label{eqB34}
\end{eqnarray}
For constant density flow, we use (\ref{eqC28}), and the gradient term is inconsequential to the dynamics.
The governing equations are the continuity equation (\ref{eqB29}), the vortex line conservation equation (\ref{eqB30}), and the vortex line velocities (\ref{eqB31}) and (\ref{eqB32}), and the electric field is (\ref{eqB34}). These equations comprise a closed system for the variables $\vbf_{n,p}$ and $\Bbf $. 

We now perturb the equations about the equilibrium described in \S\ref{sec2b}.  The equilibrium velocities are (\ref{eq12}) and (\ref{eq13}). 
To zeroth order the vortex line equations of motion give
\begin{eqnarray}
  \vbf_{Ln0}&=&\frac{\jbf_{p0}}{\rho_p}-\mathcal{B}_n \hat{\omegabf}_n \times \left(\frac{\jbf_{n0}}{\rho_n}-\frac{\jbf_{p0}}{\rho_p}\right) +\left(1-\mathcal{B}'_n\right) \left(\frac{\jbf_{n0}}{\rho_n}-\frac{\jbf_{p0}}{\rho_p}\right) \,,  \label{eqB35} \\
 \vbf_{Lp0}&=&\frac{\jbf_{p0}}{\rho_p}  \,, \label{eqB36}
\end{eqnarray}
where the equilibrium mass currents are
\begin{eqnarray}
  \jbf_{n0}&=&R \left[ \rho_n \Omega_n-\rho_{np} \Delta\Omega \right] \hat{y}+\Delta v_z \hat{z} \,,\\
  \jbf_{p0}&=&R \left[ \rho_p \left(\Omega_n-\Delta \Omega \right)+\rho_{np} \Delta\Omega \right] \hat{y} \,,
\end{eqnarray}
and equilibrium vorticity is 
\begin{eqnarray}
  \omegabf_{n0}&=&2 \Omega_n \hat{z} \,, \\ 
  \omegabf_{p0}&=&\frac{e B_{0y}}{m c} \hat{y}  + \left[ 2 \left(\Omega_n -\Delta \Omega \right)+\frac{e B_{0z}}{m c}\right] \hat{z} \,.
\end{eqnarray}

Denoting perturbed quantities by $\delta$, the perturbed vorticity conservation equation is
\begin{eqnarray}
  \frac{\partial \delta \omegabf_x}{\partial t}+\nabla \times \left(  \delta \omegabf_x \times \vbf_{L0x}+\omegabf_{0x}\times \delta\vbf_{Lx}\right)=0\,, \label{eqB41}
\end{eqnarray}
where the perturbed vortex line velocities are
\begin{eqnarray}
  \delta \vbf_{Ln}&=& \frac{\delta \jbf_{p}}{\rho_p} +\left(1-\mathcal{B}_n'\right)\left( \frac{\delta \jbf_{n}}{\rho_n}-\frac{\delta \jbf_{p}}{\rho_p}  - \hat{z}\times \frac{\delta \Tbf_n}{|\omegabf_{n0}|} \right)  \nonumber \\
&-&\mathcal{B}_n \left[\hat{z}\times \left(\frac{\delta \jbf_{n}}{\rho_n}-\frac{\delta \jbf_{p}}{\rho_p}\right)+\delta\hat{\omegabf}_n\times \left(\frac{\jbf_{n0}}{\rho_n}-\frac{\jbf_{p0}}{\rho_p}\right) +  \frac{\delta \Tbf_n}{|\omegabf_{n0}|} \right] \,, \label{eqB42}  \\
  \delta \vbf_{Lp}&=& \frac{\delta \jbf_{p}}{\rho_p} -\left(1-\mathcal{B}_p'\right)\hat{\omegabf}_{p0} \times \frac{\delta \Tbf_p}{|\omegabf_{p0}|} - \mathcal{B}_p \frac{\delta \Tbf_p}{|\omegabf_{p0} |}  \label{eqB43} \,,
\end{eqnarray}
and 
\begin{eqnarray}
  \delta \Tbf_x &=& -\nu_x \left( \omegabf_{x0}\cdot\nabla \right)\delta\hat{\omegabf}_x\,, \\
 \delta\hat{\omegabf}_x&=& \frac{1}{|\omegabf_{x0}|}\left[\delta \omegabf_x - \hat{\omegabf}_{x0} \left(\hat{\omegabf}_{x0}\cdot  \delta \omegabf_x \right)\right]\,.
\end{eqnarray} 
The perturbed induction equation is 
\begin{eqnarray}
  \frac{\partial \delta\Bbf }{\partial t} &=& \nabla \times \left\{ \vbf_{Lp0} \times \delta\Bbf +\delta\vbf_{Lp} \times \Bbf_0   - \frac{mc}{e} \left[ \left( \nabla\times \vbf_{p0} \right) \times \left( \delta \vbf_{Lp}-\frac{\delta \jbf_p}{\rho_p} \right) \right.\right.\nonumber \\
&+& \left( \nabla\times \delta \vbf_p \right) \times \left( \vbf_{Lp0}-\frac{\jbf_{p0}}{\rho_p} \right)+ \frac{\rho_n}{\rho_p}\omegabf_{n0} \times \left( \delta \vbf_{Ln}-\frac{ \delta \jbf_n}{\rho_n} \right)+ \frac{\rho_n}{\rho_p}\delta\omegabf_{n} \times \left( \vbf_{Ln0}-\frac{\jbf_{n0}}{\rho_n} \right)  \nonumber \\
&-& \left.\left.\delta \Tbf_p-\frac{\rho_n}{\rho_p}\delta \Tbf_n + \frac{ \eta }{ \rho_p} \nabla^2 \frac{\delta \jbf_p}{\rho_p} \right] \right\} \,. \label{eqB46}
\end{eqnarray}

At this point, the restoring force for deformations of the vortex lattice from its equilibrium can be included by making the replacement
\begin{eqnarray}
  \delta \Tbf_n &=& -\nu_n \left( \omegabf_{n0}\cdot\nabla \right)\delta\hat{\omegabf}_n + \frac{\kappa |\omegabf_{n0}|}{8 \pi} \left[2 \nabla_{\perp}\left(\nabla\cdot \xibf \right)-\nabla_\perp^2  \xibf  \right]\,, \label{eqB47}
\end{eqnarray} 
where the vortex line displacement vector $ \xibf $ is related to the vorticity perturbation by
\begin{eqnarray}
  \delta \hat{\omegabf}_n &=& \hat{\omegabf}_{n0}\cdot \nabla  \xibf \,.
\end{eqnarray} 
Equation (\ref{eqB47}) was derived by \citet{bay83} and describes Tkachenko oscillations \citep{tka66}, but is only valid for the linear theory.

The dispersion relation is obtained by satisfying the continuity equations using (\ref{eq19}) and (\ref{eq20}).  The $x$ and $y$ components of (\ref{eqB41}) and (\ref{eqB46}) give a matrix system of six equations in the unknowns
$\psi_{nx}$, $\psi_{ny}$, $\psi_{px}$, $\psi_{py}$, $A_x$ and
$A_y$. 
The complete dispersion relation is extremely lengthy, and we do not present it here.

\section{Neglecting Kelvin waves and magnetic field evolution} \label{secB}

The equations in \S\ref{secA} can be simplified for our stability analysis by observing that Kelvin waves and magnetic field evolution occur over timescales much longer than those of interest for the unstable modes.  
In this section, we present scaling arguments demonstrating which terms can be neglected.  
We then present the simplest set of equations appropriate for studying oscillation modes in the outer core of neutron stars.   

Kelvin waves occur in superfluids and type II superconductors, modifying the inertial mode frequencies.
The frequency of Kelvin waves depends on the tension parameters, given by (\ref{eqA24}) and (\ref{eqA25}).  
Using the numbers in \S\ref{secAa} gives
\begin{eqnarray}
  \nu_n &=& 4\times 10^{-3} \,{\rm cm^2\,s^{-1}}\,, \nonumber \\
 \nu_p &=& 10^{-4} \,{\rm cm^2\,s^{-1}}\,. \label{eqC1}
\end{eqnarray}
Comparing the tension forces in $\Tbf_p$ and $\Tbf_n$ that give Kelvin waves with the Coriolis force gives 
\begin{eqnarray}
\frac{\nu_n}{\Omega_n R^2}&=&6\times 10^{-17}\, \left(\frac{\nu_n}{4\times 10^{-3} \,{\rm cm^2\,s^{-1}}}\right) \left(\frac{R}{10^6\,{\rm cm}}\right)^{-2} \left(\frac{\Omega_n}{20\pi\, {\rm rad\,s^{-1}}}\right)^{-1}\,,   \label{eqC2} \\
 \frac{\nu_p}{\Omega_n R^2}&=&2\times 10^{-18}\, \left(\frac{\nu_p}{1\times 10^{-4} \,{\rm cm^2\,s^{-1}}}\right) \left(\frac{R}{10^6\,{\rm cm}}\right)^{-2} \left(\frac{\Omega_n}{20\pi\, {\rm rad\,s^{-1}}}\right)^{-1}\,.\label{eqC3}
\end{eqnarray}
Therefore the contribution from Kelvin waves is small.
The restoring force for Tkachenko modes \citep{tka66,bay83} scales as $\kappa/(\Omega_{n} R^2)$ [see Equation (\ref{eqB47})], and therefore enters at the same order of magnitude as the neutron vortex tension.

To neglect Kelvin waves in the superfluid, we take $\nu_n=0$.  However, in a superconductor, the flux tube tension $\Tbf_p$ produces both Kelvin waves and the superconducting equivalent to Alfv\'en waves, the vortex-cyclotron modes with frequency $\sqrt{H_{c1} B_0/4\pi\rho_p}\, $ \citep{eas79a,men98}, where
\begin{eqnarray}
 H_{\rm cl} = 4\times 10^{14}  \left(\frac{\rho}{3\times 10^{14} \,{\rm g\,cm^{-3}}}\right)\left(\frac{x_p}{0.1}\right)\left(\frac{\nu_p }{10^{-4} \,{\rm cm^2\,s^{-1}}}\right)^{-1/2} \,{\rm G} \,. \label{eqC4}
\end{eqnarray}
The flux tube tension parameter $\nu_p$ is relevant for both Kelvin waves {\it and} the vortex-cyclotron waves through $H_{c1}$, and therefore cannot be assumed to be zero.  
Assuming the neutron and proton--electron fluids have comparable rotation rates, the ratio of the rotation term to the magnetic field term in (\ref{eqA2}) is $e B_0/2 \Omega_p mc \sim 10^{-14}$, as demonstrated by (\ref{eq3}).
The contributions to the magnetic field in (\ref{eqA9}) from the London field and entrained neutron currents are negligible, giving $\Bbf=n_{vp} \phi_0 \hat{\bbf}$. 
Therefore, we approximate $\omegabf_p \sim e \Bbf/mc$.  
This neglects the rotational contribution to $\omegabf_p$ in (\ref{eqA2}), and using (\ref{eqA30}), the tension force in the proton--electron fluid reduces to (\ref{eq8}).

Next, we consider the evolution of the magnetic field through the mutual friction forces appearing in (\ref{eqB20}).  Using the definition (\ref{eqA2}), the electric field is
\begin{eqnarray}
  \Ebf&=&-\frac{\jbf_p}{c \rho_p} \times \Bbf  - \frac{\mathcal{ B}_p}{c \rho_p } \Bbf \times \left[ \hat{\omegabf}_p \times \left( \nabla  \times \rho_p\nu_p \hat{\omegabf}_p  \right)\right] - \frac{\mathcal{B}_p'}{c \rho_p }  \Bbf \times \left( \nabla  \times \rho_p \nu_p \hat{\omegabf}_p  \right)  \nonumber \\
&-& \frac{m \mathcal{ B}_p}{e \rho_p } \left(\nabla \times \vbf_p\right) \times \left[ \hat{\omegabf}_p \times \left( \nabla  \times \rho_p\nu_p \hat{\omegabf}_p  \right)\right] - \frac{m \mathcal{B}_p' }{e \rho_p }  \left(\nabla \times \vbf_p\right) \times \left( \nabla  \times \rho_p \nu_p \hat{\omegabf}_p  \right)  \nonumber \\
&-& \frac{m \mathcal{ B}_n}{e \rho_p }  \hat{\omegabf}_n \times \left[ \omegabf_n \times \left( \nabla  \times \rho_n \nu_n \hat{\omegabf}_n  \right)\right]-\frac{m \mathcal{B}_n' }{e \rho_p}  \left[ \omegabf_n \times \left(\nabla  \times \rho_n \nu_n \hat{\omegabf}_n  \right)\right] \nonumber \\
&-& \frac{m \rho_n \mathcal{ B}_n }{e \rho_p}\hat{\omegabf}_n \times \left[\omegabf_n \times \left( \frac{\jbf_n}{\rho_n}- \frac{\jbf_p}{\rho_p}  \right)\right] -\frac{m \rho_n \mathcal{B}_n' }{e \rho_p} \left[ \omegabf_n \times \left( \frac{\jbf_n}{\rho_n}- \frac{\jbf_p}{\rho_p}   \right)\right]\nonumber \\
&-&\frac{m s}{e \rho_p} \nabla T +\frac{m}{e \rho_p} \nabla_j  T^e_{ij}\,. \label{eqC5} 
\end{eqnarray}
The first term in (\ref{eqC5}) is the dominant term in the induction equation and scales as $\Omega_n R B_0/c$.
This term comprises the electric field in the ideal MHD limit.
The second and third terms in (\ref{eqC5}), typically written in terms of $H_{c1}$ using (\ref{eqA30}), describe processes analogous to conventional MHD.
The term parameterized by the coefficient $\mathcal{B}_p$ describes dissipative forces that produce entropy according to (\ref{eqB24}), and describes an effect analogous to ohmic diffusion.
The contributions from $\mathcal{B}'_p$ are dissipation-less and parameterize a process analogous to Hall diffusion (see, \eg, \citealt{gra15,pas17b}).  
To assess the relevance of these terms, we evaluate the mutual friction coefficients $\mathcal{B}_p$ and $\mathcal{B}'_p$ in a manner analogous to $\mathcal{B}_n$ and $\mathcal{B}'_n$ in \S\ref{sec3}.
In terms of the scattering coefficients, the mutual friction coefficients are
\begin{eqnarray}
  \mathcal{B}_p&=&  \frac{ \mathcal{R}_{p}}{1+\mathcal{R}_{p}^2} \,, \nonumber \\
  \mathcal{B}'_p&=& \frac{\mathcal{R}_p^2}{1+\mathcal{R}_{p}^2} \,, 
\end{eqnarray}
where the scattering coefficient is related to the scattering time $\tau_{sp}$ by $\mathcal{R}_{p} = ( |\omegabf_p| \tau_{sp})^{-1}$.
The relaxation time for the electron distribution function due to relativistic electron scattering from a flux tube and a magnetized neutron vortex was calculated by \citet{alp84a}, \citet{har86} and \citet{jon87}, yielding
\begin{eqnarray}
   \mathcal{R}_{p}&=&  \frac{3 \pi  e^2 \phi_0^2}{64 m_n c E_{Fe} \Lambda \kappa } \,, \label{eqC7} 
\end{eqnarray}
Comparing (\ref{eqC7}) with (\ref{eq26}) and using (\ref{eq35z}), we obtain the approximate expression
\begin{eqnarray}
   \mathcal{R}_{p}&=&0.011 \left(\frac{m_p}{m_p^*}\right)^{1/2} \left(\frac{x_p^{1/6}}{1-x_p}\right)\left(\frac{\rho}{3\times 10^{14}\,{\rm g\,cm^{-3}}}\right)^{1/6}\,. \label{eqC8}
\end{eqnarray}
For typical neutron star numbers, we find $\mathcal{R}_x \ll 1$, and therefore  
\begin{eqnarray}
\mathcal{B}_{p} &\approx&\mathcal{R}_p= 10^{-2} \,, \nonumber \\
  \mathcal{B}'_{p} &\approx&\mathcal{R}_p^2= 10^{-4} \,,  \,. \label{eqC9} 
\end{eqnarray}
The relative sizes for the second and third terms of $\Ebf$ in (\ref{eqC5}) compared with the first term are are $\nu_p\mathcal{B}_p/(R^2 \Omega_n)$ and $\nu_p\mathcal{B}'_p/(R^2 \Omega_n)$ respectively, giving
\begin{eqnarray}
\frac{\nu_p \mathcal{B}_p}{R^2 \Omega_n} &=& 3\times 10^{-20} \left( \frac{\nu_p}{ 10^{-4} \,{\rm cm^2\,s^{-1}}} \right)  \left( \frac{\mathcal{R}_{p}}{ 10^{-2}} \right) \left( \frac{R }{10^{6} \, {\rm cm} } \right)^{-2}\left(\frac{\Omega_n}{20\pi \,{\rm rad\,s^{-1}}}\right)^{-1} \,, \label{eqC10} \\
 \frac{\nu_p\mathcal{B}'_p}{R^2 \Omega_n} &=& 4\times 10^{-22}  \left( \frac{\nu_p}{ 10^{-4} \,{\rm cm^2\,s^{-1}}} \right)  \left( \frac{\mathcal{R}_{p}}{ 10^{-2}} \right)^2 \left( \frac{R }{10^{6} \, {\rm cm} } \right)^{-2}\left(\frac{\Omega_n}{20\pi\, {\rm rad\,s^{-1}}}\right)^{-1}\,.  \label{eqC11} 
\end{eqnarray}
Therefore the terms associated with the evolution of the magnetic field are extremely small and are neglected for the study of inertial modes considered in this paper. 

Now consider the fourth and fifth terms in (\ref{eqC5}).  
Assuming the rotational velocities of the neutron and proton--electron fluids are comparable, the second and third terms are of order $2\Omega_n m c/(e B_0)\sim 10^{-14}$; see Equation (\ref{eq3}).
Therefore the fourth and fifth terms are negligible compared with the magnetic field evolution terms.
The sixth and seventh terms scale similarly and are also negligible.

The eighth and ninth terms in (\ref{eqC5}) contain the velocity difference between neutron and proton--electron fluids and produce processes analogous to ambipolar diffusion, that is, the rearrangement of the magnetic field resulting from the scattering of the proton--electron fluid with a neutral species, \ie, neutrons.  In contrast with classic ambipolar diffusion, there are dissipative and nondissipative contributions.
The term proportional to $\mathcal{B}_n$ is dissipative, producing entropy according to (\ref{eqB24}), while the term proportional to $\mathcal{B}'_n$ is nondissipative.
The relative sizes of each term depends compared with the first term in (\ref{eqC5}) are $\rho_n mc \Delta \Omega  \mathcal{B}_n/\rho_p e B_0$ and $\rho_n mc \Delta \Omega  \mathcal{B}'_n/\rho_p e B_0$ respectively.
In the pinning regime, the mutual friction coefficients are given by (\ref{eq33}) and (\ref{eq34}), yielding
\begin{eqnarray}
  \frac{\rho_n mc  \Delta \Omega \mathcal{B}_n }{\rho_p e B_0 } &=& 2\times 10^{-27}  \left( \frac{x_p}{ 0.1} \right)^{-1}  \left( \frac{n_{vp}/n_{vn}}{ 8\times 10^{13}} \right)^{-1} \left( \frac{\Omega_n }{20\pi  \, {\rm rad\,s^{-1}} } \right)^{-1}\left(\frac{\tau_{sd}}{10 \, {\rm kyr} }\right)^{-1}\,,  \\
 \frac{\rho_n mc \Delta \Omega  \mathcal{B}'_n }{\rho_p e B_0 } &=& 2\times 10^{-16} \left( \frac{x_p}{ 0.1} \right)^{-1}  \left( \frac{n_{vp}/n_{vn}}{ 8\times 10^{13}} \right)^{-1} \left( \frac{\Omega_n }{20\pi  \, {\rm rad\,s^{-1}} } \right)^{-1}\left(\frac{ \Delta \Omega_{crit} }{0.1 \, {\rm rad\,s^{-1}} }\right) \,. 
\end{eqnarray} 
Therefore the rearrangement of the magnetic field due to the dragging of flux tubes by the vortex lines may be the dominant mechanism for magnetic field evolution.
Note, however, that the equations presented in this paper only approximate pinning effects by defining (\ref{eq24a}) and (\ref{eq24}). These estimates may change with the development of a more rigorous incorporation of pinning into the hydrodynamics.  This will be considered in future work.

We now list the final equations neglecting Kelvin waves and magnetic field evolution terms.  
The continuity equation for the entrained neutron and proton condensates is
\begin{eqnarray}
 \frac{\partial \rho_x}{\partial t}+\nabla\cdot\jbf_x&=&0\,,
\end{eqnarray}
where the mass currents are defined as
\begin{eqnarray}
 \jbf_n&=&\rho_{nn}\vbf_n +\rho_{np} \vbf_p\,, \\
 \jbf_p&=&\rho_{pp} \vbf_p+\rho_{np} \vbf_n\,.
\end{eqnarray}
The momentum equations for the neutron and proton--electron fluids are 
\begin{eqnarray}
  \frac{\partial \vbf_n}{\partial t}  + \left(\nabla \times \vbf_n\right) \times \frac{\jbf_n}{\rho_n} &=&-\nabla \left(\tilde{\mu}_n +\frac{1}{2} v_n^2\right)  +\Fbf_n\,,  \label{eqC16} \\
\frac{\partial \vbf_p}{\partial t}+ \left(\nabla \times \vbf_p\right)\times \frac{\jbf_p}{\rho_p}  &=&-\nabla \left(\tilde{\mu}_p +\frac{1}{2} v_p^2\right) -\frac{s}{\rho_p}\nabla T-\Tbf_p -\frac{\rho_n}{\rho_p}\Fbf_n +\frac{1}{\rho_p} \nabla_j T^e_{ij} \,,  \label{eqC17} 
\end{eqnarray}
where the flux tube tension is
\begin{eqnarray}
\Tbf_p= \frac{\Bbf}{4\pi \rho_p} \times \nabla \times \left( H_{c1} \hat{\bbf}\right)\,.
\end{eqnarray}
The energy functional takes the form
\begin{equation}
  \mathcal{E}\left(\rho_n,\rho_p,s,\omegabf_n,\omegabf_p\right)=\mathcal{E}_0 (\rho_n,\rho_p,s)-\frac{\rho_{np}}{2}\left(\vbf_n-\vbf_p\right)^2+ \frac{ H_{c1}  |\Bbf |}{4\pi} \,, \label{eqC19}
\end{equation}
and the first law of thermodynamics is
\begin{eqnarray}
 \drm \mathcal{E}&=& T \drm s+ \tilde{\mu}_n \drm \rho_n +\tilde{\mu}_p \drm\rho_p-\rho_{np} \left(\vbf_n-\vbf_p\right) \cdot \drm \left(\vbf_n-\vbf_p\right) + \frac{ H_{c1} \hat{\bbf} }{4 \pi}  \cdot \drm\Bbf \,.
\end{eqnarray}
Note that $\rho_{np}$ and $H_{c1}$ are functions of $\rho_n$ and $\rho_p$ when calculating $\tilde{\mu}_x$.
The induction equation is
\begin{eqnarray}
  \frac{\partial \Bbf}{\partial t}&=&\nabla \times \left(\frac{\jbf_p}{\rho_p} \times \Bbf \right)\,. \label{eqC21} 
\end{eqnarray}
Combining (\ref{eqC16})--(\ref{eqC21}), we obtain the equation for energy conservation: 
\begin{eqnarray}
 &&\frac{\partial}{\partial t}\left[\frac{1}{2} \rho_n v_n^2+\frac{1}{2}\rho_p v_p^2+\mathcal{E}  \right] \nonumber \\
 &&\,\,+\nabla\cdot\left[\jbf_n\left(\frac{1}{2}v_n^2+\tilde{\mu}_n\right) +\jbf_p\left(\frac{1}{2}v_p^2+\tilde{\mu}_p +\frac{sT}{\rho_p}  \right) - \frac{j_{pj}}{\rho_p} T^{vp}_{ij} - \frac{j_{pj}}{\rho_p} T^e_{ij}\right]=0\,,
\end{eqnarray} 
where the stress tensor for the flux tubes is
\begin{eqnarray}
  T^{vp}_{ij}= \frac{H_{c1} |\Bbf| }{4\pi}  \left( \hat{b}_i \hat{b}_j-\delta_{ij}\right)\,,
\end{eqnarray}
which is the result obtained by \citet{eas77}.
The entropy equation is
\begin{eqnarray}
  T \left[\frac{\partial s }{\partial t}+ \nabla \cdot \left(  s \frac{\jbf_p}{\rho_p}  \right)\right]&=& \left(\nabla_i \frac{j_{pj}}{\rho_p} \right) T^e_{ij}  + \rho_n \mathcal{B}_n|\omegabf_n| \left[ \Ubf_n^2 - \left(\hat{\omegabf}_n\cdot \Ubf_n \right)^2 \right]    \,, 
\end{eqnarray}
where
\begin{eqnarray}
  \Ubf_n&=& \frac{\jbf_n}{\rho_n}- \frac{\jbf_p}{\rho_p}  \,.
\end{eqnarray}
The momentum conservation law is 
\begin{equation}
 \frac{\partial }{\partial t}\left(\rho_n \vbf_n+\rho_p \vbf_p\right)+\nabla_j\cdot\left(j_{nj} v_{ni}+ j_{pj}v_{pi}-T^{vp}_{ij}-T^e_{ij}+ p\delta_{ij} \right)=0\,.
\end{equation}

For constant density flow, the equations in \S\ref{sec2b} are recovered using the results (\ref{eqC27})--(\ref{eqC29}).

\section{Thermal Activation} \label{secD}

In this paper, we have assumed that the mutual friction coefficients
in $\mathcal{B}_n$ and $\mathcal{B}'_n$ are independent of the
rotational states of the neutron and proton fluids. The fraction of
unpinned vorticity depends on temperature, the details of the pinning interaction and the relative velocity between the two condensates \citep{lin14}; in
particular, thermal effects give rise to nonlinear dependence of the
mutual friction coefficients on the velocity differences, the
consequences of which we study in this appendix.

Recall that the mutual friction coefficients defined in (\ref{eq7}) are \citep{lin14},
\begin{eqnarray}
 \mathcal{B}_n&=&  \frac{e^{-\beta A} \mathcal{R}_n}{1+\mathcal{R}_n^2} \,, \label{eqD1} \\
1-\mathcal{B}'_n&=& \frac{e^{-\beta A}}{1+\mathcal{R}_n^2} \label{eqD2} \,.  
\end{eqnarray}
where ${\rm e}^{-\beta A}$ is the fraction of unpinned vorticity, $\beta^{-1}=k_B T$, $k_B$ is Boltzmann's constant, $T$ is the temperature, and $A$ is the activation energy for unpinning. 
The activation energy required to unpin a vortex is derived from the force balance on a vortex filament and depends on the magnetic energy between a vortex and a flux tube $E_p$, a dimensionless vortex tension $\mathcal{T}$, the relative velocity between the fluids, and the critical velocity for unpinning $R \Omega_{crit}$ as \citep{lin14}
\begin{equation}
  A\left(\vert
\hat{\omegabf}_n\times (\vbf_n-\vbf_p)\vert \right)=5.1 E_p \mathcal{T}^{1/2}\left(1-\frac{\vert
\hat{\omegabf}_n\times (\vbf_n-\vbf_p)\vert }{R \Omega_{crit}} \right)^{5/4}\,. \label{eqD3}
\end{equation}
Equation (\ref{eqD3}) shows that the mutual friction coefficients (\ref{eqD1}) and (\ref{eqD2})  depend on $\vert \hat{\omegabf}_n\times (\vbf_n-\vbf_p)\vert$ through $A$; \ie, they are nonconstant coefficients. 
Thermal effects give the mutual friction of (\ref{eq7})
nonlinear dependence on the velocity difference between the neutron
and proton condensates.

Consider the equilibrium studied in \S\ref{sec2b}.  To leading order, the mutual friction coefficients (\ref{eqD1})--(\ref{eqD3}) are
\begin{eqnarray}
 \mathcal{B}_{n0} &=&  \frac{e^{ - \beta A_0} \mathcal{R}_n}{1+\mathcal{R}_n^2}\,, \label{eqD4} \\
 1-\mathcal{B}'_{n0} &=& \frac{e^{ - \beta A_0}}{1+\mathcal{R}_n^2} \label{eqD5} \,,
\end{eqnarray}
where
\begin{equation}
  A_0\left( \Delta \Omega \right)=5.1 E_p \mathcal{T}^{1/2}\left(1-\frac{\Delta \Omega }{\Delta\Omega_{crit} } \right)^{5/4}\,. \label{eqD6} 
\end{equation}
Using the result (\ref{eqD4}), and approximating $\mathcal{R}_n \ll 1$, the equation for the equilibrium lag (\ref{eq14z}) becomes
\begin{eqnarray}
   \Delta\Omega = \left(4\tau_{sd}  \mathcal{R}_n \right)^{-1}  e^{   5.1 E_p \beta\mathcal{T}^{1/2}\left(1-\frac{\Delta \Omega }{\Delta\Omega_{crit} } \right)^{5/4}}   \,.\label{eqD7}
\end{eqnarray}
Equation (\ref{eqD7}) must be solved for the equilibrium lag $\Delta \Omega$ using typical pulsar and pinning parameters. 
In the outer core we expect \citep{lin14}
\begin{eqnarray}
   E_p \beta \mathcal{T}^{1/2} = 4\times 10^2 \, \left(\frac{E_p}{100\,{\rm MeV}}\right)\left(\frac{k_B T}{10\,{\rm keV}}\right)^{-1}\left(\frac{\mathcal{T}}{0.12}\right)^{1/2} \,. \label{eqD8}
\end{eqnarray}
For the pulsar parameters in \S\ref{sec3}, (\ref{eqD7}) and (\ref{eqD8}) give $\Delta\Omega = 0.996 \Delta \Omega_{crit}$.  Therefore, the equilibrium lag is approximately $\Delta \Omega\approx \Delta \Omega_{crit}$, and the scaling arguments used to derive the estimates  (\ref{eq32a}) and (\ref{eq32b}) are applicable to the equilibrium mutual friction coefficients (\ref{eqD4}) and (\ref{eqD5}).

Perturbing the mutual friction force (\ref{eq7}) with the mutual friction coefficients given by (\ref{eqD1})--(\ref{eqD3}) yields 
\begin{eqnarray}
  \delta \Fbf_n&=& \mathcal{B}_{n0} R\Delta \Omega \hat{x} \times \left(\nabla \times \delta \vbf_n \right) -  \mathcal{B}_{n0} \Delta v_z \hat{z}\times \left[ \hat{z}\times \left(\nabla \times \delta \vbf_n \right) \right] +\mathcal{B}_{n0}  2\Omega_n \hat{z}\times\left[ \hat{z}\times \left(\delta \vbf_n -\delta \vbf_p\right) \right] \nonumber \\
&+& \mathcal{B}_{n0} \hat{z} \times \delta \Tbf_n -\mathcal{ B}_{n0}   \alpha \left[2\Omega_n\left(\delta \vbf_n-\delta \vbf_p\right) -\Delta v_z \left(\nabla\times \delta\vbf_n\right)  \right]\cdot\hat{y} \, \hat{y} \nonumber \\
&+& \mathcal{B}'_n \left(\nabla \times \delta \vbf_n \right)\times \left( R\Delta \Omega \hat{y} + \Delta v_z \hat{z} \right)  +  \mathcal{B}'_n 2\Omega_n \hat{z} \times \left(\delta \vbf_n -\delta \vbf_p\right) \nonumber \\
&+& \mathcal{ B}'_n \delta \Tbf_n -\left( \mathcal{B}'_{n0}-1\right) \alpha \left[2\Omega_n\left(\delta \vbf_n-\delta \vbf_p\right) -\Delta v_z \left(\nabla\times \delta\vbf_n\right)  \right]\cdot\hat{y} \,   \hat{x} \,. \label{eqD9}
\end{eqnarray}
where we define the dimensionless parameter
\begin{eqnarray}
  \alpha &=& 6.4  E_p \beta \mathcal{T}^{1/2} \left(\frac{\Delta \Omega}{\Delta\Omega_{crit}} \right) \left(1-\frac{\Delta \Omega}{\Delta\Omega_{crit}}\right)^{\frac{1}{4}}\,. \label{eqD10}
\end{eqnarray}
The terms containing $\alpha$ correspond to perturbations of the non-constant mutual friction coefficients.  Equation (\ref{eq15}) is recovered for $\alpha =0$.
The remaining perturbation equations presented in \S\ref{sec2b} are unchanged.

The magnitude of $\alpha$ compares the size of the perturbation terms arising from the nonconstant mutual friction coefficients with the perturbation terms assuming constant coefficients.  To calculate $\alpha$, we approximate $\Delta \Omega\approx \Delta \Omega_{crit}$ as before and use (\ref{eqD7}) to obtain
\begin{eqnarray}
   \alpha  &=& 4.6 \left( E_p \beta \mathcal{T}^{1/2} \right)^{4/5} \left[ \ln \left( 4 \tau_{sd} \,\Delta \Omega_{crit} \mathcal{R}_n \right)\right]^{1/5} \,. 
\end{eqnarray}
The dependence on the logarithmic factor is weak, and therefore
\begin{eqnarray}
   \alpha  \simeq 6 \times 10^3 \, \left(\frac{E_p}{100\,{\rm MeV}}\right)^{4/5}\left(\frac{k_B T}{10\,{\rm keV}}\right)^{-4/5}\left(\frac{\mathcal{T}}{0.12}\right)^{2/5} \,. \label{eqD11}
\end{eqnarray}
As $\alpha \gg 1$, the nonconstant mutual friction coefficients may have a significant impact on the instability results.

We now revisit the results in the paper by considering the full dispersion relation derived in \S\ref{sec2b} with the mutual friction force (\ref{eqD9}) accounting for thermal activation.  Examining the solutions to the dispersion relation numerically, we find that thermal activation introduces no new instabilities.  Only the Donnelly--Glaberson instability is significantly modified by thermal activation and we repeat the analysis of \S\ref{sec4bb} here. 
The dispersion relation (\ref{eq85z}) generalizes to
\begin{eqnarray}
\omega^2+\left( B_r + i B_i\right)\omega +C =0\,, \label{eqD15}
\end{eqnarray}
where 
\begin{eqnarray}
B_r&=&-\left(2+\alpha\right) \left(1-\mathcal{B}'_{n0}\right) k_z \Delta v_z  \,,  \nonumber \\
B_i&=& 2 \mathcal{B}_{n0} \left( \nu_n k_z^2 +2\Omega_n+\alpha  \Omega_n\right) \,, \nonumber \\
C&=&\left[\mathcal{B}_{n0}^2+ \left(1-\mathcal{B}'_{n0}\right)^2 \right]\left[ \left(1+\alpha\right)\left(\Delta v_z k_z\right)^2 - \left(\nu_n k_z^2 +2\Omega_n \right)\left(\nu_n k_z^2 +2\Omega_n + 2\alpha \Omega_n  \right)\right]\,, 
\end{eqnarray}
are all real numbers.
After separating out the real and imaginary parts, the unstable solution to (\ref{eqD15}) can be written as
\begin{eqnarray}
  \omega&=&-\frac{B_r}{2} - \frac{1}{2\sqrt{2}}\sqrt{\sqrt{\left(B_r^2-B_i^2-4C\right)^2+\left(2 B_r B_i\right)^2}+\left(B_r^2-B_i^2-4C\right)} \nonumber \\
&-& i \left[ \frac{B_i}{2} - \frac{1}{2\sqrt{2}}\sqrt{\sqrt{\left(B_r^2-B_i^2-4C\right)^2+\left(2 B_r B_i\right)^2}-\left(B_r^2-B_i^2-4C\right)}  \right] \,. \label{eqD20}
\end{eqnarray}
For instability, we require that the imaginary component of (\ref{eqD20}) is positive, which occurs for $C>0$.  
This gives values of $k_z^2$ between the two solutions:
\begin{eqnarray}
  k_\pm^2 = \frac{\left(1+\alpha\right) \Delta v_z^2-2 \left(2+\alpha  \right) \Omega_n \nu_n \pm\sqrt{\left[\left(1+\alpha\right) \Delta v_z^2-2 \left(2+\alpha  \right) \Omega_n \nu_n\right]^2-16\left(1+\alpha\right) \Omega_n^2\nu_n^2}}{2\nu_n^2} \,. \label{eqD21}
\end{eqnarray}
For real and distinct bounds, the square root term in (\ref{eqD21}) must give a real number, which happens when
\begin{eqnarray}
  \Delta v_z \geq \sqrt{2 \Omega_n \nu_n} \left(1+\frac{1}{\sqrt{1+\alpha }}\right) \,. \label{eqD22}
\end{eqnarray}
The results (\ref{eqD15})--(\ref{eqD22}) generalize the Donnelly--Glaberson instability to account for the thermal activation of pinned vorticity.  For $\alpha =0$, the classic Donnelly--Glaberson instability conditions in \S\ref{sec4bb} are recovered. 

We now evaluate the critical wavenumber for this instability, accounting for thermal activation.
In the outer core of a neutron star, $\alpha \gg 1$;  see Equation (\ref{eqD11}).  
In this limit, the instability condition (\ref{eqD22}) gives half that of (\ref{eq106c}).
Therefore, the critical wobble angle (\ref{eq122f}) required to excite the Donnelly--Glaberson instability is virtually unaffected by thermal activation.

In the limit $\Delta v_z \gg \sqrt{2 \Omega_n \nu_n }$, the lower bound $k_-$ is unchanged from the classic result (\ref{eq123a}).
The upper bound is approximately $k_+=\Delta v_z \sqrt{1+\alpha} /\nu_n$, which for $\alpha\gg 1$ gives
\begin{eqnarray}
  k_+ R <  10^{18} \left(\frac{\Delta v_z}{\Omega_n R} \right) \left(\frac{\alpha}{6\times 10^3}\right)^{1/2} \left(\frac{\nu_n}{4\times 10^{-3} \,{\rm cm^2\,s^{-1}}}\right)^{-1}  \left(\frac{\Omega_n }{20 \pi \,{\rm rad\,s^{-1}}}\right)\left(\frac{R}{10^6\,{\rm cm}}\right)^{2} \,. \label{eqD23}
\end{eqnarray}
This upper bound is increased by a factor of $ \sqrt{1+\alpha}$ by thermal activation, a factor of 100.

\begin{figure*}[ht]
\centering
\includegraphics[width=.4\linewidth]{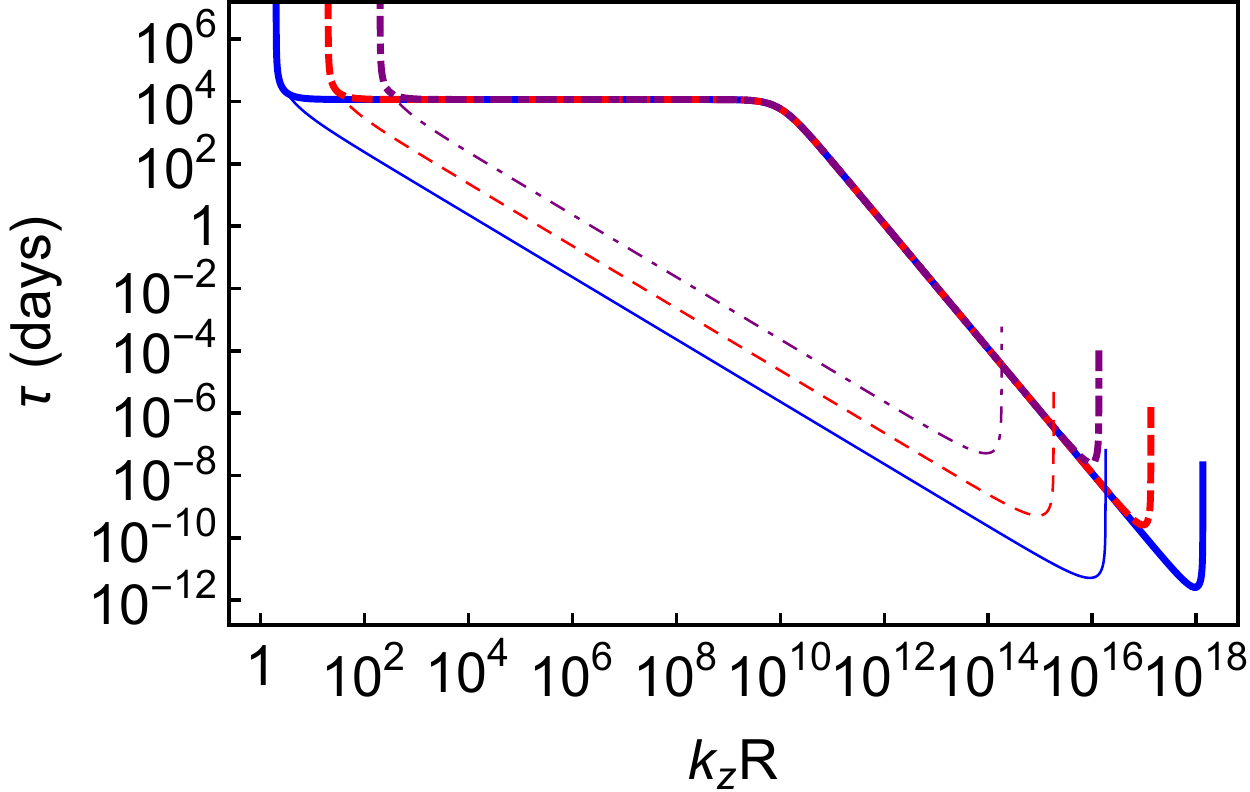} 
\caption{\footnotesize Growth time of the Donnelly--Glaberson instability as a function of dimensionless wavenumber. Thick curves correspond to $\Delta v_z=0.01 \Omega_n R$ (solid), $0.1 \Omega_n R$ (dashed), and $ \Omega_n R$ (dot-dashed).  The thin lines show the corresponding Donnelly--Glaberson instability in the absence of vortex slippage ($\alpha=0$; see (\ref{eqD10}) ). Thermal activation increases the growth time, which is constant in the regime where the hydrodynamic approximation is valid.}
\label{fig7}
\end{figure*}
In Figure \ref{fig7}, we plot the growth time of the unstable mode (\ref{eqD20}).  
This figure demonstrates the modification to the Donnelly--Glaberson instability growth time arising from the thermal activation of pinned vorticity.  
The growth time is plotted for three values of $\Delta v_z$: $\Omega_n R$ (solid), $0.1 \Omega_n R$ (dashed), and $0.01 \Omega_n R$ (dot-dashed).  The thick curves show the growth time including the thermal activation, while the growth times for $\alpha=0$ are shown as thin curves for comparison.

Figure \ref{fig7} shows that the growth time of the Donnelly--Glaberson instability is slowed by thermal activation.  The growth time is independent of $\Delta v_z$; all curves overlap in the unstable range.  For low wavenumber, the growth time is independent of wavenumber.
These features are understood in the limit $\alpha \gg 1$.  Expanding the solution (\ref{eqD20}), the growth time in the unstable range is approximately
\begin{eqnarray}
 \tau \approx \left[ 2\mathcal{B}_{n0} \left( \Omega_n+ \frac{2\Omega_n+\nu_n k_z^2}{\alpha} \right) \right]^{-1}\,. \label{eqD24}
\end{eqnarray}
The growth time (\ref{eqD24}) is approximately independent of wavenumber for $\nu_n k_z^2\ll \alpha \Omega_n$, or wavenumbers satisfying
\begin{eqnarray}
  k_z R \ll 10^{10}  \left(\frac{\alpha}{6\times 10^3}\right)^{1/2}  \left(\frac{\nu_n}{4\times 10^{-3} \,{\rm cm^2\,s^{-1}}}\right)^{-1/2}  \left(\frac{\Omega_n }{20 \pi \,{\rm rad\,s^{-1}}}\right)^{1/2} \left(\frac{R}{10^6\,{\rm cm}}\right)  \,. \label{eqD25}
\end{eqnarray}
For low wavenumbers given by (\ref{eqD25}), equation (\ref{eqD24}) gives the growth time
\begin{eqnarray}
  \tau &=&  30 \,  \left(\frac{\Omega_n }{20 \pi \,{\rm rad\,s^{-1}}}\right)^{-1} \left( \frac{\Delta \Omega_{crit}}{0.1 \,{\rm rad\,s^{-1}}} \right)\left(\frac{\tau_{sd}}{10\,{\rm kyr}}\right) \,  {\rm yrs} \,, \label{eqD26}
\end{eqnarray}
where we have used the scaling relations for the mutual friction coefficient (\ref{eq32a}).
For high wavenumbers, the growth time is approximately
\begin{eqnarray}
  \tau &=&  10^{-7} \,  \left(\frac{\alpha}{6\times 10^3} \right) \left(\frac{k_z R}{ 10^{12} } \right)^{-2}  \left( \frac{\Delta \Omega_{crit}}{0.1 \,{\rm rad\,s^{-1}}} \right)\left(\frac{\tau_{sd}}{10\,{\rm kyr}}\right) \,  {\rm s} \,. 
\end{eqnarray}
Recall that the hydrodynamic approximation breaks down for wavenumbers greater than $2\pi /d_n$, given by (\ref{eq94z}).
Therefore the growth time in the hydrodynamic regime is given by constant and given by (\ref{eqD25}). At high wavenumbers the instability occurs on the level of individual vortices moving under thermal activation up to wavenumbers given by (\ref{eqD23}).


\begin{thebibliography}{}
\expandafter\ifx\csname natexlab\endcsname\relax\def\natexlab#1{#1}\fi
\providecommand{\url}[1]{\href{#1}{#1}}
\providecommand{\dodoi}[1]{doi:~\href{http://doi.org/#1}{\nolinkurl{#1}}}
\providecommand{\doeprint}[1]{\href{http://ascl.net/#1}{\nolinkurl{http://ascl.net/#1}}}
\providecommand{\doarXiv}[1]{\href{https://arxiv.org/abs/#1}{\nolinkurl{https://arxiv.org/abs/#1}}}

\bibitem[{{Akg{\"u}n} {et~al.}(2006){Akg{\"u}n}, {Link}, \&
  {Wasserman}}]{akg06}
{Akg{\"u}n}, T., {Link}, B., \& {Wasserman}, I. 2006, Monthly Notices of the
  Royal Astronomical Society, 365, 653,
  \dodoi{10.1111/j.1365-2966.2005.09745.x}

\bibitem[{{Alpar}(1977)}]{alp77}
{Alpar}, M.~A. 1977, The Astrophysical Journal, 213, 527,
  \dodoi{10.1086/155183}

\bibitem[{{Alpar} {et~al.}(1993){Alpar}, {Chau}, {Cheng}, \& {Pines}}]{alp93}
{Alpar}, M.~A., {Chau}, H.~F., {Cheng}, K.~S., \& {Pines}, D. 1993, The
  Astrophysical Journal, 409, 345, \dodoi{10.1086/172668}

\bibitem[{{Alpar} {et~al.}(1996){Alpar}, {Chau}, {Cheng}, \& {Pines}}]{alp96}
---. 1996, The Astrophysical Journal, 459, 706, \dodoi{10.1086/176935}

\bibitem[{{Alpar} {et~al.}(1984{\natexlab{a}}){Alpar}, {Langer}, \&
  {Sauls}}]{alp84a}
{Alpar}, M.~A., {Langer}, S.~A., \& {Sauls}, J.~A. 1984{\natexlab{a}}, The
  Astrophysical Journal, 282, 533, \dodoi{10.1086/162232}

\bibitem[{{Alpar} {et~al.}(1986){Alpar}, {Nandkumar}, \& {Pines}}]{alp86}
{Alpar}, M.~A., {Nandkumar}, R., \& {Pines}, D. 1986, The Astrophysical
  Journal, 311, 197, \dodoi{10.1086/164765}

\bibitem[{{Alpar} {et~al.}(1984{\natexlab{b}}){Alpar}, {Pines}, {Anderson}, \&
  {Shaham}}]{alp84b}
{Alpar}, M.~A., {Pines}, D., {Anderson}, P.~W., \& {Shaham}, J.
  1984{\natexlab{b}}, The Astrophysical Journal, 276, 325,
  \dodoi{10.1086/161616}

\bibitem[{{Anderson} \& {Itoh}(1975)}]{and75b}
{Anderson}, P.~W., \& {Itoh}, N. 1975, Nature, 256, 25,
  \dodoi{10.1038/256025a0}

\bibitem[{{Andersson}(2003)}]{and03}
{Andersson}, N. 2003, Classical and Quantum Gravity, 20, 105

\bibitem[{{Andersson} {et~al.}(2004){Andersson}, {Comer}, \& {Prix}}]{and04}
{Andersson}, N., {Comer}, G.~L., \& {Prix}, R. 2004, Monthly Notices of the
  Royal Astronomical Society, 354, 101,
  \dodoi{10.1111/j.1365-2966.2004.08166.x}

\bibitem[{{Andersson} {et~al.}(2013){Andersson}, {Glampedakis}, \&
  {Hogg}}]{and13}
{Andersson}, N., {Glampedakis}, K., \& {Hogg}, M. 2013, Physical Review D, 87,
  063007, \dodoi{10.1103/PhysRevD.87.063007}

\bibitem[{{Andreev} \& {Bashkin}(1975)}]{and75a}
{Andreev}, A.~F., \& {Bashkin}, E.~P. 1975, Soviet Journal of Experimental and
  Theoretical Physics, 69, 319

\bibitem[{{Arzoumanian} {et~al.}(1994){Arzoumanian}, {Nice}, {Taylor}, \&
  {Thorsett}}]{arz94}
{Arzoumanian}, Z., {Nice}, D.~J., {Taylor}, J.~H., \& {Thorsett}, S.~E. 1994,
  The Astrophysical Journal, 422, 671, \dodoi{10.1086/173760}

\bibitem[{{Avogadro} {et~al.}(2007){Avogadro}, {Barranco}, {Broglia}, \&
  {Vigezzi}}]{avo07}
{Avogadro}, P., {Barranco}, F., {Broglia}, R.~A., \& {Vigezzi}, E. 2007,
  Physics Review C, 75, 012805, \dodoi{10.1103/PhysRevC.75.012805}

\bibitem[{{Barenghi} {et~al.}(1983){Barenghi}, {Donnelly}, \& {Vinen}}]{bar83}
{Barenghi}, C.~F., {Donnelly}, R.~J., \& {Vinen}, W.~F. 1983, Journal of Low
  Temperature Physics, 52, 189, \dodoi{10.1007/BF00682247}

\bibitem[{{Baym} \& {Chandler}(1983)}]{bay83}
{Baym}, G., \& {Chandler}, E. 1983, Journal of Low Temperature Physics, 50, 57,
  \dodoi{10.1007/BF00681839}

\bibitem[{{Baym} {et~al.}(1969){Baym}, {Pethick}, \& {Pines}}]{bay69a}
{Baym}, G., {Pethick}, C., \& {Pines}, D. 1969, Nature, 224, 673,
  \dodoi{10.1038/224673a0}

\bibitem[{{Boynton} {et~al.}(1972){Boynton}, {Groth}, {Hutchinson}, {Nanos},
  {Partridge}, \& {Wilkinson}}]{boy72}
{Boynton}, P.~E., {Groth}, E.~J., {Hutchinson}, D.~P., {et~al.} 1972, The
  Astrophysical Journal, 175, 217, \dodoi{10.1086/151550}

\bibitem[{{Braithwaite}(2009)}]{bra09}
{Braithwaite}, J. 2009, Monthly Notices of the Royal Astronomical Society, 397,
  763, \dodoi{10.1111/j.1365-2966.2008.14034.x}

\bibitem[{{Braithwaite} \& {Nordlund}(2006)}]{bra06}
{Braithwaite}, J., \& {Nordlund}, {\AA}. 2006, Astronomy and Astrophysics, 450,
  1077, \dodoi{10.1051/0004-6361:20041980}

\bibitem[{{Chamel} \& {Haensel}(2006)}]{cha06}
{Chamel}, N., \& {Haensel}, P. 2006, Physical Review C, 73, 045802,
  \dodoi{10.1103/PhysRevC.73.045802}

\bibitem[{{Chandler} \& {Baym}(1986)}]{cha86}
{Chandler}, E., \& {Baym}, G. 1986, Journal of Low Temperature Physics, 62,
  119, \dodoi{10.1007/BF00681323}

\bibitem[{{Chau} {et~al.}(1992){Chau}, {Cheng}, \& {Ding}}]{cha92}
{Chau}, H.~F., {Cheng}, K.~S., \& {Ding}, K.~Y. 1992, The Astrophysical
  Journal, 399, 213, \dodoi{10.1086/171917}

\bibitem[{{Cheng}(1987{\natexlab{a}})}]{che87a}
{Cheng}, K.~S. 1987{\natexlab{a}}, The Astrophysical Journal, 321, 805,
  \dodoi{10.1086/165673}

\bibitem[{{Cheng}(1987{\natexlab{b}})}]{che87b}
---. 1987{\natexlab{b}}, The Astrophysical Journal, 321, 799,
  \dodoi{10.1086/165672}

\bibitem[{{Cordes} \& {Downs}(1985)}]{cor85}
{Cordes}, J.~M., \& {Downs}, G.~S. 1985, The Astrophysical Journal Supplement
  Series, 59, 343, \dodoi{10.1086/191076}

\bibitem[{{Cordes} \& {Helfand}(1980)}]{cor80c}
{Cordes}, J.~M., \& {Helfand}, D.~J. 1980, The Astrophysical Journal, 239, 640,
  \dodoi{10.1086/158150}

\bibitem[{{Cutler} \& {Lindblom}(1987)}]{cut87}
{Cutler}, C., \& {Lindblom}, L. 1987, The Astrophysical Journal, 314, 234,
  \dodoi{10.1086/165052}

\bibitem[{{Cutler} {et~al.}(2003){Cutler}, {Ushomirsky}, \& {Link}}]{cut03}
{Cutler}, C., {Ushomirsky}, G., \& {Link}, B. 2003, The Astrophysical Journal,
  588, 975, \dodoi{10.1086/368308}

\bibitem[{{D'Alessandro} {et~al.}(1995){D'Alessandro}, {McCulloch}, {Hamilton},
  \& {Deshpande}}]{dal95}
{D'Alessandro}, F., {McCulloch}, P.~M., {Hamilton}, P.~A., \& {Deshpande},
  A.~A. 1995, Monthly Notices of the Royal Astronomical Society, 277, 1033,
  \dodoi{10.1093/mnras/277.3.1033}

\bibitem[{{Ding} {et~al.}(1993){Ding}, {Cheng}, \& {Chau}}]{din93}
{Ding}, K.~Y., {Cheng}, K.~S., \& {Chau}, H.~F. 1993, The Astrophysical
  Journal, 408, 167, \dodoi{10.1086/172577}

\bibitem[{{Dodson} {et~al.}(2002){Dodson}, {McCulloch}, \& {Lewis}}]{dod02}
{Dodson}, R.~G., {McCulloch}, P.~M., \& {Lewis}, D.~R. 2002, The Astrophysical
  Journal, 564, L85, \dodoi{10.1086/339068}

\bibitem[{{Donati} \& {Pizzochero}(2006)}]{don06}
{Donati}, P., \& {Pizzochero}, P.~M. 2006, Physics Letters B, 640, 74,
  \dodoi{10.1016/j.physletb.2006.07.047}

\bibitem[{{Donnelly}(2005)}]{don05}
{Donnelly}, R.~J. 2005, {Quantized Vortices in Helium II} (Cambridge University
  Press), 364

\bibitem[{{Easson}(1979)}]{eas79a}
{Easson}, I. 1979, The Astrophysical Journal, 228, 257, \dodoi{10.1086/156842}

\bibitem[{{Easson} \& {Pethick}(1977)}]{eas77}
{Easson}, I., \& {Pethick}, C.~J. 1977, Physical Review D, 16, 275,
  \dodoi{10.1103/PhysRevD.16.275}

\bibitem[{{Epstein} \& {Baym}(1988)}]{eps88}
{Epstein}, R.~I., \& {Baym}, G. 1988, The Astrophysical Journal, 328, 680,
  \dodoi{10.1086/166325}

\bibitem[{{Espinoza} {et~al.}(2011){Espinoza}, {Lyne}, {Stappers}, \&
  {Kramer}}]{esp11}
{Espinoza}, C.~M., {Lyne}, A.~G., {Stappers}, B.~W., \& {Kramer}, M. 2011,
  Monthly Notices of the Royal Astronomical Society, 414, 1679,
  \dodoi{10.1111/j.1365-2966.2011.18503.x}

\bibitem[{{Flanagan}(1990)}]{fla90}
{Flanagan}, C.~S. 1990, Nature, 345, 416, \dodoi{10.1038/345416a0}

\bibitem[{{Glaberson} {et~al.}(1974){Glaberson}, {Johnson}, \&
  {Ostermeier}}]{gla74}
{Glaberson}, W.~I., {Johnson}, W.~W., \& {Ostermeier}, R.~M. 1974, Physical
  Review Letters, 33, 1197, \dodoi{10.1103/PhysRevLett.33.1197}

\bibitem[{{Glampedakis} \& {Andersson}(2009)}]{gla09}
{Glampedakis}, K., \& {Andersson}, N. 2009, Physical Review Letters, 102,
  141101, \dodoi{10.1103/PhysRevLett.102.141101}

\bibitem[{{Glampedakis} {et~al.}(2008){Glampedakis}, {Andersson}, \&
  {Jones}}]{gla08}
{Glampedakis}, K., {Andersson}, N., \& {Jones}, D.~I. 2008, Physical Review
  Letters, 100, 081101, \dodoi{10.1103/PhysRevLett.100.081101}

\bibitem[{{Glampedakis} {et~al.}(2011){Glampedakis}, {Andersson}, \&
  {Samuelsson}}]{gla11}
{Glampedakis}, K., {Andersson}, N., \& {Samuelsson}, L. 2011, Monthly Notices
  of the Royal Astronomical Society, 410, 805,
  \dodoi{10.1111/j.1365-2966.2010.17484.x}

\bibitem[{{Graber} {et~al.}(2015){Graber}, {Andersson}, {Glampedakis}, \&
  {Lander}}]{gra15}
{Graber}, V., {Andersson}, N., {Glampedakis}, K., \& {Lander}, S.~K. 2015,
  Monthly Notices of the Royal Astronomical Society, 453, 671,
  \dodoi{10.1093/mnras/stv1648}

\bibitem[{{Greenspan}(1968)}]{gre68}
{Greenspan}, H.~P. 1968, {The Theory of Rotating Fluids (Cambridge Monographs
  on Mechanics and Applied Mathematics)} (Cambridge University Press)

\bibitem[{{Greenspan} \& {Howard}(1963)}]{gre63}
{Greenspan}, H.~P., \& {Howard}, L.~N. 1963, Journal of Fluid Mechanics, 17,
  385, \dodoi{10.1017/S0022112063001415}

\bibitem[{{Greenstein}(1970)}]{gre70}
{Greenstein}, G. 1970, Nature, 227, 791, \dodoi{10.1038/227791a0}

\bibitem[{{Gusakov} \& {Kantor}(2013)}]{gus13}
{Gusakov}, M.~E., \& {Kantor}, E.~M. 2013, Physical Review D, 88, 101302,
  \dodoi{10.1103/PhysRevD.88.101302}

\bibitem[{{Haber} {et~al.}(2016){Haber}, {Schmitt}, \& {Stetina}}]{hab16}
{Haber}, A., {Schmitt}, A., \& {Stetina}, S. 2016, Physical Review D, 93,
  025011, \dodoi{10.1103/PhysRevD.93.025011}

\bibitem[{{Hall}(1960)}]{hal60}
{Hall}, H.~E. 1960, Advances in Physics, 9, 89,
  \dodoi{10.1080/00018736000101169}

\bibitem[{{Hall} \& {Vinen}(1956{\natexlab{a}})}]{hal56a}
{Hall}, H.~E., \& {Vinen}, W.~F. 1956{\natexlab{a}}, Royal Society of London
  Proceedings Series A, 238, 204

\bibitem[{{Hall} \& {Vinen}(1956{\natexlab{b}})}]{hal56b}
---. 1956{\natexlab{b}}, Royal Society of London Proceedings Series A, 238, 215

\bibitem[{{Harvey} {et~al.}(1986){Harvey}, {Ruderman}, \& {Shaham}}]{har86}
{Harvey}, J.~A., {Ruderman}, M.~A., \& {Shaham}, J. 1986, Physical Review D,
  33, 2084, \dodoi{10.1103/PhysRevD.33.2084}

\bibitem[{{Hills} \& {Roberts}(1977)}]{hil77}
{Hills}, R.~N., \& {Roberts}, P.~H. 1977, Archive for Rational Mechanics and
  Analysis, 66, 43, \dodoi{10.1007/BF00250851}

\bibitem[{{Hobbs} {et~al.}(2006){Hobbs}, {Lyne}, \& {Kramer}}]{hob06b}
{Hobbs}, G., {Lyne}, A., \& {Kramer}, M. 2006, Chinese Journal of Astronomy and
  Astrophysics Supplement, 6, 169

\bibitem[{{Hobbs} {et~al.}(2010){Hobbs}, {Lyne}, \& {Kramer}}]{hob10}
{Hobbs}, G., {Lyne}, A.~G., \& {Kramer}, M. 2010, Monthly Notices of the Royal
  Astronomical Society, 402, 1027, \dodoi{10.1111/j.1365-2966.2009.15938.x}

\bibitem[{{Janssen} \& {Stappers}(2006)}]{jan06}
{Janssen}, G.~H., \& {Stappers}, B.~W. 2006, Astronomy \& Astrophysics, 457,
  611, \dodoi{10.1051/0004-6361:20065267}

\bibitem[{{Jones}(1987)}]{jon87}
{Jones}, P.~B. 1987, Monthly Notice of the Royal Astronomical Society, 228,
  513, \dodoi{10.1093/mnras/228.3.513}

\bibitem[{{Jones}(1990)}]{jon90}
---. 1990, Monthly Notice of the Royal Astronomical Society, 246, 364

\bibitem[{{Jones}(1991)}]{jon91}
---. 1991, The Astrophysical Journal, 373, 208, \dodoi{10.1086/170038}

\bibitem[{{Kantor} \& {Gusakov}(2014)}]{kan14}
{Kantor}, E.~M., \& {Gusakov}, M.~E. 2014, Monthly Notice of the Royal
  Astronomical Society, 442, L90, \dodoi{10.1093/mnrasl/slu061}

\bibitem[{{Khalatnikov}(1965)}]{kha65}
{Khalatnikov}, I.~M. 1965, {Introduction to the Theory of Superfluidity}
  ({Benjamin, New York})

\bibitem[{{Larson} \& {Link}(2002)}]{lar02}
{Larson}, M.~B., \& {Link}, B. 2002, Monthly Notices of the Royal Astronomical
  Society, 333, 613, \dodoi{10.1046/j.1365-8711.2002.05439.x}

\bibitem[{{Link}(2007)}]{lin07}
{Link}, B. 2007, Astrophysics and Space Science, 308, 435,
  \dodoi{10.1007/s10509-007-9315-0}

\bibitem[{{Link}(2009)}]{lin09}
---. 2009, Physical Review Letters, 102, 131101,
  \dodoi{10.1103/PhysRevLett.102.131101}

\bibitem[{{Link}(2012{\natexlab{a}})}]{lin12b}
---. 2012{\natexlab{a}}, Monthly Notices of the Royal Astronomical Society,
  422, 1640, \dodoi{10.1111/j.1365-2966.2012.20740.x}

\bibitem[{{Link}(2012{\natexlab{b}})}]{lin12a}
---. 2012{\natexlab{b}}, Monthly Notices of the Royal Astronomical Society,
  421, 2682, \dodoi{10.1111/j.1365-2966.2012.20498.x}

\bibitem[{{Link}(2014)}]{lin14}
---. 2014, The Astrophysical Journal, 789, 141,
  \dodoi{10.1088/0004-637X/789/2/141}

\bibitem[{{Link} \& {Epstein}(1996)}]{lin96}
{Link}, B., \& {Epstein}, R.~I. 1996, The Astrophysical Journal, 457, 844,
  \dodoi{10.1086/176779}

\bibitem[{{Liu} {et~al.}(2011){Liu}, {Verbiest}, {Kramer}, {Stappers}, {van
  Straten}, \& {Cordes}}]{liu11}
{Liu}, K., {Verbiest}, J.~P.~W., {Kramer}, M., {et~al.} 2011, Monthly Notices
  of the Royal Astronomical Society, 417, 2916,
  \dodoi{10.1111/j.1365-2966.2011.19452.x}

\bibitem[{{Lyne} {et~al.}(2010){Lyne}, {Hobbs}, {Kramer}, {Stairs}, \&
  {Stappers}}]{lyn10}
{Lyne}, A., {Hobbs}, G., {Kramer}, M., {Stairs}, I., \& {Stappers}, B. 2010,
  Science, 329, 408, \dodoi{10.1126/science.1186683}

\bibitem[{{Mastrano} \& {Melatos}(2005)}]{mas05}
{Mastrano}, A., \& {Melatos}, A. 2005, Monthly Notices of the Royal
  Astronomical Society, 361, 927, \dodoi{10.1111/j.1365-2966.2005.09219.x}

\bibitem[{{McCulloch} {et~al.}(1990){McCulloch}, {Hamilton}, {McConnell}, \&
  {King}}]{mcc90}
{McCulloch}, P.~M., {Hamilton}, P.~A., {McConnell}, D., \& {King}, E.~A. 1990,
  Nature, 346, 822, \dodoi{10.1038/346822a0}

\bibitem[{{McCulloch} {et~al.}(1987){McCulloch}, {Klekociuk}, {Hamilton}, \&
  {Royle}}]{mcc87}
{McCulloch}, P.~M., {Klekociuk}, A.~R., {Hamilton}, P.~A., \& {Royle}, G.~W.~R.
  1987, Australian Journal of Physics, 40, 725

\bibitem[{{Melatos}(2012)}]{mel12}
{Melatos}, A. 2012, The Astrophysical Journal, 761, 32,
  \dodoi{10.1088/0004-637X/761/1/32}

\bibitem[{{Melatos} \& {Link}(2014)}]{mel14}
{Melatos}, A., \& {Link}, B. 2014, Monthly Notices of the Royal Astronomical
  Society, 437, 21, \dodoi{10.1093/mnras/stt1828}

\bibitem[{{Melatos} \& {Peralta}(2007)}]{mel07}
{Melatos}, A., \& {Peralta}, C. 2007, The Astrophysical Journal Letters, 662,
  L99, \dodoi{10.1086/518598}

\bibitem[{{Melatos} {et~al.}(2008){Melatos}, {Peralta}, \& {Wyithe}}]{mel08}
{Melatos}, A., {Peralta}, C., \& {Wyithe}, J.~S.~B. 2008, The Astrophysical
  Journal, 672, 1103, \dodoi{10.1086/523349}

\bibitem[{{Melatos} \& {Warszawski}(2009)}]{mel09}
{Melatos}, A., \& {Warszawski}, L. 2009, The Astrophysical Journal, 700, 1524,
  \dodoi{10.1088/0004-637X/700/2/1524}

\bibitem[{{Mendell}(1991{\natexlab{a}})}]{men91a}
{Mendell}, G. 1991{\natexlab{a}}, The Astrophysical Journal, 380, 515,
  \dodoi{10.1086/170609}

\bibitem[{{Mendell}(1991{\natexlab{b}})}]{men91b}
---. 1991{\natexlab{b}}, The Astrophysical Journal, 380, 530,
  \dodoi{10.1086/170610}

\bibitem[{{Mendell}(1998)}]{men98}
---. 1998, Monthly Notices of the Royal Astronomical Society, 296, 903

\bibitem[{{Migdal}(1959)}]{mig59}
{Migdal}, A. 1959, Nuclear Physics A, 13, 655,
  \dodoi{10.1016/0029-5582(59)90264-0}

\bibitem[{{Passamonti} {et~al.}(2017){Passamonti}, {Akg{\"u}n}, {Pons}, \&
  {Miralles}}]{pas17b}
{Passamonti}, A., {Akg{\"u}n}, T., {Pons}, J.~A., \& {Miralles}, J.~A. 2017,
  Monthly Notices of the Royal Astronomical Society, 469, 4979,
  \dodoi{10.1093/mnras/stx1192}

\bibitem[{{Passamonti} {et~al.}(2016){Passamonti}, {Andersson}, \&
  {Ho}}]{pas16}
{Passamonti}, A., {Andersson}, N., \& {Ho}, W.~C.~G. 2016, Monthly Notices of
  the Royal Astronomical Society, 455, 1489, \dodoi{10.1093/mnras/stv2149}

\bibitem[{{Peralta} \& {Melatos}(2009)}]{per09}
{Peralta}, C., \& {Melatos}, A. 2009, The Astrophysical Journal Letters, 701,
  L75, \dodoi{10.1088/0004-637X/701/2/L75}

\bibitem[{{Peralta} {et~al.}(2006){Peralta}, {Melatos}, {Giacobello}, \&
  {Ooi}}]{per06a}
{Peralta}, C., {Melatos}, A., {Giacobello}, M., \& {Ooi}, A. 2006, The
  Astrophysical Journal, 651, 1079, \dodoi{10.1086/507576}

\bibitem[{{Peralta} {et~al.}(2008){Peralta}, {Melatos}, {Giacobello}, \&
  {Ooi}}]{per08}
---. 2008, Journal of Fluid Mechanics, 609, 221,
  \dodoi{10.1017/S002211200800236X}

\bibitem[{{Peralta}(2007)}]{per07}
{Peralta}, C.~A. 2007, PhD thesis, University of Melbourne, Australia

\bibitem[{{Qiao} {et~al.}(2003){Qiao}, {Xue}, {Xu}, {Wang}, \& {Xiao}}]{qia03}
{Qiao}, G.~J., {Xue}, Y.~Q., {Xu}, R.~X., {Wang}, H.~G., \& {Xiao}, B.~W. 2003,
  Astronomy \& Astrophysics, 407, L25, \dodoi{10.1051/0004-6361:20031055}

\bibitem[{{Radhakrishnan} \& {Manchester}(1969)}]{rad69}
{Radhakrishnan}, V., \& {Manchester}, R.~N. 1969, Nature, 222, 228,
  \dodoi{10.1038/222228a0}

\bibitem[{{Ruderman} {et~al.}(1998){Ruderman}, {Zhu}, \& {Chen}}]{rud98}
{Ruderman}, M., {Zhu}, T., \& {Chen}, K. 1998, The Astrophysical Journal, 492,
  267, \dodoi{10.1086/305026}

\bibitem[{{Shternin} \& {Yakovlev}(2008)}]{sht08}
{Shternin}, P.~S., \& {Yakovlev}, D.~G. 2008, Physical Review D, 78, 063006,
  \dodoi{10.1103/PhysRevD.78.063006}

\bibitem[{{Sidery} \& {Alpar}(2009)}]{sid09b}
{Sidery}, T., \& {Alpar}, M.~A. 2009, Monthly Notices of the Royal Astronomical
  Society, 400, 1859, \dodoi{10.1111/j.1365-2966.2009.15575.x}

\bibitem[{{Sidery} {et~al.}(2008){Sidery}, {Andersson}, \& {Comer}}]{sid08}
{Sidery}, T., {Andersson}, N., \& {Comer}, G.~L. 2008, Monthly Notices of the
  Royal Astronomical Society, 385, 335,
  \dodoi{10.1111/j.1365-2966.2007.12805.x}

\bibitem[{{Srinivasan} {et~al.}(1990){Srinivasan}, {Bhattacharya}, {Muslimov},
  \& {Tsygan}}]{sri90}
{Srinivasan}, G., {Bhattacharya}, D., {Muslimov}, A.~G., \& {Tsygan}, A.~J.
  1990, Current Science, 59, 31

\bibitem[{{Stairs} {et~al.}(2000){Stairs}, {Lyne}, \& {Shemar}}]{sta00}
{Stairs}, I.~H., {Lyne}, A.~G., \& {Shemar}, S.~L. 2000, Nature, 406, 484,
  \dodoi{10.1038/35020010}

\bibitem[{{Tkachenko}(1966)}]{tka66}
{Tkachenko}, V.~K. 1966, Soviet Journal of Experimental and Theoretical
  Physics, 23, 1049

\bibitem[{{Urama} {et~al.}(2006){Urama}, {Link}, \& {Weisberg}}]{ura06}
{Urama}, J.~O., {Link}, B., \& {Weisberg}, J.~M. 2006, Monthly Notices of the
  Royal Astronomical Society, 370, L76,
  \dodoi{10.1111/j.1745-3933.2006.00192.x}

\bibitem[{{van Eysden}(2014)}]{van14}
{van Eysden}, C.~A. 2014, The Astrophysical Journal, 789, 142,
  \dodoi{10.1088/0004-637X/789/2/142}

\bibitem[{{van Eysden} \& {Melatos}(2010)}]{van10}
{van Eysden}, C.~A., \& {Melatos}, A. 2010, Monthly Notices of the Royal
  Astronomical Society, 409, 1253, \dodoi{10.1111/j.1365-2966.2010.17387.x}

\bibitem[{{van Hoven} \& {Levin}(2008)}]{van08}
{van Hoven}, M., \& {Levin}, Y. 2008, Monthly Notices of the Royal Astronomical
  Society, 391, 283, \dodoi{10.1111/j.1365-2966.2008.13881.x}

\bibitem[{{Warszawski} \& {Melatos}(2008)}]{war08}
{Warszawski}, L., \& {Melatos}, A. 2008, Monthly Notices of the Royal
  Astronomical Society, 390, 175, \dodoi{10.1111/j.1365-2966.2008.13662.x}

\bibitem[{{Wong} {et~al.}(2001){Wong}, {Backer}, \& {Lyne}}]{won01}
{Wong}, T., {Backer}, D.~C., \& {Lyne}, A.~G. 2001, The Astrophysical Journal,
  548, 447, \dodoi{10.1086/318657}

\end{thebibliography}
\end{document}